\documentclass[12pt]{article}

\usepackage[left=1.5cm,right=1.5cm,top=1.9cm,bottom=1.9cm]{geometry}

\usepackage{graphicx,psfrag} 
\graphicspath{{images}}
\usepackage[figurename=Fig.]{caption}
\usepackage{amsmath,amsfonts,amssymb,amsthm,mathtools} 
\usepackage{polynom}
\usepackage{xfrac}
\usepackage{mathrsfs} 
\usepackage{array}
\usepackage{floatrow}
\usepackage{floatflt}
\usepackage[normalem]{ulem}
\usepackage[inline,sizes]{enumitem}
\usepackage[bottom]{footmisc}
\usepackage{framed}

\usepackage{multicol}
\usepackage{icomma} 
\usepackage{epigraph}
\usepackage{indentfirst} 
\usepackage{cite}

\usepackage{fancyhdr}
\numberwithin{equation}{section}
\makeatletter
\def\Appendix{\appendix
  \def\@seccntformat##1{Appendix~\csname the##1\endcsname.~~}}
\makeatother

\usepackage[usenames]{color}
\usepackage{colortbl}
\usepackage{tikz}
\usepackage{cancel}
\tikzset{elliparc/.style args={#1:#2:#3}{%
		insert path={++(#1:#3) arc (#1:#2:#3)}}} 
\tikzset{->-/.style={decoration={
			markings,
			mark=at position .5 with {\arrow{>}}},postaction={decorate}}}
	
\usepackage{tikzsymbols}
\usepackage{pgf}
\usepackage{pgfplots}
\pgfplotsset{compat=1.18}
\usepgfplotslibrary{fillbetween} 
\usepackage{tkz-euclide}
\usetikzlibrary{calc,intersections,shapes.arrows,arrows.meta,patterns,angles,quotes}
\usepackage{xcolor-material}
\usetikzlibrary{fit}
\usetikzlibrary{patterns} 
\usepackage{circledsteps}

\usepackage{subcaption} 

\usepackage[hypertexnames=false]{hyperref}
\hypersetup{
    colorlinks=true,
	linkcolor=myblue,
	filecolor=magenta,      
	urlcolor=myblue,
	citecolor=blue,
}

\usepackage{footnotehyper} 

\makeatletter
\newcommand*{\rom}[1]{\expandafter\@slowromancap\romannumeral #1@}
\makeatother

\renewcommand{\Im}{\mathop{\mathrm{Im}}\nolimits}
\renewcommand{\Re}{\mathop{\mathrm{Re}}\nolimits}
\newcommand{\Res}{\mathop{\mathrm{Res}}\nolimits}

\newcommand{\gh}{\text{gh}}

\definecolor{mygreen} {RGB} {56, 140, 70}

\definecolor{myblue} {RGB} {45, 112, 179}

\let\oldforall\forall
\renewcommand{\forall}{\oldforall\,}
\let\oldexists\exists
\renewcommand{\exists}{\oldexists\,}

\newcommand{\cc}{\text{c.\,c.}}

\begin{document}
\title{\textbf{On correlation numbers $V_{0,4}$ and $V_{1,1}$\\ in Virasoro Minimal String Theory}\vspace*{0.3cm}}
\date{}
\author{Dmitry Khromov$^{1,2,3}$ and Alexey Litvinov$^{1}$
\\[\medskipamount]
\parbox[t]{0.85\textwidth}{\normalsize\it\centerline{1. Skolkovo Institute of Science and Technology, 121205 Moscow, Russia}}
\\
\parbox[t]{0.85\textwidth}{\normalsize\it\centerline{2. Landau Institute for Theoretical Physics, 142432 Chernogolovka, Russia}}
\\
\parbox[t]{0.85\textwidth}{\normalsize\it\centerline{3. Moscow Institute of Physics and Technology, 141700 Dolgoprudny, Russia}}}
\maketitle
\begin{abstract}
     We study correlation numbers in Virasoro minimal string \cite{Collier:2023cyw}. Using analytic properties of correlation functions in spacelike and timelike Liouville theories, we verify exact expressions for correlation numbers for the four punctured sphere and the once punctured torus conjectured in \cite{Collier:2023cyw}. 
\end{abstract}
\tableofcontents

\section{Introduction}
In Polyakov's approach \cite{Polyakov:1981rd,Knizhnik:1988ak,David:1988hj,Distler:1988jt}, bosonic string theory is formulated as a conformal field theory coupled to two-dimensional quantum gravity. In the case of $D$ free bosons for critical dimension $D=26$, gravity essentially decouples, so one can study strings without having to solve the two-dimensional quantum gravity. At $D\neq26$ due to the conformal anomaly a Liouville mode emerges. Formally, the action of the resulting theory, known as either non-critical string theory or $2D$ quantum gravity, is a direct sum of three theories, a given CFT (sometimes referred to as matter CFT), Liouville CFT and $BC$ ghost system, which interact only through the central charge balance 
\begin{equation}\label{2D-gravity-action}
S=S_{\textrm{CFT}}+S_{\textrm{L}}+S_{\textrm{ghosts}},\quad
c_{\textrm{CFT}}+c_{\textrm{L}}+c_{\textrm{ghosts}}=0.
\end{equation}

Physical observables in the theory \eqref{2D-gravity-action} correspond to BRST cohomology. The simplest class of observables with ghost number zero, are obtained from primary fields $\Phi_{\Delta}(x)$ of the matter CFT dressed by Liouville primaries $V_{1-\Delta}(x)$, producing $(1,1)$ forms
\begin{equation}\label{U-integrated}
    U_{\Delta}(x)=\Phi_{\Delta}(x)V_{1-\Delta}(x).
\end{equation}
and then integrated over the worldsheet. Then, as is common in string theory, the goal is to calculate the amplitudes, or correlation numbers
\begin{equation}\label{correlation-numbers}
    V_{g,n}(\Delta_1,\dots,\Delta_n)=\int_{\mathcal{M}_{g,n}}Z_{\textrm{gh}}\langle U_{\Delta}(x_1)\dots U_{\Delta}(x_n)\rangle_g,
\end{equation}
where $\langle\dots\rangle_g$ is the correlation function on a Riemann surface of the genus $g$ and $Z_{\textrm{gh}}$ is a contribution from the ghost sector.

The most well-known example of \eqref{2D-gravity-action} is the minimal string theory, also known as minimal Liouville gravity. In this theory, the matter content consists of a single minimal model of CFT \cite{Belavin:1984vu}. This theory is believed to be dual to a double-scaling limit of a certain matrix model and most of analytic results for correlation numbers \eqref{correlation-numbers} have been obtained within this duality on the matrix model side (see \cite{DiFrancesco:1993cyw} for a review). Starting from \cite{Zamolodchikov:2005fy} and after the discovery of higher equations of motion in Liouville CFT \cite{Zamolodchikov:2003yb}, it became possible to perform analytical computations on the field theory side \cite{Belavin:2005yj,Artemev:2022rng,Artemev:2022sfi}. The results of \cite{Belavin:2005yj,Artemev:2022rng,Artemev:2022sfi} were tested against predictions from the matrix model side, and it was noticed that they coincide only when the number of conformal blocks is specific.

Another model of 2D quantum gravity has been proposed in the seminal paper \cite{Collier:2023cyw} and is called the Virasoro Minimal String (VMS). In this case, compared to the case mentioned above (minimal string theory/minimal Liouville gravity), the spectrum of the underlying matter CFT, the so-called timelike Liouville CFT \cite{Zamolodchikov:2005fy,Ribault:2015sxa}, is continuous. In \cite{Collier:2023cyw} a remarkably simple closed formula for the amplitudes \eqref{correlation-numbers} in this theory, called volumes, has been conjectured. The volumes \eqref{correlation-numbers}, being polynomials in $\Delta_k$'s, are very similar to Mirzakhani's celebrated formulae for the volumes of the moduli space of curves with geodesic boundaries \cite{Mirzakhani:2006fta}. Moreover, the authors of \cite{Collier:2023cyw} managed to derive a generalization of Mirzakhani recursion relations, which effectively computes these volumes. However, the way they arrived at their results is non-direct and uses the conjectural duality of VMS to matrix models rather than field-theoretic arguments. In two special cases the exact results (here $c$ is the central charge of the Liouville CFT and $P_k$'s are the Liouville CFT momenta) are as follows
\begin{equation}\label{two-exact-results}
\begin{gathered}
    V_{0,4}(\Delta_1,\Delta_2,\Delta_3,\Delta_4)=
    \frac{c-13}{24}+\sum_{k=1}^4P_k^2,\qquad
    V_{1,1}(\Delta_1)=\frac{1}{24}
    \left(\frac{c-13}{24}+P_1^2\right),\\
    c=1+6Q^2,\quad \Delta_k=\frac{Q^2}{4}+P_k^2.
\end{gathered}    
\end{equation}
These expressions were checked numerically in \cite{Collier:2023cyw} with high precision using standard methods of conformal field theory.

In the current notes, we verify \eqref{two-exact-results} using CFT analytic methods. The basic idea is to treat the VMS amplitudes as analytic functions of the external momenta $P_k$, making analytic continuation from the region where the amplitudes are defined. In particular, for $(g,n)=(0,4)$ we show that despite the fact that the integrand in \eqref{correlation-numbers} is a complicated meromorphic function of $P_k$, the residues are exact forms and thus do not contribute to $V_{0,4}$.  Then, assuming polynomial growth of the volume at large momenta,  we demonstrate that it is a polynomial of degree $1$ and compute the coefficients of this polynomial in two exactly solvable cases. The volume $V_{1,1}$ can be obtained from the volume $V_{0,4}$ by means of the well known relation between $4-$point correlation functions on a sphere and $1-$point correlation function on a torus \cite{Fateev:2009me,Hadasz:2009sw}.

The main tool used in our derivation is the higher equations of motion in Liouville CFT \cite{Zamolodchikov:2003yb}, also used in \cite{Belavin:2005yj,Artemev:2022rng,Artemev:2022sfi}, and a variant of higher equations of motion in timelike Liouville CFT \cite{Ribault:2014hia}. We note that similar analytic approach has been used in recent series of papers devoted to studies of another model of quantum gravity~--- Complex Liouville String \cite{Collier:2024kmo,Collier:2024kwt}.

The structure of this paper is as follows. In Section \ref{sec:Liouville}, we review both spacelike (conventional) and timelike Liouville conformal field theories (CFTs). Particular emphasis will be placed on the analytic properties of the correlation functions of primary fields in these theories. In section \ref{sec:VMS}, after introducing the main actor of this paper~--- Virasoro Minimal string, we provide our main result: the analytic computation of two volumes \eqref{two-exact-results}.  In section \ref{sec:Numerics} we provide numerical tests of a certain conjecture formulated in section \ref{sec:VMS}. In Section \ref{sec:Conclusion} we make short concluding remarks. In appendices we collect needed information about the special functions used in the main text.
\section{Liouville theory}\label{sec:Liouville}
There are several ways to define Liouville field theory (LFT)  \cite{Zamolodchikov:1995aa,Ribault:2014hia,Guillarmou:2020wbo}. Here we will focus on the bootstrap approach \cite{Belavin:1984vu}. The idea behind this approach is to use the operator product expansion (OPE) and consistency conditions to constrain correlation functions. In the case of conformal field theories, the constraints are amplified by the conformal symmetry. In two dimensions, they are strong enough to provide a way to, in principle, compute all correlation functions in a given CFT. 
\subsection{Conformal bootstrap for Liouville field theory}
In order to use the conformal bootstrap approach to define Liouville field theory, we first postulate that LFT is a conformal field theory with the central charge $c$ parametrized as
\begin{equation}
	c = 1+6Q^2, \qquad Q = b+\frac{1}{b}, \quad b\in\mathbb{C}.
\end{equation} 
The symmetry algebra of the theory is the product of two Virasoro algebras, holomorphic and anti-holomorphic. The commutation relations for the generators of these algebras are as follows:
\begin{equation}
\begin{aligned}
&[L_n, L_m] = (n-m)L_{n+m}+\frac{c}{12}(n^3-n)\delta_{n+m,0},\\
&[\bar{L}_n, \bar{L}_m] = (n-m)\bar{L}_{n+m}+\frac{c}{12}(n^3-n)\delta_{n+m,0},
\end{aligned} \qquad n, m\in\mathbb{Z}.
\end{equation}

The fields in LFT are combined into conformal families. A conformal family consists of a primary field $V_{P}(z)$ of dimension $\Delta(P)$ and descendant fields. The primary field satisfies
\begin{equation}
    L_0 V_{P}(z) = \Delta(P)V_{P}(z), \qquad L_{n}V_P(z) = 0, \quad \text{for}\quad n>0. 
\end{equation}
Descendant fields are obtained from $V_{P}(z)$ by acting on it with creation operators $L_{-n}$ and $\bar{L}_{-n}$ with $n>0$. Correlation functions of descendant fields may be obtained from correlation functions of the corresponding primary fields by acting with some differential operators \cite{Belavin:1984vu}. 

The conformal dimensions of the primary fields in the spectrum of LFT are
\begin{equation}
\Delta(P) = \bar{\Delta}(P) = \frac{Q^2}{4} + P^2, \qquad P\in\mathbb{R}.
\end{equation}
Clearly, $\Delta(P) = \Delta(-P)$. We demand that the two fields $V_P(z)$ and $V_{-P}(z)$ represent the same primary field. It follows that they must satisfy the reflection relation
\begin{equation}\label{eq:reflection}
	V_P(z) = R_P V_{-P}(z), \qquad \text{where}\qquad R_{P}R_{-P} = 1.
\end{equation}

Using the conformal symmetry, one can derive the coordinate dependence of $2$-point and $3$-point correlation functions of primary fields:
\begin{equation}\label{eq:23point}
\begin{aligned}
\langle V_{P_1}(z_1)V_{P_2}(z_2)\rangle &= \frac{N(P_1)}{|z_1-z_2|^{4\Delta_1}}(\delta(P_1+P_2) + R_{P_1}\delta(P_1-P_2)), \quad \\
\langle V_{P_1}(z_1)V_{P_2}(z_2)V_{P_3}(z_3)\rangle &= \frac{C(P_1, P_2, P_3)}{|z_1-z_2|^{2(\Delta_1+\Delta_2-\Delta_3)}|z_1-z_3|^{2(\Delta_1+\Delta_3-\Delta_2)}|z_2-z_3|^{2(\Delta_2+\Delta_3-\Delta_1)}}.
\end{aligned}
\end{equation}
Here $N(P)$ and $R_P$ depend only on normalization of the fields and $C(P_1, P_2, P_3)$ is a ``dynamical''~quantity, i.\,e. it depends on the theory.

In order to compute correlation functions with more than $3$ field insertions, one needs the notion of the operator algebra. Namely, the product of any two local operators\footnote{In spite of the name, the operators in correlation functions can be commuted freely.} in LFT can be expressed as a sum over the local operators in the spectrum. For the primary fields this operator product expansion (OPE) has the form\footnote{As we will see later the contour of integration in timelike Liouville theory \cite{Ribault:2015sxa} is $\mathbb{R}+i\epsilon$}:
\begin{equation}\label{eq:OPE}
V_{P_1}(z)V_{P_2}(w) = \frac12 \int_{\mathbb{R}} dP\, C^P_{P_1,P_2} |z-w|^{2(\Delta(P) - \Delta(P_1)-\Delta(P_2))} 
\sum\limits_{\boldsymbol{\lambda}, \boldsymbol{\bar{\lambda}}}
(z-w)^{|\boldsymbol{\lambda}|}(\bar{z}-\bar{w})^{|\boldsymbol{\bar{\lambda}}|}
\beta_{\boldsymbol{\lambda}}(P)\beta_{\boldsymbol{\bar{\lambda}}}(P) V^{\boldsymbol{\lambda}, \boldsymbol{\bar{\lambda}}}_{P}(w).
\end{equation}
Here the sum goes over all integer partitions:
\begin{equation}
\boldsymbol{\lambda} = \{\lambda_1, \lambda_2, \ldots\}, \quad \lambda_1\geq\lambda_2\geq\ldots\geq1, \quad 
\boldsymbol{\bar{\lambda}} = \{\bar{\lambda}_1, \bar{\lambda}_2, \ldots\}, \quad \bar{\lambda}_1\geq\bar{\lambda}_2\geq\ldots\geq1, 
\end{equation}
$|\boldsymbol{\lambda}|$, $|\boldsymbol{\bar{\lambda}}|$ are sizes of the partitions and $V^{\boldsymbol{\lambda}, \boldsymbol{\bar{\lambda}}}_{P}$ are descendant fields in the conformal family of $V_{P}$:
\begin{equation}
V^{\boldsymbol{\lambda}, \boldsymbol{\bar{\lambda}}}_{P} \equiv L_{-\boldsymbol{\lambda}}\bar{L}_{-\boldsymbol{\bar{\lambda}}} V_{P} \equiv (L_{-\lambda_1}L_{-\lambda_2}\ldots)(\bar{L}_{-\bar{\lambda}_1}\bar{L}_{-\bar{\lambda}_2}\ldots)V_{P}.
\end{equation}
The empty partition $\boldsymbol{\lambda} = \boldsymbol{\bar{\lambda}} = \varnothing$ corresponds to the primary field itself, $\beta_{\varnothing}(P) = 1$. The coefficients~$\beta_{\boldsymbol{\lambda}}(P)$, $\beta_{\boldsymbol{\bar{\lambda}}}(P)$ are fixed unambiguously by the conformal symmetry \cite{Belavin:1984vu}. They are rational functions of $P$, $P_1$ and $P_2$ and may have poles only in variable $P$ in the following points (corresponding to degenerate fields):
\begin{equation}\label{eq:betapoles}
P = \pm P_{m,n} \equiv \pm \frac{i}2 \left(mb +nb^{-1} \right), \qquad m,n\in\mathbb{Z}_{> 0}.
\end{equation}
The only thing that is not fixed by the conformal symmetry are the structure constants $C_{P_1, P_2}^P$. From \eqref{eq:reflection}, \eqref{eq:23point} and \eqref{eq:OPE} it follows that
\begin{equation}
    C_{P_1, P_2}^{P} = \frac{C(-P, P_1, P_2)}{N(P)}, \quad R_P = \frac{C(P, P_1, P_2)}{C(-P, P_1, P_2)}.
\end{equation}

Now we consider the $4$-point correlation function 
\[
\langle V_{P_1}(z_1) V_{P_2}(z_2)V_{P_3}(z_3) V_{P_4}(z_4) \rangle.
\]
Using the conformal invariance, we can set three of the four coordinates to $0$, $1$ and $\infty$:
\begin{multline}
\langle V_{P_1}(z_1) V_{P_2}(z_2) V_{P_3}(z_3) V_{P_4}(z_4)\rangle 
=
\\
=
|z_{14}|^{-4\Delta_1} |z_{23}|^{2(\Delta_4-\Delta_1-\Delta_2-\Delta_3)}|z_{24}|^{2(\Delta_1+\Delta_3-\Delta_2-\Delta_4)}|z_{34}|^{2(\Delta_1+\Delta_2-\Delta_3-\Delta_4)}
\times 
\\
\times
\langle V_{P_1}(z) V_{P_2}(0) V_{P_3}(1) V_{P_4}(\infty)\rangle, \quad \text{where} \quad z = \frac{z_{12}z_{34}}{z_{14}z_{32}}, \quad z_{ij} = z_i - z_j.
\end{multline}
Here
\begin{equation}\label{eq:4pointinfty}
\langle V_{P_1}(z) V_{P_2}(0) V_{P_3}(1) V_{P_4}(\infty)\rangle = \lim\limits_{\zeta\to\infty}|\zeta|^{4\Delta_4} \langle V_{P_1}(z) V_{P_2}(0) V_{P_3}(1) V_{P_4}(\zeta)\rangle.    
\end{equation}
Using the operator product expansion \eqref{eq:OPE} we can reduce this 4-point function to 3-point functions. If we use it for $V_{P_1}(z) V_{P_2}(0)$ we get the s-channel decomposition:
\begin{equation}\label{eq:s-channel}
\langle V_{P_1}(z) V_{P_2}(0) V_{P_3}(1) V_{P_4}(\infty)\rangle = \frac12\int_\mathbb{R}dP\,\frac{1}{N(P)}\, C(P_1, P_2, -P)C(P, P_3, P_4) \left|\mathfrak{F}^{(b)}(\{P_k\},P|z)\right|^2.
\end{equation}
Here $\{P_k\} = \{P_1, P_2, P_3, P_4\}$ are called ``external momenta'', $P$ is called ``internal momentum'', function $\mathfrak{F}$ is called s-channel conformal block and by $\left|\mathfrak{F}^{(b)}(\{P_k\},P|z)\right|^2$ we denote not the modulus squared of the conformal block, but the product $\mathfrak{F}^{(b)}(\{P_k\},P|z)\mathfrak{F}^{(b)}(\{P_k\},P|\bar{z})$.

We could have used OPE for $V_{P_1}(z)V_{P_3}(1)$. We would then obtain the t-channel decomposition:
\begin{equation}\label{eq:t-channel}
\langle V_{P_1}(z) V_{P_2}(0) V_{P_3}(1) V_{P_4}(\infty)\rangle = \frac12\int_\mathbb{R}dP\,\frac{1}{N(P)}\, C(P_1, P_3, -P)C(P, P_2, P_4) \left|\mathfrak{F}^{(b)}_{t}(\{P_k\},P|z)\right|^2,
\end{equation}
where $\mathfrak{F}_t$ is t-channel conformal block:
\begin{equation}
\mathfrak{F}^{(b)}_{t}(P_1, P_2, P_3, P_4,P|z) = \mathfrak{F}^{(b)}(P_1, P_3, P_2, P_4,P|1-z)
\end{equation}

We demand that the two formulas \eqref{eq:s-channel} and \eqref{eq:t-channel} give the same result. This condition is called crossing symmetry. It should be viewed as a set of constraints on 3-point function $C$ for each $\{P_k\}$. We could also have used the OPE of $V_{P_1}(z)V_{P_4}(\infty)$, however it would not bring any independent conditions. 

The constraints implied by crossing symmetry are very hard to solve analytically in the case of arbitrary $P_k\in\mathbb{R}$. However, as described in the next subsections, they are believed to have a unique smooth solution in case of $b\in\mathbb{R}$ and $b\in i\mathbb{R}$. We will refer to the LFT with $b\in\mathbb{R}$ as ``spacelike LFT'' and to the LFT with $b\in i\mathbb{R}$ as ``timelike LFT'' \cite{Zamolodchikov:2005fy}. Then, using \eqref{eq:OPE} one can in principle compute any correlation function in spacelike or timelike LFT by reducing the number of field insertions in it to $3$.
\subsection{Degenerate fields in Liouville field theory}\label{subsec:DegenerateFields}

In order to find a candidate for the solution of the crossing symmetry constraints we introduce the second operator algebra, that of degenerate fields. Degenerate fields are primary fields that have a vanishing descendant. By Kac-Feigin-Fuchs theorem \cite{Kac:1978ge,Feigin:1981st} for general $b$ these fields $V_{m,n}$ have conformal dimensions
\begin{equation}
\Delta_{m,n} = \Delta(P_{m,n}), \quad P_{m,n} = \frac{i}{2} \left(m b + n b^{-1}\right), \quad m, n\in\mathbb{Z}_{>0}.    
\end{equation}
As one can see, for all $b\in\mathbb{C}\setminus i\mathbb{R}$ the momenta $P_{m,n}$ do not lie in the spectrum of LFT, they a priori form a different operator algebra than the one described above\footnote{As we will see below, in spacelike LFT this operator algebra may be obtained from the one described above by analytic continuation. However, in timelike LFT that is not the case.}. For $b\in i\mathbb{R}$ in spite of $P_{m,n}$ being in the spectrum, we still consider the operator algebra of degenerate fields to be independent.

The vanishing descendant of $V_{m,n}$ has the form $D_{m,n}\bar{D}_{m,n} V_{m,n}$. Here $D_{m,n}$ is a graded polynomial of~$L_{-n}$:
\begin{equation}\label{eq:Dmnb}
    D_{m,n}^{(b)} = L_{-1}^{mn} + c_1(b) L_{-1}^{mn-2}L_{-2} + \ldots
\end{equation}
For the cases $(m,n) = (2,1)$ and $(m,n) = (1,2)$ it has the form
\begin{equation}
D_{2,1}^{(b)} = L_{-1}^2 + b^2 L_{-2}, \qquad D_{1,2}^{(b)} = L_{-1}^2 + b^{-2}L_{-2}.     
\end{equation}
Because of these vanishing descendants, correlation functions with degenerate fields satisfy BPZ differential equations \cite{Belavin:1984vu}. A $4$-point function with degenerate field $V_{m,n}(z)$ satisfies an ordinary differential equation of order $mn$. This equation has $mn$ independent solutions which correspond to $mn$ conformal families in OPE of $V_{m,n}$ with another field\footnote{There may be fewer conformal families in this OPE if the other field is also degenerate.}. This means that the operator algebra of degenerate fields has the form
\begin{equation}\label{eq:VmnVrsOPE}
V_{m,n}(z) V_{r,s}(w) = \sum_{k,l}C_{m,n;\, r,s}^{k,l} |z-w|^{2(\Delta_{k,l}-\Delta_{m,n}-\Delta_{r,s})} [V_{k,l}(w) + \ldots],    
\end{equation}
where by $\ldots$ we denote the contribution of descendant fields and the sum goes over the set
\begin{equation}\label{klTwoDegenerateFieldsFusion}
k\in\{|r-m|+1, |r-m|+3, \ldots, r+m-3, r+m-1\}, \quad l\in\{|s-n|+1, |s-n|+3, \ldots, s+n-3, s+n-1\}.
\end{equation}
As one can see from this OPE, one only needs the fields $V_{2,1}$ and $V_{1,2}$ to obtain all the fields $V_{m,n}$. 

The reason we introduced degenerate fields is to consider correlation functions with both degenerate fields and primary operators in the spectrum. By the same reason described above, OPE of $V_{m,n}$ with a field $V_{P}$ in the spectrum has the form
\begin{equation}\label{eq:VmnOPE}
V_{m,n}(z) V_{P}(w) = \sum_{k,l}C_{m,n}^{k,l}(P) |z-w|^{2(\Delta(P+P_{k,l})-\Delta(P)-\Delta_{m,n})} [V_{P + P_{k,l}}(w) + \ldots],   
\end{equation}
where the sum goes over the set
\begin{equation}\label{klOneDegenerateFieldsFusion}
k\in\{-m+1, -m+3, \ldots, m-3, m-1\}, \quad l\in\{-n+1, -n+3, \ldots, n-3, n-1\}.
\end{equation}
The resulting momenta $P + P_{k, l}$ may not lie in the spectrum. Hence, correlation functions with~$V_{P+P_{k,l}}$ should be thought of as analytic continuations of correlation functions with momenta in the spectrum.

We now turn to 4-point functions with degenerate fields $V_{1,2}$ and $V_{2,1}$:
\[
\langle V_{2,1}(z)V_{P_2}(0)V_{P_3}(1)V_{P_4}(\infty)\rangle, \quad
\langle V_{1,2}(z)V_{P_2}(0)V_{P_3}(1)V_{P_4}(\infty)\rangle.
\]
Because of the presence of the degenerate fields, there are only 2 conformal blocks in each decomposition of these correlation functions, moreover, they can be expressed in terms of hypergeometric functions. This allows us to reduce the crossing symmetry constraint to \cite{Teschner:1995yf}
\begin{equation}\label{eq:degcrossing}
\begin{aligned}
\frac{C_{2,1}^{1,0}(P_2+ib)C(P_2+ib,P_3,P_4)}{C_{2,1}^{-1,0}(P_2)C(P_2, P_3, P_4)} &=  \frac{\prod_{\pm_1, \pm_2}\gamma(\frac12-\frac{b^2}2+ib(P_2\pm_1P_3\pm_2P_4))}{\gamma(-b^2+2ibP_2)\gamma(1-b^2+2ibP_2)},\\
\frac{C_{1,2}^{0,1}(P_2+ib^{-1})C(P_2+ib^{-1},P_3,P_4)}{C_{1,2}^{0,-1}(P_2)C(P_2, P_3, P_4)} &= \frac{\prod_{\pm_1, \pm_2}\gamma(\frac12-\frac{b^{-2}}2+ib^{-1}(P_2\pm_1P_3\pm_2P_4))}{\gamma(-b^{-2}+2ib^{-1}P_2)\gamma(1-b^{-2}+2ib^{-1}P_2)}.
\end{aligned}
\end{equation} 
Here $\gamma(z) = \Gamma(z)/\Gamma(1-z)$. Functions on the left-hand side of these equations depend on normalization of $V_{P}$ and $V_{m,n}$. 

If $b^2\in\mathbb{C}\setminus\mathbb{R}$, $b$ and $b^{-1}$ are linearly independent as vectors in $\mathbb{R}^2$. Therefore one can produce many solutions by multiplying by functions that are periodic in $b$ and $b^{-1}$. In the remaining two cases $b\in\mathbb{R}$ and $b\in i\mathbb{R}$ equations \eqref{eq:degcrossing} have unique solutions provided that normalization is fixed\footnote{In both cases we also demand that $b^2\notin\mathbb{Q}$.}. These solutions are also believed to solve the general crossing symmetry equations with arbitrary fields. This has been checked numerically \cite{Zamolodchikov:1995aa,Ribault:2015sxa}. We will discuss the solutions in the next two subsections.
\subsection{Spacelike Liouville field theory} \label{subsec:SpacelikeLFT}
In this subsection we discuss LFT with $b\in\mathbb{R}$. We refer to it as ``spacelike LFT''. For such $b$ equations \eqref{eq:degcrossing} have a unique solution \cite{Teschner:1995yf}. We choose field normalization in such a way that the solution takes the form \cite{Dorn:1992at,Dorn:1994xn,Zamolodchikov:1995aa,Teschner:1995yf}
\begin{equation}
C(P_1,P_{2},P_{3}) = \frac{\Upsilon_b'(0)\prod_{k=1}^{3}\Upsilon_b(-2iP_{k})}{\prod_{\pm_1,\pm_2}\Upsilon_b(Q/2+iP_1\pm_1 iP_{2}\pm_2 iP_3)},
\quad 
N(P) = 1.
\end{equation}
In this normalization, we have
\begin{equation}
R_P = \frac{\Upsilon_b(-2iP)}{\Upsilon_b(2iP)}, \qquad  C_{P_1, P_2}^{P} = C(-P, P_1, P_2).
\end{equation}
The 3-point function has the following zeroes and poles in the variable $P_1$:
\begin{equation}\label{3pointpoles}
\begin{aligned}
\text{Poles}\colon& & &P_1 \pm_2 P_2 \pm_3 P_3  =   \pm P_{2m-1, 2n-1}, & m,n\in\mathbb{Z}_{> 0}.\\ 
\text{Zeroes}\colon& & &P_1 = \pm P_{m,n},\quad P_1 = 0, \quad P_1 = -\frac{imb}{2},\quad  P_1 = -\frac{inb^{-1}}{2},  &m,n\in\mathbb{Z}_{>0}.
\end{aligned}
\end{equation}
Normalization of degenerate fields is such that their structure constants have the form
\begin{equation}\label{eq:DegenerateStructureConstants}
\begin{gathered}
C_{2,1}^{1,0}(P) = C_{1,2}^{0,1}(P) = 1, 
\quad 
C_{2,1}^{-1,0}(P) = b^{2b^2+2}\frac{\gamma(2ibP)}{\gamma(2ibP+b^2+1)},\\ 
C_{1,2}^{0,-1}(P) = b^{-2b^{-2}-2}\frac{\gamma(2ib^{-1}P)}{\gamma(2ib^{-1}P_2+b^{-2}+1)}. 
\end{gathered}
\end{equation}
\subsubsection{Analytic continuation of OPE and degenerate fields} \label{subsec:SpacelikeDegenerateFields}
\begin{figure}[H]
\begin{tikzpicture}[trig format=rad,>=latex, declare function={
		f(\n) = \n/(\b*2) + \b/(2);
	}]
	\begin{axis}[
		width=16cm, height=9cm,
		axis lines=center,
		xlabel={$\Re P$},
		ylabel={$\Im P$},
		x label style={anchor=north},
		y label style={anchor=east},
		xmin=-5.5, xmax=5.5,
		ymin=-2.75,ymax=2.75,
		xtick = {0,\Pm, \Pp,- \Pm, -\Pp},
		xticklabels ={$0$, $P_2-P_1$, $P_2 + P_1$, $-P_2+P_1$, $-P_2 - P_1$},
		ytick = {{\b/2+1/(2*\b)}, {-\b/2-1/(2*\b)}},
		yticklabels ={$\frac{iQ}{2}$, $-\frac{iQ}{2}$},
		extra x ticks={0},
		extra x tick labels = {$0$},
		x tick label style={xshift={(\ticknum==0)*(-.8em)}},
		y tick label style={yshift={(\ticknum==0)*(1.2em)},yshift={(\ticknum==1)*(-1.2em)}},
		axis equal image,
		]

		\def\b{1.556}
		\def\Pm{1.7}
		\def\Pp{4.1}
		\def\crossSize{.05}

        \draw(-\Pp-.1,2.5) node[left]{$\Circled{1}$};
		\draw(-\Pm-.1,2.5) node[left]{$\Circled{2}$};
        \draw(\Pm-.1,2.5) node[left]{$\Circled{3}$};
        \draw(\Pp-.1,2.5) node[left]{$\Circled{4}$};
        \draw(-\Pp-.1,-2.5) node[left]{$\Circled{5}$};
        \draw(-\Pm-.1,-2.5) node[left]{$\Circled{6}$};
        \draw(\Pm-.1,-2.5) node[left]{$\Circled{7}$};
        \draw(\Pp-.1,-2.5) node[left]{$\Circled{8}$};
        
		\draw[dashed](5.5, {\b/2+1/(2*\b)}) -- (-5.5, {\b/2+1/(2*\b)});
		\draw[dashed](5.5, {-\b/2-1/(2*\b)}) -- (-5.5, {-\b/2-1/(2*\b)});
		
		\pgfplotsinvokeforeach{1,...,7}{
			\draw[fill=red, color=red] (axis cs: 0,{#1/(2*\b) + \b/(2)}) circle (1pt);
			\draw[fill=red, color=red] (axis cs: 0,{#1/(2*\b) + 2*\b/(2)}) circle (1pt);
			\draw[fill=red, color=red] (axis cs: 0,{#1/(2*\b) + 3*\b/(2)}) circle (1pt);
			\draw[fill=red, color=red] (axis cs: 0,{-#1/(2*\b) - \b/(2)}) circle (1pt);
			\draw[fill=red, color=red] (axis cs: 0,{-#1/(2*\b) - 2*\b/(2)}) circle (1pt);
			\draw[fill=red, color=red] (axis cs: 0,{-#1/(2*\b) - 3*\b/(2)}) circle (1pt);
		}
		
		\pgfplotsinvokeforeach{1,...,7}{
			\draw[fill=black] (axis cs: \Pm,{#1/(2*\b) + \b/(2)}) circle (1pt);
			\draw[fill=black] (axis cs: \Pm,{#1/(2*\b) + 2*\b/(2)}) circle (1pt);
			\draw[fill=black] (axis cs: \Pm,{#1/(2*\b) + 3*\b/(2)}) circle (1pt);
			\draw[fill=black] (axis cs: \Pm,{-#1/(2*\b) - \b/(2)}) circle (1pt);
			\draw[fill=black] (axis cs: \Pm,{-#1/(2*\b) - 2*\b/(2)}) circle (1pt);
			\draw[fill=black] (axis cs: \Pm,{-#1/(2*\b) - 3*\b/(2)}) circle (1pt);
		}
		
		\pgfplotsinvokeforeach{1,...,7}{
			\draw[fill=black] (axis cs: -\Pm,{#1/(2*\b) + \b/(2)}) circle (1pt);
			\draw[fill=black] (axis cs: -\Pm,{#1/(2*\b) + 2*\b/(2)}) circle (1pt);
			\draw[fill=black] (axis cs: -\Pm,{#1/(2*\b) + 3*\b/(2)}) circle (1pt);
			\draw[fill=black] (axis cs: -\Pm,{-#1/(2*\b) - \b/(2)}) circle (1pt);
			\draw[fill=black] (axis cs: -\Pm,{-#1/(2*\b) - 2*\b/(2)}) circle (1pt);
			\draw[fill=black] (axis cs: -\Pm,{-#1/(2*\b) - 3*\b/(2)}) circle (1pt);
		}
		
		\pgfplotsinvokeforeach{1,...,7}{
			\draw[fill=black] (axis cs: \Pp,{#1/(2*\b) + \b/(2)}) circle (1pt);
			\draw[fill=black] (axis cs: \Pp,{#1/(2*\b) + 2*\b/(2)}) circle (1pt);
			\draw[fill=black] (axis cs: \Pp,{#1/(2*\b) + 3*\b/(2)}) circle (1pt);
			\draw[fill=black] (axis cs: \Pp,{-#1/(2*\b) - \b/(2)}) circle (1pt);
			\draw[fill=black] (axis cs: \Pp,{-#1/(2*\b) - 2*\b/(2)}) circle (1pt);
			\draw[fill=black] (axis cs: \Pp,{-#1/(2*\b) - 3*\b/(2)}) circle (1pt);
		}
		
		\pgfplotsinvokeforeach{1,...,7}{
			\draw[fill=black] (axis cs: -\Pp,{#1/(2*\b) + \b/(2)}) circle (1pt);
			\draw[fill=black] (axis cs: -\Pp,{#1/(2*\b) + 2*\b/(2)}) circle (1pt);
			\draw[fill=black] (axis cs: -\Pp,{#1/(2*\b) + 3*\b/(2)}) circle (1pt);
			\draw[fill=black] (axis cs: -\Pp,{-#1/(2*\b) - \b/(2)}) circle (1pt);
			\draw[fill=black] (axis cs: -\Pp,{-#1/(2*\b) - 2*\b/(2)}) circle (1pt);
			\draw[fill=black] (axis cs: -\Pp,{-#1/(2*\b) - 3*\b/(2)}) circle (1pt);
		}

	\end{axis}
\end{tikzpicture}
\caption{Structure of poles of the spacelike Liouville thee-point function $C(-P,P_1,P_2)$ (shown by black bullets) for real momenta $P_2>P_1>0$ and of OPE coefficients $\beta_{\boldsymbol{\lambda}}(P)$ (shown by red bullets). The poles of OPE coefficients are exactly canceled by the zeroes of three-point function (provided that the correlation function is at least $4$-point). The contour of integration is the real axis.}\label{C-poles}
\end{figure}
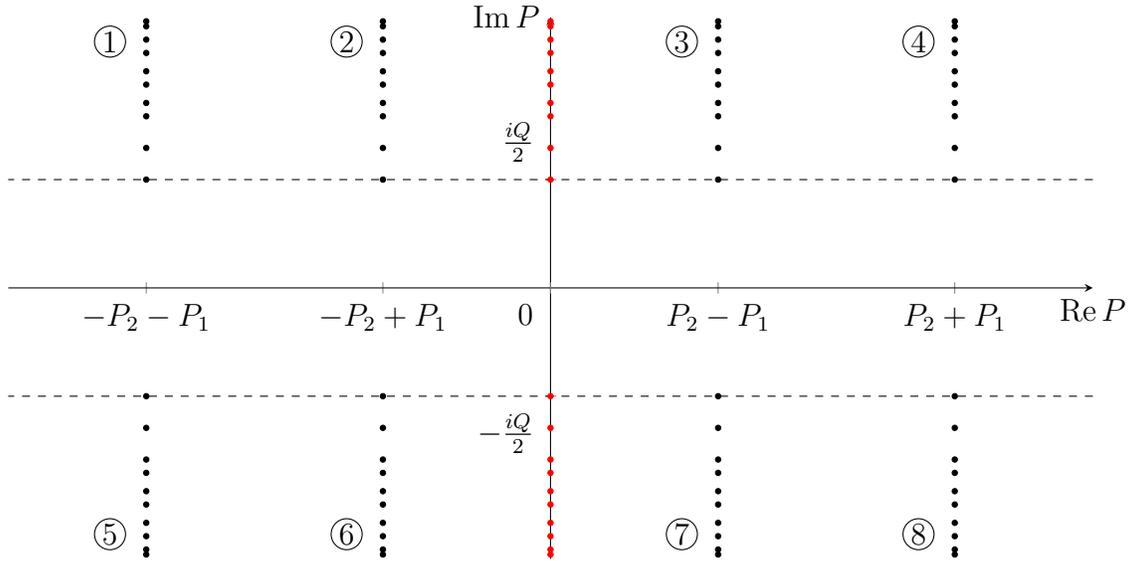

We now consider analytic continuation of OPE \eqref{eq:OPE} in the momentum $P_1$ away from the spectrum $\mathbb{R}$ with the other external momentum $P_2\in\mathbb{R}$ fixed. The integrand is a meromorphic function of $P_1$ and $P$. It has poles in the $P$ complex plane that originate from the structure constant $C^{P}_{P_1, P_2} = C(-P, P_1, P_2)$ and from the OPE coefficients $\beta_{\boldsymbol\lambda}(P)$, $\beta_{\boldsymbol{\bar{\lambda}}}(P)$. Poles of the former form $8$ series that correspond to choosing different signs in front of $P_1$, $P_2$ and $P_{2m-1, 2n-1}$ (see fig \ref{C-poles}):
\begin{equation}\label{eq:StructureConstantPoles}
P =  \pm_1 P_1 \pm_2 P_2 \pm P_{2m-1, 2n-1}, \qquad m,n\in\mathbb{Z}_{> 0}.
\end{equation}
OPE coefficients $\beta_{\boldsymbol\lambda}(P)$ and $\beta_{\boldsymbol{\bar{\lambda}}}(P)$ have simple poles at $P = \pm P_{m,n}$, $m,n\in\mathbb{Z}_{> 0}$. The double poles from the product of OPE coefficients are reduced to simple poles by the zeroes of $\Upsilon(2iP)$ in the numerator of $C(-P,P_1,P_2)$. If after performing the OPE we are left with a correlation function of at least $3$ fields, we will also have $\Upsilon(-2iP)$ in the numerator from the normalization of $V_{P}$. In this case the poles of the OPE coefficients will be fully canceled and the only poles in the $P$ complex plane are \eqref{eq:StructureConstantPoles}. From now on we will be considering this case.

If $\pm_1\Re P_1 \pm_2 P_2$ are pairwise distinct, no two poles have the same position (see fig~\ref{C-poles}). As $P_1$ moves away from $\mathbb{R}$, some poles may cross the contour of integration in $P$. This happens when $|\Im P_1|> \frac{Q}{2}$. In order to ensure analytic dependence of the correlation functions on $P_1$, one has to either deform the contour of integration or add to it discrete contributions in the form of small circles around the poles.  

If $\pm_1\Re P_1 \pm_2 P_2$ are not pairwise distinct, the positions of two poles from different series may coincide. If the contour of integration is pinched by the two poles, the contour cannot be deformed to avoid both poles, and the integral may develop a singularity. In order for this situation to occur, the pinching poles must lie on the different sides of the real axis when $\Im P_1 = 0$. This means that the poles have different signs in front of $P_{2m-1, 2n-1}$ in equation \eqref{eq:StructureConstantPoles}. We consider the case when $P_2 \neq 0$ and $\Im P_1>0$. Since $P_2\in\mathbb{R}$, we are left with two possibilities: 

\paragraph{Pinching poles are from series $\Circled{1}$ and $\Circled{6}$ and from $\Circled{3}$ and $\Circled{8}$.} \label{item:pinching1} i.\,e. we are equating the positions of the following poles:
\begin{equation}
P =   P_1 \pm P_2 - P_{2m-1,2n-1} = - P_1 \pm P_2 + P_{2m'-1, 2n'-1}.  
\end{equation}
This leads to pinching of the contour at
\begin{equation}
P = P_2 + P_{k,l}\quad \text{and} \quad P = - P_2 + P_{k,l}, \qquad \text{when} \qquad P_{1} =  P_{r,s}.
\end{equation}
Here $r = m+m'-1\in\mathbb{Z}_{>0}$, $s = n+n'-1\in\mathbb{Z}_{>0}$, and $k = m'-m$, $l = n'-n$. For a fixed external momenta $P_1 = P_{r,s}$ we have
\begin{equation}\label{FusionKLSet}
k\in\{-r+1, -r+3, \ldots, r-3, r-1\}, \quad l\in\{-s+1, -s+3, \ldots, s-3, s-1\},
\end{equation}
and so the contour is pinched in $2rs$ places. 

In order to define the limit $P_1 \to P_{r,s}$, we deform the integration contour so that it receives discrete contributions in the form of small circles around the poles above the contour. When $P_1 = P_{r,s}$ the structure constant $C^P_{P_1, P_2}$ with $P\neq \pm P_2$ vanishes due to the factor $\Upsilon_b(-2iP_1)$ in the numerator. Because of that, the OPE does not have a pole, but the only nonzero contributions come from the discrete terms:
\begin{multline}
V_{P_{r,s}}(z) V_{P_2}(w) = \pi i \sum_{\pm} \sum_{k,l}\lim\limits_{P_1\to P_{r,s}}\Res\limits_{P = -P_1 \pm P_2 + P_{r+k, s+l}}  C^P_{P_1,P_2}
|z-w|^{2(\Delta(\pm P_2 + P_{k,l}) - \Delta(P_1)-\Delta(P_2))} 
\times
\\
\times
\sum\limits_{\boldsymbol{\lambda}, \boldsymbol{\bar{\lambda}}}
(z-w)^{|\boldsymbol{\lambda}|}(\bar{z}-\bar{w})^{|\boldsymbol{\bar{\lambda}}|}
\beta_{\boldsymbol{\lambda}}(\pm P_2 + P_{k,l})\beta_{\boldsymbol{\bar{\lambda}}}(\pm P_2+P_{k,l}) V^{\boldsymbol{\lambda}, \boldsymbol{\bar{\lambda}}}_{\pm P_2 + P_{k,l}}(w)
\end{multline}
We get $rs$ discrete terms\footnote{Actual number of discrete terms is $2rs$, if we count $V_P$ and reflected field $V_{-P}$ separately.} with the same momenta as in \eqref{eq:VmnOPE}. This means that up to a normalization factor we can identify the field $V_{P_{r,s}}$ with the degenerate field $V_{r,s}$. 

The simplest example is $r = s = 1$. In this case the OPE coefficients $\beta_{\boldsymbol{\lambda}}(\pm P_2)$, $\beta_{\boldsymbol{\bar{\lambda}}}(\pm P_2)$ vanish for all non-empty partitions, and we get
\begin{equation}
V_{P_{1,1}}(z)V_{P_2}(w) = \pi \left[V_{P_2}(w) + \frac{\Upsilon_b(-2iP_2)}{\Upsilon_b(2iP_2)}V_{-P_2}(w) \right] = 2\pi V_{P_2}(w).
\end{equation} 
Due to this property, we will call $V_{P_{1,1}}/2\pi$ or $V_{1,1}$ an identity operator.

In case of $r = 2$, $s = 1$ one can check that we recover the structure constants of $V_{2,1}$ from \eqref{eq:DegenerateStructureConstants}:
\begin{equation}
\pi i \lim\limits_{P_1\to P_{2,1}}\Res\limits_{P = -P_1+ P_2 + P_{3, 1}}  C^P_{P_1,P_2} = \pi C_{2,1}^{1,0}(P_2), \qquad 
\pi i \lim\limits_{P_1\to P_{2,1}}\Res\limits_{P = -P_1 + P_2 + P_{1, 1}}  C^P_{P_1,P_2} = \pi C_{2,1}^{-1,0}(P_2).
\end{equation}

Note that we could choose a different normalization of the primary fields that does not contain $\Upsilon_b(-2iP_1)$ in the numerator of the structure constant. In this case we would have had poles when one of the fields is degenerate.

\paragraph{Pinching poles are from series $\Circled{1}$ and $\Circled{8}$ or from $\Circled{2}$ and $\Circled{7}$.} i.\,e. we are equating the positions of poles
\begin{equation}
P = (P_1 \pm P_2) - P_{2m-1,2n-1} = -(P_1\pm P_2)+ P_{2m'-1,2n'-1}.
\end{equation}
This leads to pinching poles at
\begin{equation}
P = P_{k,l}, \qquad \text{when} \qquad  P_1 \pm P_2 =   P_{r,s}.
\end{equation}
Here $r$, $s$ and $k$, $l$ are the same as above.

If either $k$ or $l$ is zero, the pole is canceled by a zero of either $\Upsilon(-2iP)$ or $\Upsilon(2iP)$ in the numerator. This results in finite contributions to the OPE. If both $k$ and $l$ are zero, then both $\Upsilon(-2iP)$ and $\Upsilon(2iP)$ are zero and the discrete contribution vanishes. If $k$ and $l$ are both nonzero, pinching leads to divergence of the discrete term. Note that for any $r$, $s$ the divergent terms, if they are present, come in pairs: from $P = P_{k,l}$ and $P = P_{k,-l}$. Below we show that despite this divergence of a single discrete term, the total contribution from each pair is finite, so the OPE does not have a pole. Moreover, we will show that the integrand in \eqref{eq:OPE} at $P = P_{k,l}$ is equal to minus the integrand at $P = P_{k,-l}$ for arbitrary external momenta~$P_1$ and $P_2$.

At $P = P_{k,-l}$ the integrand is equal to
\begin{equation}\label{Pk-lResidue}
	|z-w|^{2(\Delta_{k,-l} - \Delta(P_1)-\Delta(P_2))}
	\sum\limits_{\boldsymbol{\lambda}, \boldsymbol{\bar{\lambda}}}
	(z-w)^{|\boldsymbol{\lambda}|}(\bar{z}-\bar{w})^{|\boldsymbol{\bar{\lambda}}|}\beta_{\boldsymbol{\lambda}}(P_{k,-l})\beta_{\boldsymbol{\bar{\lambda}}}(P_{k,-l}) V^{\boldsymbol{\lambda}, \boldsymbol{\bar{\lambda}}}_{k,-l}(w) C^{P_{k,-l}}_{P_1,P_2}.
\end{equation}  
At $P = P_{k,l}$ we have
\begin{equation} 
	|z-w|^{2(\Delta_{k,l} - \Delta(P_1)-\Delta(P_2))}\sum\limits_{\boldsymbol{\lambda}, \boldsymbol{\bar{\lambda}}}
	(z-w)^{|\boldsymbol{\lambda}|}(\bar{z}-\bar{w})^{|\boldsymbol{\bar{\lambda}}|}
	\lim\limits_{P \to P_{k,l}} \left[ 
	\beta_{\boldsymbol{\lambda}}(P)\beta_{\boldsymbol{\bar{\lambda}}}(P) V^{\boldsymbol{\lambda}, \boldsymbol{\bar{\lambda}}}_{P}(w)C^{P}_{P_1,P_2} 
	\right].
\end{equation}
Momentum $P_{k,l}$ corresponds to a degenerate field, hence the structure constant and normalization of $V_{P}$ vanish. Because of that the only nonzero contribution to the limit $P \to P_{k,l}$ comes from $\beta_{\boldsymbol{\lambda}}$, $\beta_{\boldsymbol{\bar{\lambda}}}$ that have a pole, i.\,e. from descendants of $V_{P}^{\boldsymbol{\lambda}, \boldsymbol{\bar{\lambda}}}$ of the form
\[
V_{P}^{\boldsymbol{\lambda}, \boldsymbol{\bar{\lambda}}}(w)
=
L_{-\boldsymbol\mu}\bar{L}_{-\boldsymbol{\bar{\mu}}}D_{k,l}^{(b)}\bar{D}_{k,l}^{(b)}V_{P}(w),
\]
where $\boldsymbol{\mu}, \boldsymbol{\bar{\mu}}$ are arbitrary partitions. In the limit $P \to P_{k,l}$ we have
\begin{multline}
	\lim\limits_{P \to P_{k,l}} \left[ 
	\beta_{\boldsymbol{\lambda}}(P)\beta_{\boldsymbol{\bar{\lambda}}}(P) V^{\boldsymbol{\lambda}, \boldsymbol{\bar{\lambda}}}_{P}(w)C^{P}_{P_1,P_2} 
	\right]
	=
	\\
	=
	\Res\limits_{P = P_{k,l}}  \beta_{\boldsymbol{\lambda}}(P)\Res\limits_{P = P_{k,l}}  \beta_{\boldsymbol{\bar{\lambda}}}(P)
	\cdot
	L_{-\boldsymbol\mu}\bar{L}_{-\boldsymbol{\bar{\mu}}}D_{k,l}^{(b)}\bar{D}_{k,l}^{(b)}V'_{k,l}(w)\cdot \left.\frac{\partial}{\partial P} C^{P}_{P_1, P_2} \right|_{P = P_{k,l}}.
\end{multline}
The residue of an OPE coefficient is given by \cite{Zamolodchikov:1985ie}
\begin{equation}
	\Res\limits_{P = P_{k,l}}  \beta_{\boldsymbol{\lambda}}(P)
	= 
	-\beta_{\boldsymbol{\mu}}(P_{k,-l}) \cdot\frac12\prod_{p,q} 
	\left(P_1 + P_2 - P_{p,q}\right)
	\left(P_1 - P_2 - P_{p,q}\right)
	\prod_{a,b}(2P_{a,b})^{-1}
\end{equation} 
with
\begin{equation}
	\begin{aligned}
		&p\in\{-k+1, -k+3, \ldots, k-3, k-1\}, \quad q\in\{-l+1, -l+3, \ldots, l-3, l-1\},\\
		&a\in\{-k+1, -k+2, \ldots, k-1, k\}, \quad b\in\{-l+1, -l+2, \ldots, l-1, l\}, \quad (a,b)\neq(0,0).
	\end{aligned}
\end{equation}
Then, after using higher equations of motion (see \eqref{eq:SpacelikeHEM} below), we get
\begin{multline}\label{PklResidue}
	|z-w|^{2(\Delta_{k,-l} - \Delta(P_1)-\Delta(P_2))}\sum\limits_{\boldsymbol{\mu}, \boldsymbol{\bar{\mu}}}
	(z-w)^{|\boldsymbol{\mu}|}(\bar{z}-\bar{w})^{|\boldsymbol{\bar{\mu}}|}\beta_{\boldsymbol{\lambda}}(P_{k,-l}) \beta_{\boldsymbol{\bar{\lambda}}}(P_{k,-l})V^{\boldsymbol{\mu}, \boldsymbol{\bar{\mu}}}_{k,-l}(w)
	\times
	\\
	\times
	\frac{i}2\Upsilon'(2iP_{k,l})B_{k,l}
	\prod_{a,b}(2P_{a,b})^{-2}
	\prod_{p,q} 
	\left(P_1 + P_2 - P_{p,q}\right)^2
	\left(P_1 - P_2 - P_{p,q}\right)^2
	\cdot \lim\limits_{P\to P_{k,l}}\frac{C^{P}_{P_1,P_2}}{\Upsilon(2iP)}.
\end{multline}

From the identity \eqref{UsefulUpsilon} it follows that
\begin{equation}
\frac{\Upsilon_b(2iP_{k,-l})}{C^{P_{k,-l}}_{P_1,P_2}}\lim\limits_{P\to P_{k,l}}\frac{C^{P}_{P_1,P_2}}{\Upsilon(2iP)}
=
\prod_{p,q} \frac{1}{(P_1 + P_2-P_{p,q})^2}\frac1{(P_1 - P_2 - P_{p,q})^2}
\end{equation}
One can also show that
\begin{equation}\label{HEMConstantRelation}
	\frac{i}2 \frac{\Upsilon'(2iP_{k,l})}{\Upsilon(2iP_{k,-l})}B_{k,l} 
	=
	\frac{\Upsilon'(2iP_{k,l})}{\Upsilon(2iP_{k,-l})}\frac{\Upsilon'(-2iP_{k,l})}{\Upsilon(-2iP_{k,-l})} 
	=
	-\prod_{a,b}(2P_{a,b})^2,
\end{equation}
From above two formulas it follows that for any $P_1$, $P_2$ and $k,l \in\mathbb{Z}_{>0}$:
\begin{multline}\label{InterestingOPEFact}
|z-w|^{2(\Delta_{k,-l} - \Delta(P_1)-\Delta(P_2))}
	\sum\limits_{\boldsymbol{\lambda}, \boldsymbol{\bar{\lambda}}}
	(z-w)^{|\boldsymbol{\lambda}|}(\bar{z}-\bar{w})^{|\boldsymbol{\bar{\lambda}}|}\beta_{\boldsymbol{\lambda}}(P_{k,-l})\beta_{\boldsymbol{\bar{\lambda}}}(P_{k,-l}) V^{\boldsymbol{\lambda}, \boldsymbol{\bar{\lambda}}}_{k,-l}(w) C^{P_{k,-l}}_{P_1,P_2}
=
\\
=
-|z-w|^{2(\Delta_{k,l} - \Delta(P_1)-\Delta(P_2))}\sum\limits_{\boldsymbol{\lambda}, \boldsymbol{\bar{\lambda}}}
	(z-w)^{|\boldsymbol{\lambda}|}(\bar{z}-\bar{w})^{|\boldsymbol{\bar{\lambda}}|}
	\lim\limits_{P \to P_{k,l}} \left[ 
	\beta_{\boldsymbol{\lambda}}(P)\beta_{\boldsymbol{\bar{\lambda}}}(P) V^{\boldsymbol{\lambda}, \boldsymbol{\bar{\lambda}}}_{P}(w)C^{P}_{P_1,P_2} 
	\right]
\end{multline}
Hence, when $P_1\pm P_2 = P_{r,s}$, the contributions to the residue from $P = P_{k,l}$ and $P = P_{k,-l}$ cancel. One can also show that by taking the corresponding residues and using \eqref{UsefulUpsilon}, \eqref{UsefulDeriv} and \eqref{HEMConstantRelation}.

\subsubsection{Higher equations of motion}\label{subsubsec:HEMSpacelike}
In \ref{subsec:SpacelikeDegenerateFields} we identified degenerate fields $V_{m,n}$ in spacelike LFT with primary fields $V_{P_{m,n}}$. By definition of degenerate fields, they satisfy
\begin{equation}
D_{m,n}^{(b)}\bar{D}_{m,n}^{(b)}V_{m,n}(z) = 0, \qquad m,n\in\mathbb{Z}_{>0}. 
\end{equation}
In \cite{Zamolodchikov:2003yb} Al. Zamolodchikov discovered the following ``higher equations of motion'' (HEM) in spacelike LFT:
\begin{equation}\label{eq:SpacelikeHEM}
D_{m,n}^{(b)}\bar{D}_{m,n}^{(b)}V'_{m,n}(z) = B_{m,n}^{(b)} V_{m,-n}(z), \qquad \text{where} \qquad V'_{m,n}(z) \equiv \left.\frac{\partial}{\partial P} V_{P}(z)\right|_{P = P_{m,n}}. 
\end{equation}
Here by $V_{m,-n}(z)$ we denote the (not degenerate) field $V_{P_{m,-n}}(z)$. The constant $B_{m,n}^{(b)}$ can be obtained by checking the equality on the level of $3$-point functions:
\begin{equation}
B_{m,n}^{(b)} = \prod\limits_{k,l}(P_1 +P_2 + P_{k,l})^2 (P_1 -P_2 + P_{k,l})^2 \cdot \frac{\left.\partial C(P, P_1, P_2)/\partial P\right|_{P = P_{m,n}}}{C(P_{m,-n},P_1,P_2)}
\end{equation}
with $k, l$ from \eqref{klOneDegenerateFieldsFusion}. In our normalization it has the following form:
\begin{equation}
B_{m,n}^{(b)} =   -2i\, \frac{\Upsilon'_b(-2iP_{m,n})}{\Upsilon_b(-2iP_{m,-n})} 
= 
-2i\cdot\gamma(m-nb^{-2})b^{-1+2m-2nb^{-2}}\prod\limits_{\substack{k=1-m,\\l=1-n,\\(k,l)\neq(0,0)}}^{\substack{n-1\\m-1}} (kb+lb^{-1}).
\end{equation}

These equations are valid in all points $z$, except for the insertion points of other fields. The full equations may contain contact terms, which we will not write explicitly here.

We also note that HEM manifest themselves on the level of conformal blocks:
\begin{equation}\label{eq:ConformalBlocksHEM}
\mathcal{D}_{m,n}^{(b)}\mathfrak{F}^{(b)}(P_{m,n},P_2,P_3,P_4,P|z) = \mathfrak{F}^{(b)}(P_{m,-n},P_2,P_3,P_4,P|z)\prod_{p,q}(P-P_2-P_{p,q})(P+P_2-P_{p,q}),
\end{equation}
where
\begin{equation}
    p\in\{-m+1,-m+3, \ldots,m-3,m-1\}, \quad q\in\{-n+1,-n+3, \ldots,n-3,n-1\}.
\end{equation}

\subsubsection{Analytic structure of general \texorpdfstring{$N$}{N}-point correlation function} \label{subsubsec:SpacelikeAnalyticStructure}
Consider the correlation function of $N$ primary fields:
\[
\langle V_{P_1}(z_1)V_{P_2}(z_2)\ldots V_{P_N}(z_N)\rangle.
\]
We are interested in its analytical properties in the momentum $P_1$. We choose the momenta $P_2, \ldots P_N\in\mathbb{R}$ in such a way that $\sum_{i = 2}^{N}(\pm_i P_i)$ are pairwise distinct for all choices of signs. Below we argue that in this case the only poles of the correlation function in $P_1$ are simple poles, which are known as ``screening poles'':
\begin{equation}\label{eq:npointpoles}
P_1 +\sum_{i=2}^N \pm_i P_i= \pm i\left(\frac{(N-2)Q}{2} + (r-1)b+(s-1)b^{-1}\right), 
\qquad
r,s\in\mathbb{Z}_{>0}.
\end{equation}
We argue this by induction. The statement holds for $N = 3$ as can be seen from \eqref{3pointpoles}. We assume that it's true for $N-1$ fields and consider a correlation function with $N$ fields. Using OPE of $V_{P_1}$ and $V_{P_2}$ we get
\begin{multline}
\langle V_{P_1}(z_1) V_{P_2}(z_2) \ldots V_{P_{N}}(z_{N}) \rangle 
=
\\
=
\int_{C} dP\, |z_1-z_2|^{2(\Delta(P) - \Delta(P_1)-\Delta(P_2))} C(-P, P_1, P_2) \left[\langle V_{P}(z_2) V_{P_3}(z_3)\ldots V_{P_{N}}(z_{N}) \rangle + \ldots \right].
\end{multline}
Correlation functions of descendant fields can be obtained from the correlation function of $V_{P}$ by acting on them with differential operators that depend polynomially on $P_1$. This action does not produce singularities in $P_1$ or $P$. Therefore, the resulting integral may exhibit a pole in $P_1$ only when the contour of integration is pinched by two poles in the $P$ plane. 

When $P_1\in\mathbb{R}$ the two poles must lie on different sides of the real axis. Position of one of the poles must depend on $P_1$, so it is a pole of the structure constant $C(-P, P_1, P_2)$. The second pole may come either from $C(-P, P_1, P_2)$ or from the correlation function $\langle V_{P}(z_1) V_{P_3}(z_3)\ldots V_{P_{N}}(z_{N}) \rangle$. The first case involves only the fields that participate in OPE and was studied in \ref{subsec:SpacelikeDegenerateFields}, so we are left with the case when one of the pinching poles comes from the structure constant $C(-P, P_1, P_2)$ and the other one comes from $\langle V_{P}(z_1) V_{P_3}(z_3)\ldots V_{P_{N}}(z_{N}) \rangle$. The pinching occurs when
\begin{equation}\label{P1PplaneScreeningCondition}
P_1 \pm_2 P_2 \mp P_{2m'-1,2n'-1} 
=
\pm i\left(\frac{(N-3)Q}{2} + (m-1)b+(n-1)b^{-1}\right)-\sum_{i=3}^{N} \pm_i P_i, 
\qquad
m, m', n, n'\in\mathbb{Z}_{>0}.
\end{equation}
This corresponds to \eqref{eq:npointpoles} for a correlation function with $N$ fields and $r = m+m'-1$, $s = n+n'-1$. For fixed $P_1$ that satisfies \eqref{P1PplaneScreeningCondition} the contour of integration in $P$ is pinched in $2rs$ places:
\begin{equation}
P = P_1 \pm_2 P_2\mp P_{r+k,s+l}\quad \text{and}\quad P = -P_1\mp_2P_2\pm P_{r+k,s+l}
\end{equation}
with
\begin{equation}
k\in\{-r+1, -r+3, \ldots, r-3, r-1\}, \quad l\in\{-s+1, -s+3, \ldots, s-3, s-1\},
\end{equation}

If $\sum_{k = 2}^{n}(\pm_k \Re P_k)$ are not pairwise distinct or if $P_2, \ldots P_{n}\in\mathbb{C}$, the contour may be pinched by more than two poles and the correlation function may develop higher order poles. We analyze this situation for $4$-point functions in \ref{subsubsec:Spacelike4pointAnalyticStructure}.

\subsubsection{Triality transformation for \texorpdfstring{$4$}{4}-point functions}\label{subsubsec:TrialitySpacelike}
Here we consider a transformation of $4$-point correlation functions in LFT under a certain transformation of external momenta. Namely, consider the following involution of $\{P_k\}$:
\begin{equation}\label{eq:Triality}
\begin{aligned}
& P_1 \to \tilde{P}_1 = \frac{P_1+P_2+P_3+P_4}{2}, &\qquad &P_2 \to \tilde{P}_2 = \frac{P_1+P_2-P_3-P_4}{2},\\
& P_3 \to \tilde{P}_3 = \frac{P_1-P_2+P_3-P_4}{2}, &\qquad &P_4 \to \tilde{P}_4 = \frac{P_1-P_2-P_3+P_4}{2}.
\end{aligned}
\end{equation}
The defining property of the transformation is that it permutes the sums and differences of $P_{k}$:
\begin{equation}
\begin{aligned}
\sigma = \mathrm{id},\,(34): & & \{\tilde{P}_1 + \tilde{P_2},~ \tilde{P}_1-\tilde{P}_2,~ \tilde{P}_3 + \tilde{P}_4,~ \tilde{P}_3 - \tilde{P}_4\} = \{P_1 + P_2,~ P_3 + P_4,~ P_1-P_2,~ P_3-P_4\};\\[.1cm]
\sigma = (23),\,(234): & & \{\tilde{P}_1 + \tilde{P_3},~ \tilde{P}_1-\tilde{P}_3,~ \tilde{P}_2 + \tilde{P}_4,~ \tilde{P}_2 - \tilde{P}_4\} = \{P_1 + P_3,~ P_2 + P_4,~ P_1-P_3,~ P_2-P_4\};\\[.1cm]
\sigma = (24),\,(243): & & \{\tilde{P}_1 + \tilde{P_4},~ \tilde{P}_1-\tilde{P}_4,~ \tilde{P}_2 + \tilde{P}_3,~ \tilde{P}_2 - \tilde{P}_3\} = \{P_1 + P_4,~ P_2 + P_3,~ P_1-P_4,~ P_2-P_3\}.
\end{aligned}
\end{equation}
In the vicinity of one of the points $z_2$, $z_3$, $z_4$ the 4-point function is given by
\begin{equation}
\langle V_{P_1}(z) V_{P_2}(0) V_{P_3}(1) V_{P_4}(\infty)\rangle = \frac12\int_C dP\,C(P_1, P_{\sigma(2)}, -P)C(P, P_{\sigma(3)}, P_{\sigma(4)}) \left|\mathfrak{F}^{(b)}(\{P_k\},P|z)\right|^2,
\end{equation}
where $\sigma \in S_3$ is a permutation that corresponds to the choice of OPE channel. The integrand depends on $\{P_k\}$ only through the ratio of the structure constants and the product of conformal blocks. The former is:
\begin{multline}
C\left(P_{1},P_{\sigma(2)},P\right)C\left(-P,P_{\sigma(3)},P_{\sigma(4)}\right)
=
\\
=
\frac{\Upsilon_b(2iP)\Upsilon_b(-2iP) \prod_{k=1}^4 \Upsilon(-2iP_k)}{\prod_{\pm_1,\pm_2}\Upsilon_b(Q/2+iP\pm_1 (iP_{1}\pm_2 iP_{\sigma(2)}))\Upsilon_b(Q/2+iP\pm_1 (iP_{\sigma(3)}\pm_2 iP_{\sigma(4)}))}.
\end{multline}
The right-hand side is invariant under permutations of sums and differences of $P_{k}$ except for the normalization factors $\Upsilon_b(-2iP_k)$ which do not affect the integration over the internal momentum. 

Using \eqref{FblocksInTermsOfHBlocks} conformal blocks may be rewritten in terms of elliptic conformal blocks, which may be defined by the Zamolodchikov recursion representation (see equation \eqref{eq:Recursion.H} in appendix \ref{appendix:Blocks}). The block $\mathfrak{H}^{(b)}(\{P_1, P_{\sigma(k)}\}, P|q)$ depends on $P_k$ only through the recursion kernel
\begin{multline}
R_{m,n}(\{P_1, P_{\sigma(k)}\}) 
=
-\frac12\prod_{k,l}(2P_{k,l})^{-1}\times\\\times \prod_{p,q} 
\left(P_1 + P_{\sigma(2)} - P_{p,q}\right)
\left(P_1 - P_{\sigma(2)} - P_{p,q}\right)
\left(P_{\sigma(3)} + P_{\sigma(4)} - P_{p,q}\right)
\left(P_{\sigma(3)} - P_{\sigma(4)} - P_{p,q}\right).
\end{multline}
Clearly, the kernel is invariant under permutations of sums and differences of $P_k$ and therefore the elliptic conformal blocks are as well. Then the conformal blocks transform in the following way:
\begin{equation}
\mathfrak{F}^{(b)}(\{\tilde{P}_{k}\},P|z) = z^{-\frac12 (P_1-P_2-P_3-P_4)(P_1-P_2+P_3+P_4)}(1-z)^{-\frac12 (P_1-P_2-P_3-P_4)(P_1+P_2-P_3+P_4)}\mathfrak{F}^{(b)}(\{P_{k}\},P|z).
\end{equation}
Collecting this all together we have the following transformation of spacelike LFT correlation functions under \eqref{eq:Triality}:
\begin{multline}\label{TrialityCorrFuncSpacelike}
\langle V_{\tilde{P}_1}(z) V_{\tilde{P}_2}(0) V_{\tilde{P}_3}(1) V_{\tilde{P}_4}(\infty)\rangle
=
|z|^{- (P_1-P_2-P_3-P_4)(P_1-P_2+P_3+P_4)}|1-z|^{- (P_1-P_2-P_3-P_4)(P_1+P_2-P_3+P_4)}\times\\\times
\prod_{k=1}^4\frac{\Upsilon_b(-2i\tilde{P}_k)}{\Upsilon_b(-2iP_k)}
\langle V_{P_1}(z) V_{P_2}(0) V_{P_3}(1) V_{P_4}(\infty)\rangle
\end{multline}

An important property of the transformation \eqref{TrialityCorrFuncSpacelike} is that it maps correlation functions that satisfy screening conditions to correlation functions with degenerate fields. Indeed, the screening condition \eqref{P1PplaneScreeningCondition} for the $4$-point function has the form (for a particular choice of signs)
\begin{equation}
    P_1 + P_2 + P_3 + P_4 = P_{2r,2s},
\end{equation}
which means that $\tilde{P}_1 = P_{r,s}$. This provides us with an expression for the residue of the $4$-point function:
\begin{multline}
\Res\limits_{2\tilde{P}_1 = P_{2r, 2s}} \langle V_{P_1}(z) V_{P_2}(0) V_{P_3}(1) V_{P_4}(\infty)\rangle
=
|z|^{- (P_1-P_2-P_3-P_4)(P_1-P_2+P_3+P_4)}\times\\\times
|1-z|^{- (P_1-P_2-P_3-P_4)(P_1+P_2-P_3+P_4)}\frac{i\prod_{k=1}^4\Upsilon_b(-2iP_k)}{\Upsilon'(0)\prod_{k=2}^4\Upsilon_b(-2i\tilde{P}_k)}
\langle V_{r,s}(z) V_{\tilde{P}_2}(0) V_{\tilde{P}_3}(1) V_{\tilde{P}_4}(\infty)\rangle.
\end{multline}

\subsubsection{Analytic structure of the \texorpdfstring{$4$}{4}-point function} \label{subsubsec:Spacelike4pointAnalyticStructure}

Here we analyze the situations when the contour of integration in intermediate momentum is pinched by 3 poles. In particular, we analyze the following 3 cases:

\paragraph{Two degenerate fields or two screening conditions.} Without loss of generality we assume that one of the degenerate fields is~$V_{P_1} = V_{r,s}$. If we temporarily assume that other fields are not degenerate, the $4$-point function takes the form
\begin{equation} \label{OneDegenerate4point}
\langle V_{r,s}(z) V_{P_2}(0) V_{P_3}(1) V_{P_4}(\infty)\rangle = \sum_{k,l} C_{r,s}^{k,l}(P_2)C(P_2+P_{k,l}, P_3, P_4) \left|\mathfrak{F}^{(b)}(P_{r,s}, P_2, P_3, P_4,P_2+P_{k,l}|z)\right|^2
\end{equation}
with $k, l$ from \eqref{klOneDegenerateFieldsFusion}. Now, we should take the limit of this expression when another external momentum approaches $P_{r',s'}$. Because of the crossing symmetry, we may pick this momentum to be $P_2$. Then, we are computing the OPE of two degenerate fields. From equation \eqref{eq:VmnVrsOPE} we know that the result is
\begin{equation}
\langle V_{r,s}(z) V_{r',s'}(0) V_{P_3}(1) V_{P_4}(\infty)\rangle = \sum_{k,l} C_{r,s;r',s'}^{k,l}C(P_{k,l}, P_3, P_4) \left|\mathfrak{F}^{(b)}(P_{r,s}, P_{r',s'}, P_3, P_4,P_2+P_{k,l}|z)\right|^2
\end{equation}
with $k, l$ from \eqref{klTwoDegenerateFieldsFusion}. The $3$-point function on the right-hand side is zero, unless $P_{3}$ and $P_4$ satisfy
\begin{equation}
\pm_3 P_3 \pm_4 P_4 = \pm P_{2m-1,2n-1} + P_{k,l}
\end{equation}
for some $m,n\in\mathbb{Z}_{>0}$, $k,l$ and some choice of signs. If they do not, the first non vanishing term in the limit $P_{2} \to P_{r', s'}$ is the $4$-point function with a derivative of $V_{P_2}$:
\begin{multline}
\lim\limits_{P_2 \to P_{r,s}} \frac{\langle V_{r,s}(z) V_{P_2}(0) V_{P_3}(1) V_{P_4}(\infty)\rangle}{\Upsilon(-2iP_{2})} 
=
\frac{i}{2\Upsilon'(0)}\langle V_{r,s}(z) V'_{r',s'}(0) V_{P_3}(1) V_{P_4}(\infty)\rangle =
\\
= \frac{i}{2\Upsilon'(0)}\sum_{k,l} C_{r,s;r',s'}^{k,l}\left.\frac{\partial}{\partial P_2}C(P_2, P_3, P_4)\right|_{P_2 = P_{r', s'}} \left|\mathfrak{F}^{(b)}(P_{r,s}, P_{r',s'}, P_3, P_4,P_2+P_{k,l}|z)\right|^2.
\end{multline}

Because of the triality transformation this case is also essentially the same as when two screening conditions are satisfied. 

\paragraph{One degenerate field and a screening condition.} We consider what happens in the limit $P_1 \to P_{r,s}$ when external momenta satisfy
\begin{equation}
P_{r,s} + P_2 + P_3 + P_4 = P_{2r', 2s'}, \qquad r', s'\in\mathbb{Z}_{>0}.
\end{equation}
In this limit there will be terms that come from pinching the contour by three poles simultaneously. The leading order of the result will consist only of such terms. In order to define the limit we will split the three poles and then take consecutive limits. One way to do this is to start from \ref{OneDegenerate4point} and then pick the singular terms.

In the denominator of $C(P_2 + P_{k,l}, P_3, P_4)$ there are four $\Upsilon_b$ functions, one of them depends on the sum~$P_2$, $P_3$, $P_4$. 
\begin{equation}
\Upsilon_b\left(\frac{Q}{2} + i(P_2 + P_{k,l}) + iP_3 + iP_4\right) = \Upsilon_b\left(-b \frac{2r' - r+k-1}2-b^{-1} \frac{2s' - s+l-1}2\right).
\end{equation}
In the sum over $k, l$ this $\Upsilon_b$ leads to divergence of the following terms: 
\begin{equation}
\left\{
\begin{aligned} 
r-1&\geq k\geq r+1-2r'\\
s-1&\geq l\geq s+1-2s'
\end{aligned}
\right.
\quad \text{and} \quad
\left\{
\begin{aligned} 
1-r&\leq k\leq r-1-2r'\\
1-s&\leq l\leq s-1-2s'.
\end{aligned}
\right.
\end{equation}
There are $r's' + (r-r')(s-s')>0$ diverging terms if $r'>r$ and $s'>s$ and $r's'>0$ terms otherwise.

\subsection{Timelike Liouville field theory}\label{subsec:TimelikeLFT}
We now discuss the LFT with $b = i\hat{b}\in i\mathbb{R}$. It has several names in the literature including ``Timelike Liouville theory'' and ``Liouville theory with $c<1$''. We will use the first one. In order to distinguish the primary fields in spacelike and timelike LFT we will denote the ones from timelike LFT as $\Phi_{P}$. We will also decorate the quantities in timelike LFT with hats, for example 
\begin{equation}\label{eq:TimelikeDefinitions}
\hat{c} = 1+6\hat{Q}^2, 
\qquad\hat{Q} = b+b^{-1} = i\hat{b}-i\hat{b}^{-1}, 
\qquad\hat{P}_{m,n} = i\left(\frac{mb}{2} + \frac{nb^{-1}}{2}\right) = -\frac{m\hat{b}}2 + \frac{n\hat{b}^{-1}}{2}, 
\qquad \text{etc}.    
\end{equation}
For $b = i\hat{b}\in i\mathbb{R}$ functional equations \eqref{eq:degcrossing} have a unique solution \cite{Zamolodchikov:2005fy}. We choose the normalization in such a way that
\begin{equation}
\begin{gathered}
\hat{C}_{(i\hat{b})}(\hat{P}_1, \hat{P}_2, \hat{P}_3) = \frac{1}{C_{(\hat{b})}(-i\hat{P}_1, -i\hat{P}_2, -i\hat{P}_3)} = \frac{\prod_{\pm_1,\pm_2}\Upsilon_{\hat{b}}((\hat{b}+\hat{b}^{-1})/2+\hat{P}_1\pm_1 \hat{P}_{2}\pm_2 \hat{P}_3)}{\Upsilon'_{\hat{b}}(0)\prod_{k=1}^{3}\Upsilon_{\hat{b}}(\hat{b}+\hat{b}^{-1} + 2\hat{P}_{k})},\\ 
\hat{N}(\hat{P}) = \frac1{(i\hat{P})^2}.
\end{gathered}
\end{equation}
The 3-point function has the following zeroes and poles in the variable $\hat{P}_1$:
\begin{equation}
\begin{aligned}
\text{Poles} \colon& & &\hat{P}_1 = \pm\hat{P}_{m,-n},\quad \hat{P}_1 = 0,\quad  \hat{P}_1 = \frac{m\hat{b}}2, \quad \hat{P}_1 = \frac{n\hat{b}^{-1}}{2},  & m,n\in\mathbb{Z}_{>0}.\\ 
\text{Zeroes} \colon& & &\hat{P}_1 \pm_2 \hat{P}_2 \pm_3 \hat{P}_3  =  \pm \hat{P}_{2m-1, -2n+1}, & m,n\in\mathbb{Z}_{>0}.
\end{aligned}
\end{equation}
The structure constant $\hat{C}^{\hat{P}}_{\hat{P}_1, \hat{P}_2} = \hat{C}(-\hat{P}, \hat{P}_1, \hat{P}_2)/\hat{N}(\hat{P})$ inherits the zeroes and poles of the 3-point function except for the pole at $\hat{P} = 0$, which is canceled by $\hat{N}(\hat{P})$.

\subsubsection{Analytic continuation of OPE and degenerate fields} \label{subsubsec:TimelikeAnalyticStructure}
Operator product expansion \eqref{eq:OPE} in timelike LFT is slightly ill-defined. Since we have $b\in i\mathbb{R}$, the momenta that correspond to degenerate fields now lie in the spectrum: $\hat{P}_{m,n}\in\mathbb{R}$. This leads to a problem in using OPE, because the double poles of the structure constant and the poles \eqref{eq:betapoles} of coefficients $\beta_{\boldsymbol\lambda}$, $\beta_{\boldsymbol{\bar{\lambda}}}$ now lie on the real axis. Moreover, the set of all poles of the latter $\{\hat{P}_{m,n}|\, m,n\in\mathbb{Z}_{>0}\}$ is dense in the real line. This problem is solved by changing the contour of integration to $\mathbb{R}+i\varepsilon$ and taking the limit $\varepsilon\to 0$ after computing the correlation function \cite{Ribault:2015sxa}.

This prescription fixes the problem, because similarly to \eqref{InterestingOPEFact} we have the following identity:
\begin{multline}\label{TimelikeInterestingOPEFact}
|z-w|^{2(\hat{\Delta}(\hat{P}_{m,n}) - \hat{\Delta}(\hat{P}_1)-\hat{\Delta}(\hat{P}_2))}   
\sum\limits_{\boldsymbol{\lambda}, \boldsymbol{\bar{\lambda}}}
(z-w)^{|\boldsymbol{\lambda}|}(\bar{z}-\bar{w})^{|\boldsymbol{\bar{\lambda}}|}
\hspace*{-8pt}\lim\limits_{\hat{P} \to \hat{P}_{m,n}} 
\hspace*{-3pt}
\left[ 
(\hat{P} - \hat{P}_{m,n})^2
\hat{C}^{\hat{P}}_{\hat{P}_1,\hat{P}_2}
\beta_{\boldsymbol{\lambda}}(\hat{P})\beta_{\boldsymbol{\bar{\lambda}}}(\hat{P}) \Phi^{\boldsymbol{\lambda}, \boldsymbol{\bar{\lambda}}}_{\hat{P}}(w)
\right]
=
\\
=
-|z-w|^{2(\hat{\Delta}(\hat{P}_{m,-n}) - \hat{\Delta}(\hat{P}_1)-\hat{\Delta}(\hat{P}_2))} 
\sum\limits_{\boldsymbol{\lambda}, \boldsymbol{\bar{\lambda}}}
(z-w)^{|\boldsymbol{\lambda}|}(\bar{z}-\bar{w})^{|\boldsymbol{\bar{\lambda}}|}
\hspace*{-8pt}\lim\limits_{\hat{P} \to \hat{P}_{m,-n}}\hspace*{-3pt} 
\left[ (\hat{P} - \hat{P}_{m,-n})^2 
\hat{C}^{\hat{P}}_{\hat{P}_1,\hat{P}_2}
\beta_{\boldsymbol{\lambda}}(\hat{P})\beta_{\boldsymbol{\bar{\lambda}}}(\hat{P}) \Phi^{\boldsymbol{\lambda}, \boldsymbol{\bar{\lambda}}}_{\hat{P}}(w)
\right]
\end{multline}
where $\Phi^{\boldsymbol{\lambda}, \boldsymbol{\bar{\lambda}}}_{\hat{P}}(w)$ are the descendant fields. From this identity it follows that in the integral over $\hat{P}$ the singular contributions from $\hat{P}_{m,n}$ and $\hat{P}_{m,-n}$ cancel each other. When proving this identity we treat both sides as meromorphic functions in $\hat{P}$. This means that we make a cutoff in the sum: we must not include the $\boldsymbol{\lambda}$, $\boldsymbol{\bar{\lambda}}$ that correspond to the terms with poles in some fixed small neigbourhood of $\hat{P}_{m,n}$. 

Analytic continuation of operator product expansion \eqref{eq:OPE} is straightforward: since the poles of the structure  constant $\hat{C}(-\hat{P}, \hat{P}_1, \hat{P}_2)/\hat{N}(\hat{P})$ in the $\hat{P}$ complex plane do not depend on external momenta, the contour is never pinched and we do not need to deform it. This, however, means that we do not recover the discrete OPE in the limit $\hat{P}_1\to \hat{P}_{m,n}$ as we did in \ref{subsec:SpacelikeDegenerateFields}. Therefore, we cannot identify the degenerate fields with the primary fields $\Phi_{\hat{P}_{m,n}}$. In spite of that in order to simplify the notation we will denote the (not degenerate) primary field $\Phi_{\hat{P}_{m,n}}$ as $\Phi_{m,n}$. In the following we will not use the true degenerate field, so hopefully there will be no confusion.

We now consider analytic properties of a correlation function $\langle \Phi_{\hat{P}_1}(z_1)\Phi_{\hat{P}_2}(z_2) \ldots \Phi_{\hat{P}_n}(z_n)\rangle$ as a function of $\hat{P}_1$ with arbitrary fixed $\hat{P}_2$, $\hat{P}_3$, $\ldots$, $\hat{P}_n$. Since the position of the poles of the structure constant in the $\hat{P}$ complex plane do not depend on $\hat{P}_1$, the only poles of the correlation function are
\begin{equation}
\hat{P}_1= \pm\hat{P}_{m,-n}, \quad \hat{P}_1 = 0, \quad\hat{P}_1 = \frac{m\hat{b}}{2}, \quad \hat{P}_1 = \frac{n\hat{b}^{-1}}{2},  \qquad m,n\in\mathbb{Z}_{>0}.
\end{equation}
One could get rid of these poles by choosing a different normalization that does not contain the $\Upsilon_{\hat{b}}(-2\hat{P}_1)$ in the denominator of the structure constants. However, we will not do so in order to simplify the analysis in the section \ref{sec:VMS}. 

In some computations correlation functions with $\Phi_{m,n}$ and residues of correlation functions at $\hat{P} = \hat{P}_{m,-n}$ do not contribute (see section \ref{sec:VMS}) and we need the next term in the Laurent expansion of the correlation function. We introduce the following notation:
\begin{equation}\label{TimelikeDerivativeFieldsDefitnition}
\Phi_{m,-n} \equiv \Res\limits_{\hat{P} = \hat{P}_{m,-n}} \Phi_{\hat{P}}, 
\quad 
\Phi'_{m,-n} \equiv \left.\frac{\partial}{\partial\hat{P}}(\hat{P}-\hat{P}_{m,-n})\Phi_{\hat{P}}\right|_{\hat{P} = \hat{P}_{m,-n}}, 
\quad \Phi'_{m,n} \equiv\left.\frac{\partial}{\partial\hat{P}}\Phi_{\hat{P}}\right|_{\hat{P} = \hat{P}_{m,n}},
\end{equation}
so that
\begin{equation}
\begin{aligned}
& \Phi_{\hat{P}} = \frac{\Phi_{m,-n}}{\hat{P}-\hat{P}_{m,-n}} +  \Phi'_{m,-n} + O(\hat{P} -\hat{P}_{m,-n}). \\
& \Phi_{\hat{P}} = \Phi_{m,n} +  (\hat{P}-\hat{P}_{m,n})\Phi'_{m,n} + O((\hat{P} -\hat{P}_{m,n})^2).
\end{aligned}
\end{equation}
This definition of $\Phi'_{m,n}$ and $\Phi'_{m,-n}$ depends on normalization. But, if we chose a different normalization, the result would differ by a term that is proportional to $\Phi_{m,n}$ and $\Phi_{m,-n}$ respectively and by our assumption the terms do not contribute to the result.

\subsubsection{Higher equations of motion}\label{subsubsec:HEMtimelike}

There is an analogue of HEM in timelike LFT \cite{Ribault:2014hia}. The primary field $\Phi_{m,n}$ satisfies the following equation:
\begin{equation}\label{eq:TimelikeHEM}
D_{m,n}^{(ib)}\bar{D}_{m,n}^{(ib)}\Phi_{m,n} (z) = B_{m,n}^{(i\hat{b})}\Phi_{m,-n} (z),
\end{equation}
The constant $B_{m,n}^{(i\hat{b})}$ is again obtained by considering the 3-point functions:
\begin{equation}
B_{m,n}^{(i\hat{b})} = \prod\limits_{k,l}(\hat{P}_1 +\hat{P}_2 + \hat{P}_{k,l})^2 (\hat{P}_1 - \hat{P}_2 + \hat{P}_{k,l})^2 
\cdot 
\frac{\hat{C}(\hat{P}_{m,n}, \hat{P}_1, \hat{P}_2)}{\Res\limits_{\hat{P} = \hat{P}_{m,-n}}\hat{C}(\hat{P}, \hat{P}_1, \hat{P}_2)}.
\end{equation}
It is equal to
\begin{equation}
B_{m,n}^{(i\hat{b})} = 2\frac{\Upsilon'_{\hat{b}}(-2\hat{P}_{m,-n})}{\Upsilon_{\hat{b}}(-2\hat{P}_{m,n})}.
\end{equation} 
Note that if $\hat{b} = b$, we have
\begin{equation}\label{HEMConstantsRelation}
B_{m,n}^{(ib)} = iB_{m,n}^{(b)}.
\end{equation}
\subsubsection{Triality transformation} \label{subsubsec:TrialityTimelike}
For the same reason as in the spacelike LFT, correlation functions of time like LFT transform nicely under \eqref{eq:Triality} with $P_k$ replaced with $\hat{P}_k$:
\begin{multline}\label{TrialityCorrFuncTimelike}
\langle \Phi_{\tilde{P}_1}(z) \Phi_{\tilde{P}_2}(0) \Phi_{\tilde{P}_3}(1) \Phi_{\tilde{P}_4}(\infty)\rangle
=
|z|^{- (\hat{P}_1-\hat{P}_2-\hat{P}_3-\hat{P}_4)(\hat{P}_1-\hat{P}_2+\hat{P}_3+\hat{P}_4)}\times\\\times
|1-z|^{- (\hat{P}_1-\hat{P}_2-\hat{P}_3-\hat{P}_4)(\hat{P}_1+\hat{P}_2-\hat{P}_3+\hat{P}_4)}
\prod_{k=1}^4\frac{\Upsilon_b(-2\hat{P}_k)}{\Upsilon_b(-2\tilde{P}_k)}
\langle \Phi_{\hat{P}_1}(z) \Phi_{\hat{P}_2}(0) \Phi_{\hat{P}_3}(1) \Phi_{\hat{P}_4}(\infty)\rangle
\end{multline}
\section{Virasoro minimal string}\label{sec:VMS}

Virasoro minimal string is believed to have several equivalent definitions \cite{Collier:2023cyw}. We will think of it as a theory of two-dimensional Liouville quantum gravity with timelike LFT describing the matter:
\begin{equation}\label{eq:VMSdef}
\text{VMS} = \text{Spacelike LFT}~\otimes~\text{Timelike LFT}~\otimes~\text{Fadeev\,--\,Popov ghosts}.
\end{equation}
The ghost system consists of two fields $B$ and $C$ with the following action:
\begin{equation}
S_{\gh} = \int d^2z\, (C\bar\partial B + \bar{C} \partial\bar B).
\end{equation}
It is a conformal field theory with the central charge $c_{\gh} = -26$. The conformal dimensions of the fields are~$\Delta_B = 2$, $\Delta_C = -1$.

In order for the conformal anomaly to vanish, the total central charge of the three CFTs in \eqref{eq:VMSdef} should be equal to $0$. It follows that
\begin{equation}
\hat{c} = 26-c = 1-6\left(b-b^{-1}\right)^2,
\end{equation}
i.\,e. the parameter $\hat{b}\in\mathbb{R}$ used in the subsection \ref{subsec:TimelikeLFT} is equal to the parameter $b\in\mathbb{R}$ that parametrizes the central charge of spacelike LFT.

In correlation functions in quantum gravity the matter fields $\Phi_{\hat{P}}$ are dressed by spacelike Liouville primaries $V_{P}$. The total dimension of the resulting field must be equal to $1$. We choose the signs in such a way that:
\begin{equation}\label{eq:TimelikeSpacelikeMomenta}
\Delta(P) + \hat{\Delta}(\hat{P}) = 1, 
\quad \Longrightarrow\quad \hat{P} = iP.
\end{equation}

The CFTs in equation \eqref{eq:VMSdef} are defined on a Riemann surface with $g$ handles. The objects of interest are not the correlation functions themselves, but their invariant combinations known as correlation numbers (also known as quantum volumes). They are obtained from the $n$-point correlation functions by integrating over the moduli space $\mathcal{M}_{g,n}$:
\begin{equation}\label{eq:Vgndef}
V_{g,n}(P_1, \ldots, P_{n}) 
=
\int_{\mathcal{M}_{g,n}} \,  	
Z_{\mathrm{gh}}\cdot\langle V_{P_1}(z_1) \ldots V_{P_n}(z_n)\rangle \langle\Phi_{iP_1}(z_1) \ldots \Phi_{iP_n}(z_n)\rangle.
\end{equation}
Here $Z_{\gh}$ is the corresponding correlation function of ghosts, which is determined by the Riemann surface. Since the correlation functions depend only on the dimensions of the fields and do not depend on the order of the fields, $V_{g,n}(P_1, \ldots, P_n)$ is symmetric and even in each argument. 

The definition \eqref{eq:Vgndef} is very hard to work with: first one has to compute the correlation functions in the integrand, which are themselves usually written in terms of multiple integrals of complicated special functions and are almost never known in closed form, and then one has to perform the integral over the moduli space. In spite of that in \cite{Collier:2023cyw} it was conjectured that integrals are polynomials in $\{P_k\}$. In particular, 
\begin{equation}\label{eq:V04hypothesis}
V_{0,4}(P_1, P_2, P_3, P_4) = \frac{c-13}{24} + \sum_{k=1}^4P_k^2, \qquad V_{1,1}(P_1) = \frac{1}{24}\left(\frac{c-13}{24} + P_1^2\right).
\end{equation}

In this paper we consider VMS on a sphere ($g = 0$) and on a torus ($g = 1$).

\subsection{Definition and symmetries of \texorpdfstring{$V_{0,4}$}{V04}}
The first nontrivial volume on a sphere is the one with $4$ field insertions: $V_{0,4}(P_1, P_2, P_3, P_4)$. The moduli space of the sphere with $4$ marked points $\mathcal{M}_{0,4}$ is parametrized by the position $z_1$ of one of the points:~$z_1\in\mathbb{C}\setminus\{z_2, z_3, z_4\}$. Taking into account that the ghost correlation function on the sphere contains three $C\bar{C}$ fields, we have
\begin{multline}\label{eq:V04def}
V_{0,4}(P_1, P_2, P_3, P_4) 
=
\\
=
\int_{\mathbb{C}} d^2z_1\, \langle C\bar{C}(z_2)C\bar{C}(z_3)C\bar{C}(z_4)\rangle
\langle V_{P_1}(z_1) V_{P_2}(z_2) V_{P_3}(z_3) V_{P_4}(z_4)\rangle
\langle \Phi_{iP_1}(z_1) \Phi_{iP_2}(z_2) \Phi_{iP_3}(z_3) \Phi_{iP_4}(z_4)\rangle.
\end{multline}
Here we denote $d^2z_1 \equiv \mathcal{N} d \Re z_1 d \Im{z}_1 $. We include the normalization factor $\mathcal{N} = 4/\pi^2$ so that the result agrees with \eqref{eq:V04hypothesis}.

One can see that $V_{0,4}$ does not depend on the position of the fixed points. Indeed, the combined conformal dimensions of the fields and ghosts in the fixed points $z_2$, $z_3$, $z_4$ are equal to $0$ and that of the fields in $z_1$ is equal to $1$. Therefore, the integrand together with $d^2z_1$ transforms like a scalar function under conformal transformations combined with the corresponding change of the integration measure. Hence, using global conformal transformations, one can set $z_2 = 0$, $z_3 = 1$, $z_4 = \infty$\footnote{This equality is in the sense of equation \eqref{eq:4pointinfty}.}. We will denote the corresponding cross-ratio as $z$. With this choice of $z_2$, $z_3$, $z_4$ the correlation function of ghosts is equal to one and we have
\begin{equation}\label{eq:V0401infty}
V_{0,4}(P_1, P_2, P_3, P_4) 
=
\int_{\mathbb{C}} d^2z\, 
\langle V_{P_1}(z) V_{P_2}(0) V_{P_3}(1) V_{P_4}(\infty)\rangle
\langle \Phi_{iP_1}(z) \Phi_{iP_2}(0) \Phi_{iP_3}(1) \Phi_{iP_4}(\infty)\rangle.
\end{equation}
Note that because of the relation \eqref{eq:TimelikeSpacelikeMomenta} fields in at least one of the LFT correlation functions do not lie in the spectrum. Therefore this equation should be understood as an analytic continuation.

 \begin{figure}[H]
 	\begin{subfigure}[b]{0.49\linewidth}
 		\centering
 		\begin{tikzpicture}[scale=1]
 			\begin{axis}[
 				width=9cm,
 				axis lines = center,
 				xmin=-1.5, xmax=2.75, ymin=-1.7,ymax=1.7,
 				xlabel={$\Re z$},
 				ylabel={$\Im z$},
 				xtick = {-10},
 				xticklabels ={},
 				ytick = {-10},
 				yticklabels ={},
 				x label style={anchor=north},
 				y label style={anchor=east},
 				extra x ticks={0, 1, -1, .5, 2},
 				extra x tick labels = {$0$, $1$, $-1$, $\frac12$, $2$},
 				extra y ticks={1,-1},
 				extra y tick labels = {$i$, $-i$},
 				x tick label style={xshift={(\ticknum==0)*(-.8em)},xshift={(\ticknum==1)*(.4em)},xshift={(\ticknum==2)*(-.8em)},xshift={(\ticknum==3)*(-.4em)},xshift={(\ticknum==4)*(.6em)}},
 				y tick label style={yshift={(\ticknum==0)*(.6em)},yshift={(\ticknum==1)*(-.6em)}},
 				grid=none,
 				extra x tick style={grid=none},
 				extra y tick style={grid=none},
 				axis equal image,
 				]
				
 				\draw[thick](.5,-2)--(.5,2); 
 				\addplot[thick,domain=0:360,samples=100,smooth] ({cos(x)},{sin(x)});
 				\addplot[thick,domain=0:360,samples=100,smooth] ({1+cos(x)},{sin(x)});
				
 				\draw[] (axis cs:.25,.25) node[]{$\Circled{1}$};
 				\draw[] (axis cs:.75,.25) node[]{$\Circled{4}$};
 				\draw[] (axis cs:1.25,.5) node[]{$\Circled{5}$};
 				\draw[] (axis cs:-.25,.5) node[]{$\Circled{2}$};
 				\draw[] (axis cs:1.75,1.25) node[]{$\Circled{6}$};
 				\draw[] (axis cs:-.75,1.25) node[]{$\Circled{3}$};
 			\end{axis}
 		\end{tikzpicture}
 		\caption{Domains \eqref{eq:zdomains} of integration by $z$.}\label{fig:domainsz}
 	\end{subfigure}
 	\begin{subfigure}[b]{0.49\linewidth}
 		\centering
 		\begin{tikzpicture}[scale=1]
 			\begin{axis}[
 				width=9cm,
 				axis lines = center,
 				xmin=-1.5, xmax=1.5, ymin=-1.1,ymax=2,
 				xlabel={$\Re \tau$},
 				ylabel={$\Im \tau$},
 				xtick = {-10},
 				xticklabels ={},
 				ytick = {-10},
 				yticklabels ={},
 				x label style={anchor=north},
 				y label style={anchor=east},
 				extra x ticks={0, 1, -1, .5,-.5},
 				extra x tick labels = {$0$, $1$, $-1$, $\frac12$, $-\frac12$},
 				extra y ticks={1,-1},
 				extra y tick labels = {$i$, $-i$},
 				x tick label style={xshift={(\ticknum==0)*(-.8em)},xshift={(\ticknum==1)*(.4em)},xshift={(\ticknum==5)*(-.6em)},xshift={(\ticknum==2)*(-.6em)},xshift={(\ticknum==3)*(-.4em)}},
 				y tick label style={yshift={(\ticknum==0)*(-.6em)},yshift={(\ticknum==1)*(.6em)}},
 				grid=none,
 				extra x tick style={grid=none},
 				extra y tick style={grid=none},
 				axis equal image,
 				]
				
 				\draw[thick](axis cs: .5,0.866)--(axis cs: .5,2); 
 				\draw[thick](axis cs: -.5,0.866)--(axis cs: -.5,2); 
 				\draw[thick, dashed](axis cs: -.5,0)--(axis cs: -.5,.866); 
 				\draw[thick, dashed](axis cs: .5,0)--(axis cs: .5,.866); 
				
 				\addplot[thick,domain=0:360,samples=100,smooth] ({cos(x)},{sin(x)});
				
 				\draw[] (axis cs:.25,1.25) node[]{$\Circled{F}$};
 			\end{axis}
 		\end{tikzpicture}
 		\caption{Domain $F$ of integration by $\tau$.}\label{domainstau}
 	\end{subfigure}
 	\caption{}\label{fig:domains}
 \end{figure}
In some computations, especially in numeric ones, it is useful to reduce the domain of integration using global conformal transformations. Global conformal transformations that leave the set $\{0, 1, \infty\}$ invariant are $T(z) = 1-z$, $S(z) = z^{-1}$ and their compositions. Consider a partition of $\mathbb{C}$ into the following $6$ domains:
\begin{equation}\label{eq:zdomains}
	\begin{aligned}
		&1:\quad& &\Re z\leq \frac12,~~ |1-z|\leq1, \qquad &2:\quad   &|z|\leq 1,~~ |1-z|\geq 1,\\
		&3:\quad& &\Re z\leq \frac12,~~ |z|\geq1, \qquad &4:\quad   &\Re z\geq \frac12,~~ |z|\leq 1,\\
		&5:\quad& &|1-z|\leq 1,~~ |z|\geq1, \qquad &6:\quad  & \Re z \geq \frac12,~~ |1-z|\geq 1.
	\end{aligned}
\end{equation}
The domains $2$-$6$ are mapped to the domain $1$ by transformations $TST$, $ST$, $T$, $TS$, $S$ respectively. Then, for the volume we have
\begin{multline}\label{eq:zpermutations}
V_{0,4}(P_1, P_2, P_3, P_4) 
=\\=
\int_{\Circled{1}} d^2z\, 
\sum_{\sigma}\langle V_{P_1}(z) V_{P_{\sigma(2)}}(0) V_{P_{\sigma(3)}}(1) V_{P_{\sigma(4)}}(\infty)\rangle
\langle \Phi_{iP_1}(z) \Phi_{iP_{\sigma(2)}}(0) \Phi_{iP_{\sigma(3)}}(1) \Phi_{iP_{\sigma(4)}}(\infty)\rangle,
\end{multline}
where the sum is over all permutations of $\{2, 3, 4\}$. 

The integration domain $\Circled{1}$ is mapped to the fundamental domain $F$ (see fig.~\ref{domainstau}) by the following invertible transformation of variables $z \to \tau$:
\begin{equation}\label{eq:TauToZ}
\tau(z)=i\frac{{}_{2}F_{1}\left(\frac{1}{2},\frac{1}{2};1\Big| 1-z\right)}{{}_{2}F_{1}\left(\frac{1}{2},\frac{1}{2};1\Big| z\right)}.
\end{equation}
The inverse transformation has the form 
\begin{equation}\label{eq:ZToTau}
z(\tau) = \frac{\theta_2(q)^4}{\theta_3(q)^4}, \qquad \text{where} \qquad q = e^{i\pi\tau}. 
\end{equation}
After we change the coordinates and rewrite the conformal blocks in terms of elliptic conformal blocks (see appendix~\ref{appendix:Blocks}), we have
\begin{multline}\label{eq:V04H}
V_{0,4}(P_1, P_2,P_3, P_4) = \int_{F} d^2\tau\, \sum_{\sigma} 	\int_{C}dP\,   \int_{\mathbb{R}+i\varepsilon}d\hat{P} (i\hat{P})^2\,  
e^{-2\pi (P^{2}+\hat{P}^2) \Im\tau} 16^{2(P^2+\hat{P}^2)}
\times
\\ 
\times
\frac{C\left(P_{1},P_{\sigma(2)},P\right)C\left(-P,P_{\sigma(3)},P_{\sigma(4)}\right)}{C(P_{1},P_{\sigma(2)},i\hat{P})C(-i\hat{P},P_{\sigma(3)},P_{\sigma(4)})}
\Big|\mathfrak{H}^{(b)}(\{P_1, P_{\sigma(k)}\}, P|q)\Big|^{2}
\Big|\mathfrak{H}^{(ib)}(\{iP_1, iP_{\sigma(k)}\}, \hat{P}|q)\Big|^{2}. 
\end{multline}
Here the contour of integration $C$ is $\mathbb{R}$ if all external momenta lie in the spectrum. If not, as described in \ref{subsec:SpacelikeDegenerateFields}, the contour can acquire discrete contributions.

The formula \eqref{eq:V04H} allows us to easily see an additional symmetry of the $V_{0,4}(P_1, P_2, P_3, P_4)$, namely the ``triality symmetry''. We already encountered the triality transformation \eqref{eq:Triality} in \ref{subsubsec:TrialitySpacelike} and \ref{subsubsec:TrialityTimelike}. There we saw that the elliptic conformal blocks are invariant under the transformation and the structure constants are invariant, except for the normalization factors. In the formula \eqref{eq:V04H} the normalization factors from timelike and spacelike structure constants cancel, and we have
\begin{multline}\label{eq:V04TrialitySymmetry}
V_{0,4}(P_1, P_2, P_3, P_4) =\\= 
V_{0,4}\left(\frac{P_1 + P_2 + P_3 + P_4}2, \frac{P_1 + P_2 - P_3 - P_4}2,\frac{P_1 - P_2 + P_3 - P_4}2,\frac{P_1 - P_2 - P_3 + P_4}2\right).
\end{multline}

\subsection{Convergence and analytic properties of the integral over moduli}\label{subsubsec:V04convergence}

The integral in equation \eqref{eq:V0401infty} (or \eqref{eq:V04def}) may receive divergent or non-analytic contributions only from the neighbourhoods of fixed points or infinity. The behavior of the integral depends on what the external momenta are. We consider the following cases:

\paragraph{Real external momenta.}
Since the integrand together with $d^2z$ transforms like a scalar, there is no divergent contribution from infinity. Now we consider the behavior of the integrand near a fixed point. Without loss of generality, we choose the fixed point to be $0$ and consider the integral over a disk~$\mathbb{D}$ of radius $r< 1$ centered at the origin.
\begin{equation}\label{DiskV04Integral}
	\int_{\mathbb{D}} d^2z\, 
	\langle V_{P_1}(z) V_{P_2}(0) V_{P_3}(1) V_{P_4}(\infty)\rangle
	\langle \Phi_{iP_1}(z) \Phi_{iP_2}(0) \Phi_{iP_3}(1) \Phi_{iP_4}(\infty)\rangle
\end{equation}
The behaviour of the integrand depends on the contour of integration over $P$ in the correlation function of spacelike LFT. If the external momenta $\{P_k\}$ are real, the contour is $\mathbb{R}$. In this case we have
\begin{equation}\label{eq:Convergence1}
	\int_{\mathbb{R}} dP\, P^2\int_{\mathbb{R}+i\varepsilon} d\hat{P}\, |z|^{2(P^2 + \hat{P}^2 - 1)} \sim \frac{1}{|z|^{2}\,(\ln|z|)^2}, \qquad z\to 0.    
\end{equation} 
This contribution to the integral over $\mathbb{D}$ is convergent. Thus, when the external momenta $\{P_k\}$ are real, $V_{0,4}(P_1, P_2, P_3, P_4)$ is well defined.

\paragraph{Complex external momenta.}
Now we consider the analytic continuation of $V_{0,4}(P_1, P_2, P_3, P_4)$ to complex $\{P_{k}\}$. In this case, the contour of integration in the correlation function of spacelike LFT is $\mathbb{R}$ with possible additions of small circles around some points in the complex $P$ plane. If the discrete contributions are present, the respective integrals over $P$ are reduced to residues at points $P = P_0$ with $\Im P_0\neq0$. It results in divergent contributions to the integral over $\mathbb{D}$ if $\Re P_0^2<0$. In order to define the analytic continuation of \eqref{DiskV04Integral} in this case we first perform the integral over $\mathbb{D}$ with $\Re P_0^2>0$ and then analytically continue the result to the region with $\Re P_0^2<0$. 

Since $|z|<1$ when $z\in\mathbb{D}$, we use the s-channel decomposition of the correlation functions. After integrating over $\mathbb{D}$ and taking the residue at $P = P_0$ we get
\begin{multline}\label{VolumeConvergenceResidue}
2\pi i \Res\limits_{P = P_0} \left[C(P_1, P_2, P)C(-P,P_3,P_4)\right]
\times
\\
\times 
\int_{\mathbb{R}+i\varepsilon} d\hat{P} (i\hat{P})^2\frac{1}{C(P_1, P_2, -i\hat{P})C(i\hat{P},P_3,P_4)}\cdot \pi r^{2\pi(P_0^2+\hat{P}^2)} 
\left[\frac{1}{{P_0^2+\hat{P}^2}} + \ldots\right].
\end{multline}
Here inside the square brackets is the result of integration over $\mathbb{D}$ and taking the residue at $P = P_0$ of the product of 4 conformal blocks: two holomorphic and two antiholomorphic.  As a result of the integration, we now have additional ``on-shell'' poles $\hat{P} = \pm i\sqrt{P_0^2}$ from the leading term in the product. 

Using the Zamolodchikov recursion relations one can prove \cite{Artemev:2025pvk} that the product of ellipic conformal blocks together with the prefactor has the form
\begin{equation}\label{PrefactorsInProductOfBlocks}
	q^{P^2+\hat{P}^2}\mathfrak{H}^{(b)}(\{P_1, P_{\sigma(k)}\}, P|q)\mathfrak{H}^{(ib)}(\{iP_1, iP_{\sigma(k)}\}, \hat{P}|q)
	= q^{P^2 + \hat{P}^2} + \sum_{N = 1}^{\infty} (P^2+\hat{P}^2+N)q^{P^2+\hat{P}^2+N} \mathfrak{h}_N(\{P_k\}, P, \hat{P}),
\end{equation}
where the only poles of $\mathfrak{h}_N$ in $P$ and $\hat{P}$ are located at $P = P_{m,n}$ and $\hat{P} = \hat{P}_{m,n}$. The same is apparently true for $\mathfrak{F}$ blocks. Moreover, since we are integrating over a disk, only the terms that depend on $|z|^2$ contribute to the integral. This means that in the product of holomorphic and antiholomorphic blocks the relevant terms will have prefactors $(P^2+\hat{P}^2+N)^2$. This means that after integrating over the moduli the prefactors will cancel the additional poles that could have appeared in \eqref{VolumeConvergenceResidue} from the descendant fields and the integral in \eqref{VolumeConvergenceResidue} has the form
\begin{equation}
	f(P_0^2) = \int_{\mathbb{R}+i\varepsilon}d\hat{P}\,\frac{g(P, P_0^2)}{\hat{P}^2 + P_0^2},
\end{equation}
where $g(P, P_0^2)$ is entire in $P_0^2$. 

Now we analytically continue the integral in $P_0^2$ from $\Re P_0^2>0$ to $\Re P_0^2<0$. Since the positions of the poles depend on $\sqrt{P_0^2}$, we might have a branch cut at zero. The difference between $f(P_0^2)$ and  $f(e^{2\pi i}P_0^2)$ is the sum of residues at the poles. But, since the structure constants of spacelike LFT have a pole at $P = P_0$, the structure constants of timelike LFT have zeros at $\hat{P} = \pm i\sqrt{P_0^2}$ and the residues at $\hat{P} = \pm i\sqrt{P_0^2}$ vanish. 
 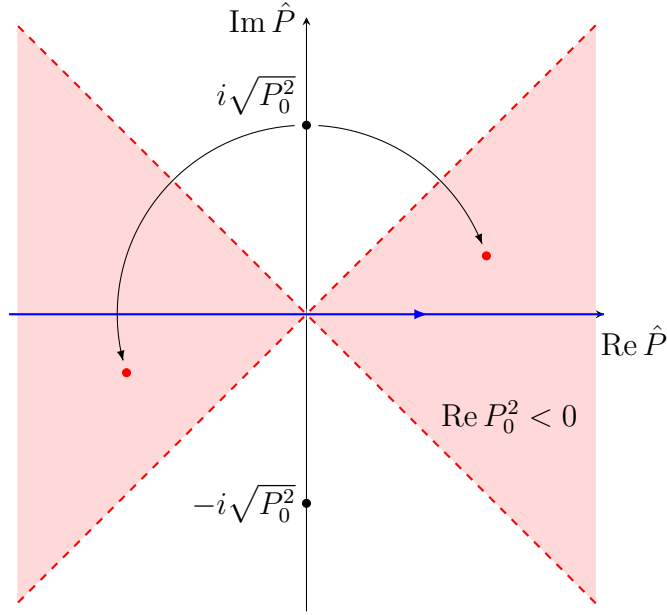
\begin{figure}[H]
 	\begin{tikzpicture}[trig format =rad, >=latex]
 		\begin{axis}[
 			width=11cm,
 			axis lines=center,
 			xlabel={$\Re \hat{P}$},
 			ylabel={$\Im \hat{P}$},
 			x label style={anchor=north, xshift=.4cm},
 			y label style={anchor=east},
 			xmin=-5.5, xmax=5.5,
 			ymin=-5.5,ymax=5.5,
 			xtick = {0},
 			xticklabels ={$0$},
 			ytick = {-10},
 			extra x ticks={0},
 			extra x tick labels = {$ $},
 			x tick label style={xshift={(\ticknum==0)*(-.8em)}},
 			axis equal image,
 			]

 			\draw[fill=black](0, 3.5) circle (1.5pt) node[above left]{$i\sqrt{P_0^2}$};
 			\draw[fill=black](0, -3.5) circle (1.5pt) node[left]{$-i\sqrt{P_0^2}$};
			
 			\draw[draw=none, fill=red!15](5.35, 5.35) -- (5.35,-5.35) -- (0,0);
 			\draw[draw=none, fill=red!15](-5.35, 5.35) -- (-5.35,-5.35) -- (0,0);
 			\draw(3.75,-1.9) node {$\Re P_0^2 <0$};
			
 			\draw[thick, dashed, red](5.35, 5.35) -- (-5.35,-5.35);
 			\draw[thick, dashed, red](5.35, -5.35) -- (-5.35,5.35);
			
 			\draw[fill=red, color=red]({3.5*cos(pi/10)}, {3.5*sin(pi/10)}) circle (1.5pt);
 			\draw[fill=red, color=red]({-3.5*cos(pi/10)}, {-3.5*sin(pi/10)}) circle (1.5pt);
			
 			\draw[blue, thick,->-](-5.5, 0) --(10,0);

 			\addplot[domain={pi/10+pi/50}:{pi/2-pi/50},samples=100,smooth, <-] ({3.5*cos(x)},{3.5*sin(x)});
			
 			\addplot[domain={pi/2+pi/50}:{pi+pi/10-pi/50},samples=100,smooth, ->] ({3.5*cos(x)},{3.5*sin(x)});
			
			
 		\end{axis}
 	\end{tikzpicture}
 	\caption{The ``on-shell'' poles are drawn for $P_0\in\mathbb{R}$. The two ways to analytically continue the discrete contributions in $P_0^2$ are shown by the two arrows. The region shaded in red is where the poles lie if~$\Re P_0^2<0$. The contour of integration is shown in blue.}\label{}
 \end{figure}

Hence, we have defined an analytic continuation of $V_{0,4}(P_1, P_2, P_3, P_4)$ for general case complex momenta and the integral over the moduli does not change the analytic structure of $V_{0,4}$.

\paragraph{Degenerate fields.} When one of the external momenta corresponds to a degenerate field, for example $P_1 = P_{m,n}$, the contour of integration in $P$ is pinched by pairs of poles and the spacelike LFT correlation function is reduced to discrete contributions. In this case as we will see in \ref{subsubsec:V04DegenerateFields}, we have terms that contain derivatives of fields. The terms are proportional to
\begin{equation}\label{DiskV04IntegralDegenerateCase1}
\int_{\mathbb{D}} d^2z\, 
\int_{\mathbb{R}+i\varepsilon} d\hat{P} (i\hat{P})^2\frac{1}{C(P_1, P_2, -i\hat{P})C(i\hat{P},P_3,P_4)}
\cdot
|z|^{2(\hat{P}^2+(P_2+P_{k,l})^2 - 1)}\left[\ln |z|^2 + \ldots\right]
\end{equation}
or
\begin{equation}
\int_{\mathbb{D}} d^2z\, 
\int_{\mathbb{R}+i\varepsilon} d\hat{P} (i\hat{P})^2
\left.\frac{\partial}{\partial P_1}\frac{1}{C(P_1, P_2, -i\hat{P})C(i\hat{P},P_3,P_4)}\right|_{P_1 = P_{m,n}}
|z|^{2(\hat{P}^2+(P_2+P_{k,l})^2 - 1)}\left[1 + \ldots\right].
\end{equation}
Here \eqref{DiskV04IntegralDegenerateCase1} corresponds to taking the derivative in the external momenta of the product of conformal blocks. Again, if $\Re (P_2+P_{k,l})^2>0$ these integrals converge. If not, we define the analytic continuation as before, by first taking the integral over $\mathbb{D}$:
\begin{equation}\label{DiskV04IntegralDegenerateCase1After}
\int_{\mathbb{R}+i\varepsilon} d\hat{P} (i\hat{P})^2\frac{2\pi|r|^{2(\hat{P}^2+(P_2+P_{k,l})^2)}}{C(P_1, P_2, -i\hat{P})C(i\hat{P},P_3,P_4)}
\left[ 
\frac{\left(-1 + 2(\hat{P}^2+(P_2+P_{k,l})^2)\ln r^2\right)}{4(\hat{P}^2+(P_2+P_{k,l})^2)^2} + \ldots
\right]
\end{equation}
or
\begin{equation}
	\int_{\mathbb{D}} d^2z\, 
	\int_{\mathbb{R}+i\varepsilon} d\hat{P} (i\hat{P})^2
	\left.\frac{\partial}{\partial P_1}\frac{1}{C(P_1, P_2, -i\hat{P})C(i\hat{P},P_3,P_4)}\right|_{P_1 = P_{m,n}}
	\left[\frac{\pi|r|^{2(\hat{P}^2+(P_2+P_{k,l})^2)}}{\hat{P}^2+(P_2+P_{k,l})^2} + \ldots\right].
\end{equation}
As before, the ``on-shell'' poles at $\hat{P} = \pm i(P_2 + P_{k,l})$ are canceled by zeroes of the structure constants. Note that in  \eqref{DiskV04IntegralDegenerateCase1After} the structure constants have double zero which cancels the double pole. The potential additional poles that could have appeared from the descendant fields are canceled by the prefactors in the product of 4 conformal blocks. Here it is important that prefactors are not just $P^2 + \hat{P}^2 + N$, but $(P^2 + \hat{P}^2 + N)^2$.

As a result of the analytic continuation we also see that in the limit $r\to 0$ all the contributions to the integral over $\mathbb{D}$ vanish. In what follows in order to avoid complications in using HEM we will consider the integral not over $\mathbb{C}$, but over $\mathbb{C}$ without small disks in the vicinity of marked points (and infinity). Because of the above mentioned property this will not change the result in the limit $r\to 0$.

\subsection{Integration by parts}\label{subsec:IntegrationByParts}

Below we prove the following ``integration by parts'' formulas for $4$-point functions:
\begin{equation}\label{eq:IBP1}
	\langle V_{m,n} D_{m,n}^{(ib)}\bar{D}^{(ib)}_{m,n} \Phi_{m,n}(z) 
	\ldots\rangle 
	=
	\partial\bar{\partial} \mathcal{H}_{m,n}\bar{\mathcal{H}}_{m,n}\langle V_{m,n}\Phi_{m,n}(z)\ldots\rangle
\end{equation}
and
\begin{multline}\label{eq:IBP2}
	\langle V_{m,n}' D_{m,n}^{(ib)}\bar{D}^{(ib)}_{m,n} \Phi_{m,n}(z) \ldots
	\rangle 
	=  
	\frac{P_{m,n}}{\hat{P}_{m,n}}
	\langle V_{m,n} D_{m,n}^{(ib)}\bar{D}^{(ib)}_{m,n} \Phi'_{m,n}(z) \ldots
	\rangle 
	+
	\langle \Phi_{m,n} D_{m,n}^{(b)}\bar{D}^{(b)}_{m,n}V_{m,n}' (z) \ldots \rangle 
	+ 
	\\
	\hfill +
	(-1)^{mn}
	\bar{\partial}\bar{\mathcal{H}}_{m,n} \langle \Phi_{m,n}D_{m,n}^{(b)}V'_{m,n}(z) \ldots \rangle 
	+
	(-1)^{mn}
	\partial\mathcal{H}_{m,n} \langle \Phi_{m,n}\bar{D}_{m,n}^{(b)}V'_{m,n}(z) \ldots \rangle
	+
	\\
	+
	\partial \bar{\partial}\mathcal{H}_{m,n}\bar{\mathcal{H}}_{m,n}\left[\langle  V'_{m,n} \Phi_{m,n}(z)\ldots\rangle -\frac{P_{m,n}}{\hat{P}_{m,n}} \langle  V_{m,n} \Phi'_{m,n}(z)\ldots\rangle\right].
\end{multline}
Here operators with $(b)$ act only on fields in spacelike LFT, while those with $(ib)$ act only on fields in timelike LFT. The $\ldots$ in correlation functions stand for the other $3$ fields, which we assume to be primary. We also assume that one of the fields is placed at infinity. The $\mathcal{H}_{m,n}$ and $\bar{\mathcal{H}}_{m,n}$ are combinations of differential operators that act on spacelike and timelike LFT correlation functions. The formulas \eqref{eq:IBP1} and \eqref{eq:IBP2} may be reformulated in a more general form as operator identities (see \cite{Belavin:2006ex}), but we refrain from doing so for clarity.

In order to prove the above formulas we first rewrite the correlation functions with $D_{m,n}^{(b)}$ in terms of correlation functions of primary fields, which are acted on by differential operators. The operators $D_{m,n}^{(b)}$, $\bar{D}_{m,n}^{(b)}$ are polynomials of Virasoro generators (see \eqref{eq:Dmnb}) that produce descendant fields. Correlation functions with the descendant fields may be rewritten as
\begin{equation}
\left\langle L_{-\boldsymbol{\lambda}} V_{P_1}(z) V_{P_2}(z_2) V_{P_3}(z_3)V_{P_4}(z_4)\right\rangle 
= 
\mathcal{L}_{-\boldsymbol{\lambda}}
\left\langle
V_{P_1}(z) V_{P_2}(z_2) V_{P_3}(z_3)V_{P_4}(z_4)\right\rangle, 
\end{equation}
where
\begin{equation}\label{eq:Ldiff1}
L_{-\boldsymbol{\lambda}} = L_{-\lambda_1}L_{-\lambda_2}\ldots, 
\quad 
\mathcal{L}_{-\boldsymbol{\lambda}} = \mathcal{L}_{-\lambda_1}\mathcal{L}_{-\lambda_2}\ldots, 
\qquad  
\mathcal{L}_{-\lambda} = \sum_{j=2}^{4} \left(\frac{(\lambda-1)\Delta_j}{(z_j-z)^{\lambda}} - \frac{\partial_j}{(z_j-z)^{\lambda-1}} \right).
\end{equation}

Correlation functions in LFT satisfy Ward identities. We will need them for $4$-point functions with 1 descendant field and 3 primary fields:
\begin{equation}\label{eq:WardIdentities}
\begin{aligned}
\sum_{i=1}^{4}\partial_i\left\langle V_{P_1}^{\boldsymbol{\lambda}, \boldsymbol{\bar{\lambda}}}(z_1) V_{P_2}(z_2)V_{P_3}(z_3)V_{P_4}(z_4)\right\rangle & = 0, \\
\sum_{i=1}^{4}(\Delta_i + z_i\partial_i)\left\langle V_{P_1}^{\boldsymbol{\lambda}, \boldsymbol{\bar{\lambda}}}(z_1) V_{P_2}(z_2)V_{P_3}(z_3)V_{P_4}(z_4)\right\rangle & = 0, \\
\sum_{i=1}^{4}(2z_i \Delta_i + z_i^2\partial_i)\left\langle V_{P_1}^{\boldsymbol{\lambda}, \boldsymbol{\bar{\lambda}}}(z_1) V_{P_2}(z_2)V_{P_3}(z_3)V_{P_4}(z_4)\right\rangle & = -\left\langle L_1V_{P_1}^{\boldsymbol{\lambda}, \boldsymbol{\bar{\lambda}}}(z_1) V_{P_2}(z_2)V_{P_3}(z_3)V_{P_4}(z_4)\right\rangle, \\
\end{aligned}
\end{equation}
These identities allow us to express the action of $z_2$, $z_3$, $z_4$ derivatives in \eqref{eq:Ldiff1} in terms of the action of $z$ derivative. After that we are free to set $z_2 = 0$, $z_3 = 1$ and $z_4 = \infty$. Setting $z_4 =\infty$ means taking the limit $\lim\limits_{z_4\to\infty}|z_4|^{2\Delta_4}$. Since the action of $L_1$ does not change the dimension of the field $V_{P_4}(z_4)$, this action is supressed by $1/z_4$ in the last identity in \eqref{eq:WardIdentities}. Hence, in the limit we have
\begin{equation}
\begin{aligned}
\partial_2 &= (z-1)\partial + \Delta_2 + \Delta_3 - \Delta_4 + \Delta_1, 
\\
\partial_3 &= -z\partial - \Delta_2 - \Delta_3 + \Delta_4 - \Delta_1, 
\\
\partial_4 &= 0.
\end{aligned}
\end{equation}
Then
\begin{equation}\label{eq:mathcalLlambda}
\mathcal{L}_{-\lambda} 
= 
a_\lambda^z\partial + a_\lambda^1\Delta_1 + a_\lambda^2\Delta_2 + a_\lambda^3\Delta_3 + a_\lambda^4\Delta_4,
\end{equation}
with
\begin{equation}
\begin{aligned}
	& a_\lambda^z = \frac{1}{(1-z)^{\lambda-1}}-\frac{1}{(1-z)^{\lambda-2}} + (-1)^{\lambda}\left(\frac{1}{z^{\lambda-2}}-\frac{1}{z^{\lambda-1}}\right), & & a_\lambda^1 = (-1)^{\lambda} \frac{1}{z^{\lambda-1}} + \frac{1}{(1-z)^{\lambda-1}}, \\
	& a_\lambda^2 = (-1)^{\lambda} \frac{\lambda-1}{z^{\lambda}} + (-1)^{\lambda} \frac{1}{z^{\lambda-1}} + \frac{1}{(1-z)^{\lambda-1}}, &
	& a_\lambda^3 = \frac{\lambda-1}{(1-z)^{\lambda}} +  \frac{1}{(1-z)^{\lambda-1}} + (-1)^{\lambda}\frac{1}{z^{\lambda-1}},
	\\
	& a_\lambda^4 = -a_\lambda^1 = (-1)^{\lambda-1} \frac{1}{z^{\lambda-1}} - \frac{1}{(1-z)^{\lambda-1}}.
\end{aligned}
\end{equation}

Now we consider the following product of correlation functions:
\[
\langle\Phi_{\hat{P}_1}(z) \ldots \rangle  \mathcal{D}^{(b)}_{m,n} \langle V_{P_1}(z) \ldots \rangle, \qquad \text{with}\qquad \Delta(P_1) + \hat{\Delta}(\hat{P}_1) = 1-mn.
\]
Here we rewrote the correlation function with $D_{m,n}^{(b)}V_{P_1}$ as a correlation function, which is acted on by the differential operator $\mathcal{D}^{(b)}_{m,n}$, using \eqref{eq:Dmnb} and \eqref{eq:mathcalLlambda}. Note that here $P_1$ is an arbitrary momenta, but the operator $\mathcal{D}^{(b)}_{m,n}$ corresponds to the singular vector that occurs at $P_1=P_{m,n}$. We want to ``integrate by parts'' each differential operator $\mathcal{L}_{-\lambda}$ in $\mathcal{D}^{(b)}_{m,n}$ one by one. As we do so we will encounter the following product of correlation functions:
\[
\langle\Phi_{\hat{P}_1}^{\boldsymbol{\nu},\varnothing}(z) \ldots \rangle  \mathcal{L}_{-\lambda} \langle V_{P_1}^{\boldsymbol{\mu},\varnothing}(z) \ldots \rangle\qquad \text{with}\qquad |\boldsymbol{\mu}| + |\boldsymbol{\nu}| +\lambda = mn.
\]
This differential operator may be integrated by parts:
\begin{equation}\label{eq:IbpOneL}
\langle\Phi_{\hat{P}_1}^{\boldsymbol{\nu},\varnothing}(z) \ldots \rangle  \mathcal{L}_{-\lambda} \langle V_{P_1}^{\boldsymbol{\mu},\varnothing}(z) \ldots \rangle
=
- 
\langle V_{P_1}^{\boldsymbol{\mu},\varnothing}(z) \ldots \rangle
\hat{\mathcal{L}}_{-\lambda}	\langle\Phi_{\hat{P}_1}^{\boldsymbol{\nu},\varnothing}(z) \ldots\rangle
+
\partial\left(a_\lambda^z \langle\Phi_{\hat{P}_1}^{\boldsymbol{\nu},\varnothing} V_{P_1}^{\boldsymbol{\mu},\varnothing}(z) \ldots \rangle \right).
\end{equation}
Here $\hat{\mathcal{L}}_{-\lambda}$ is
\begin{equation}
\hat{\mathcal{L}}_{-\lambda} = \mathcal{L}_{-\lambda}\big|_{\substack{\Delta_1 \to 1-\Delta_1-\lambda,~\Delta_2 \to 1-\Delta_2,\\ \Delta_3 \to 1-\Delta_3,~ \Delta_4\to 1-\Delta_4}}.
\end{equation}
Note such a transformation of the conformal dimensions ensures that the operator $\hat{\mathcal{L}}_{-\lambda}$ is the differential operator that corresponds to the action of $L_{-\lambda}$ on the field in timelike LFT. 

Applying equation \eqref{eq:IbpOneL} enough times one can integrate by parts the entire $\mathcal{D}_{m,n}^{(b)}$ operator. In order to clearly see that we get the $\mathcal{D}_{m,n}^{(ib)}$ in the result, we write both operators in a basis, which is symmetric with respect to reversal of the partition:
\begin{equation}
e_{\boldsymbol{\lambda}} = L_{-\boldsymbol{\lambda}} + L_{-\boldsymbol{\tilde{\lambda}}}, 
\quad \text{where} \quad 
\boldsymbol{\lambda} =  \{\lambda_1, \lambda_2, \ldots\},~ \boldsymbol{\tilde{\lambda}} = \{\ldots, \lambda_2, \lambda_1\}.
\end{equation}
For example, 
\begin{equation}
D_{4,1}^{(b)} = L_{-1}^4 + 5b^2(L_{-1}^2L_{-2} + L_{-2}L_{-1}^2) + 9b^4 L_{-2}^2 + 12b^4 (L_{-1}L_{-3} + L_{-3}L_{-1}) + 4b^2(9b^6-1)L_{-4}.
\end{equation}
Then after integration by parts each $e_{\boldsymbol{\lambda}}$ with $l(\boldsymbol{\lambda})$ generators is mapped to itself (but now acting on a timelike field), multiplied by $(-1)^{l(\boldsymbol{\lambda})}$. Now, all we need to do is to replace $b$ by $ -i\cdot ib$ and make sure that the signs that come from the $-i$ cancel all the $(-1)^{l(\boldsymbol{\lambda})}$. In \cite{Feigin:1984} Feigin and Fuchs proved that $D_{m,n}^{(b)}$ is invariant under 
\begin{equation}
b\to ib, \qquad L_{-\lambda} \to (-1)^{\lambda-1}L_{-\lambda}.
\end{equation}
Under this transformation the basis element transforms as
\begin{equation}
e_{\boldsymbol{\lambda}} \to (-1)^{|\boldsymbol{\lambda}| - l(\boldsymbol{\lambda})}e_{\boldsymbol{\lambda}} = (-1)^{mn - l(\boldsymbol{\lambda})}e_{\boldsymbol{\lambda}}.
\end{equation}
Thus, the integration by parts coupled with multiplication by $(-1)^{mn}$ leads to the same transformation of the basis elements. Hence, the coefficients depend on $b$ in such a way that the signs $(-1)^{l(\boldsymbol{\lambda})}$ are indeed canceled. This means that after the integration by parts we get $D_{m,n}^{(ib)}$ and we have 
\begin{equation}
\langle V_{P_1} D_{m,n}^{(ib)} \bar{D}_{m,n}^{(ib)}\Phi_{\hat{P}_1}(z) \ldots \rangle 
=  
(-1)^{mn}\langle \bar{D}_{m,n}^{(ib)} \Phi_{\hat{P}_1}D_{m,n}^{(b)}V_{P_1} (z) \ldots \rangle 
+ 
\partial\left( \mathcal{H}_{m,n}\langle  V_{P_1}\bar{D}_{m,n}^{(ib)}\Phi_{\hat{P}_1}(z) \ldots \rangle\right).
\end{equation}
Here $\mathcal{H}_{m,n}$ is a combination of differential operators that act on spacelike and timelike LFT correlation functions. For example, 
\begin{equation}
\mathcal{H}_{1,1} = 1, \qquad \mathcal{H}_{2,1} = \mathcal{L}_{-1}^{(ib)} - \mathcal{L}_{-1}^{(b)} + b^2 a_2^z.
\end{equation}
For each $(m,n)$ the operator $\mathcal{H}_{m,n}$ can be computed explicitly, but the closed form for it for all $(m,n)$ is unknown.

Now we repeat the same procedure for antiholomorphic operator:
\begin{multline}\label{eq:IntByPartsPreDerivative}
	\langle V_{P_1} D_{m,n}^{(ib)} \bar{D}_{m,n}^{(ib)}\Phi_{\hat{P}_1}(z) \ldots \rangle 
	=  
	\langle  \Phi_{\hat{P}_1}D_{m,n}^{(b)}\bar{D}_{m,n}^{(b)} V_{P_1} (z) \ldots \rangle 
	+
	(-1)^{mn}
	\bar{\partial}\bar{\mathcal{H}}_{m,n} \langle \Phi_{\hat{P}_1}D_{m,n}^{(b)}V_{P_1}(z) \ldots \rangle 
	+\\+
	(-1)^{mn}
	\partial\mathcal{H}_{m,n} \langle \Phi_{\hat{P}_1}\bar{D}_{m,n}^{(b)}V_{P_1}(z) \ldots \rangle
	+ \partial \bar{\partial}\mathcal{H}_{m,n}\bar{\mathcal{H}}_{m,n}\langle  V_{P_1} \Phi_{\hat{P}_1}(z)\ldots\rangle
\end{multline}
Remember that $P_1$ and $\hat{P}_1$ are such that $\Delta_1 + \hat{\Delta}_{1} = 1 - mn$. Substituting $P_1 = P_{m,n}$ we get \eqref{eq:IBP1}. Differentiating at $P_1 = P_{m,n}$ we get \eqref{eq:IBP2}.

\paragraph{Boundary terms.}\label{subsubec:BoundaryTerms} When using the above formulas in the integral over the moduli space the terms that are exact forms can be reduced to boundary terms using Stokes' theorem. The boundary consists of small circles around the marked points and infinity. As these circles shrink, the only terms that produce finite contributions are those that behave like $1/z$ or $1/\bar{z}$. Terms that do not behave like that either vanish, or require analytic continuation from the domain where they vanish. This analytic continuation was described in \ref{subsubsec:V04convergence}.  

In the neighborhood of the marked point $z_k$ the integrand contains the factor $|z-z_k|^{2\hat{P}^2}$ from the timelike LFT correlation function. The spectrum in timelike LFT is continuous and never contains discreet terms. Because of that, the integrand never behaves like $1/z$ or $1/\bar{z}$ and thus boundary terms from the marked points do not contribute.

It would be possible to obtain the desired behavior in the integrand that corresponds to a circle around infinity, if there was a term that contained both $\partial\bar{\partial}$ and logarithmic fields. Then, after differentiating the logarithm we would get $1/z$ or $1/\bar{z}$. However, in \eqref{eq:IntByPartsPreDerivative} in the term with $\partial\bar{\partial}$ we have the field $V_{P_1}\Phi_{\hat{P}_1}(z)$ with the conformal dimension $\Delta_1 + \hat{\Delta}_1 = 1-mn$. Since the total conformal dimension does not depend on $P_1$, we do not obtain logarithmic terms here.

Therefore, the boundary terms do not contribute to the integral over the moduli space.

\subsection{Analytic structure of \texorpdfstring{$V_{0,4}$}{V04}}\label{subsec:V04AnalyticStructure}
From our analysis in \ref{subsubsec:V04convergence} it follows that the integral over the moduli space does not influence the analytic structure of $V_{0,4}(P_1, P_2, P_3, P_4)$. Hence, it is inherited from the analytic structure of the correlation functions in LFT. From our analysis in subsections \ref{subsec:SpacelikeLFT} and \ref{subsec:TimelikeLFT} we know that the only potential singularities are poles, which may occur when some of the fields are degenerate or when the screening condition is satisfied \footnote{Correlation function of timelike LFT also has poles at $\hat{P} = 0$, $\hat{P} =  \frac{mb}2$, $P = \frac{nb^{-1}}{2}$, but they are canceled by zeroes of the spacelike LFT correlation function.}.  


Consider the volume at $P_1 = P_{m,n}$, where the timelike LFT correlation function has a pole, with general case $P_2$, $P_3$, $P_4$. The residue of the volume at the pole is:
\begin{equation}
\Res\limits_{P_1 = P_{m,n}} V_{0,4}(P_1, P_2, P_3, P_4) 
=
i\int d^2 z \langle V_{m,n}\Phi_{m,-n}(z) \Phi_{iP_2}V_{P_2}(0) \Phi_{iP_3}V_{P_3}(1) \Phi_{iP_4}V_{P_4}(\infty)\rangle.
\end{equation}
Applying HEM \eqref{eq:TimelikeHEM} we get
\begin{equation}
\Res\limits_{P_1 = P_{m,n}} V_{0,4}(P_1, P_2, P_3, P_4) 
=
\frac{i}{B_{m,n}^{(ib)}}
\int d^2 z \langle V_{m,n}D^{(ib)}_{m,n}\bar{D}^{(ib)}_{m,n} \Phi_{m,n}(z) \Phi_{iP_2}V_{P_2}(0) \Phi_{iP_3}V_{P_3}(1) \Phi_{iP_4}V_{P_4}(\infty)\rangle.
\end{equation}
Using the integration by parts formula \eqref{eq:IBP1} we get
\begin{equation}
\Res\limits_{P_1 = P_{m,n}} V_{0,4}(P_1, P_2, P_3, P_4) 
=
\frac{i}{B_{m,n}^{(ib)}}  
\int d^2 z \, \partial\bar{\partial} \langle \mathcal{H}_{m,n}\bar{\mathcal{H}}_{m,n}V_{m,n}\Phi_{m,n} (z) \Phi_{iP_2}V_{P_2}(0) \Phi_{iP_3}V_{P_3}(1) \Phi_{iP_4}V_{P_4}(\infty)\rangle.
\end{equation}
Now we use the Stokes' theorem to reduce the integral to boundary terms.  As we explained in \ref{subsubec:BoundaryTerms}, the boundary terms do not contribute to the integral. 

Thus, $V_{0,4}(P_1, P_2, P_3, P_4)$ does not have a pole at $P_1 = P_{m,n}$. Other potential poles are those when the screening condition is satisfied. However, we know from \eqref{eq:V04TrialitySymmetry} that this is the same as the case described above. Hence, $V_{0,4}(P_1, P_2, P_3, P_4)$ is an entire function of $P_1$ for general case $P_2, P_3, P_4$.

\subsection{Constraints on \texorpdfstring{$V_{0,4}$}{V04}}

\subsubsection{Simple cases}
For some external momenta the integrand in \eqref{eq:V04H} simplifies dramatically \cite{Fateev:2009me}. There exist two particularly simple cases, when in the recursion formula \eqref{eq:Recursion.H} for the elliptic conformal blocks all the coefficients, except for the first one, vanish:   
\begin{equation}
P_1 = \frac{ib^{-1}}{2}+\frac{ib}{4},\quad P_2 = P_3 = P_4 = \frac{ib}{4}, \qquad \text{and} \qquad P_1 = \frac{ib^{-1}}2 - \frac{ib}{4}, \quad P_2 = P_3 = P_4 = \frac{ib}{4}.  
\end{equation}
The ratio of the structure constants is simplified with the use of \eqref{UpsilonDoubleArgument}:
\begin{equation}
\frac{C\bigl(\frac{ib^{-1}}2+\frac{ib}4,\frac{ib}4,P\bigr)C\left(-P,\frac{ib}4,\frac{ib}4\right)}
{C\bigl(\frac{ib^{-1}}2+\frac{ib}4,\frac{ib}4,i\hat{P}\bigr)C\bigl(-i\hat{P},\frac{ib}4,\frac{ib}4\bigr)}
= 2^{-8P^2-8\hat{P}^2},\quad 
\frac{C\bigl(\frac{ib^{-1}}2-\frac{ib}4,\frac{ib}4,P\bigr)C\left(-P,\frac{ib}4,\frac{ib}4\right)}
{C\bigl(\frac{ib^{-1}}2-\frac{ib}4,\frac{ib}4,i\hat{P}\bigr)C\bigl(-i\hat{P},\frac{ib}4,\frac{ib}4\bigr)}
= 2^{-8P^2-8\hat{P}^2} \frac{P^2}{(i\hat{P})^2}.
\end{equation}
The volumes in these cases are computed analytically 
\begin{equation}\label{eq:SimpleCases}
\begin{aligned}
& V_{0,4}\left(\frac{ib^{-1}}{2} + \frac{ib}{4}, \frac{ib}{4}, \frac{ib}{4}, \frac{ib}{4}\right) 
=
6\int_{F} d^2\tau\,\int_{\mathbb{R}}dP\,   \int_{\mathbb{R}}d\hat{P}\, (i\hat{P})^2 
e^{-2\pi (P^{2}+\hat{P}^2) \Im\tau} = -\frac14,\\
& V_{0,4}\left(\frac{ib^{-1}}{2} - \frac{ib}{4}, \frac{ib}{4}, \frac{ib}{4}, \frac{ib}{4}\right) 
=
6\int_{F} d^2\tau\,\int_{\mathbb{R}}dP\,   \int_{\mathbb{R}}d\hat{P} \,P^2  
e^{-2\pi (P^{2}+\hat{P}^2) \Im\tau} = \frac14.
\end{aligned}
\end{equation}

\subsubsection{\texorpdfstring{$V_{0,4}$}{V04} with degenerate fields}\label{subsubsec:V04DegenerateFields}
Consider again the limit $P_1 \to P_{m,n}$ of the volume $V_{0,4}(P_1,P_2,P_3,P_4)$. In \ref{subsec:V04AnalyticStructure} we showed that $V_{0,4}$ does not have a pole in this limit. Now we consider the next order:
\begin{equation}
V_{0,4}(P_{m,n}, P_2, P_3, P_4)
=
i\int d^2 z \langle V'_{m,n}(z) \Phi_{m,-n}(z)\ldots \rangle
+
\int d^2 z \langle V_{m,n}(z) \Phi'_{m,-n}(z)\ldots \rangle,
\end{equation}
where the fields $V'_{m,n}(z)$ and $\Phi'_{m,-n}(z)$ are defined in \eqref{eq:SpacelikeHEM} and \eqref{TimelikeDerivativeFieldsDefitnition}.
In the first term we use timelike HEM \eqref{eq:TimelikeHEM}, the ``integration by parts'' formula \eqref{eq:IBP2} and then spacelike HEM \eqref{eq:SpacelikeHEM} to get the following:
\begin{multline}\label{V04IBP.1}
v_{0,4}^{m,n}(P_2, P_3, P_4) \equiv V_{0,4}(P_{m,n}, P_2, P_3, P_4)
-
V_{0,4}(P_{m,-n}, P_2, P_3, P_4)
=
\\
=
\int d^2 z_1 
\left\langle V_{m,n}\check{\Phi}'_{m,-n}(z_1)
V_{P_2}\Phi_{iP_2}C\bar{C}\left(0\right) V_{P_3}\Phi_{iP_3}C\bar{C}(1)V_{P_4}\Phi_{iP_4}C\bar{C}(\infty).
\right\rangle
\end{multline}
In the process we also used the simple relation \eqref{HEMConstantsRelation} between the constants in spacelike and timelike HEM. Here we introduced the following notation:
\begin{equation}
\check{\Phi}'_{m,-n} \equiv 
\Phi'_{m,-n} -\frac{1}{B_{m,n}^{(ib)}}\frac{\hat{P}_{m,-n}}{\hat{P}_{m,n}}D^{(ib)}_{m,n}\bar{D}^{(ib)}_{m,n}\Phi'_{m,n}.
\end{equation}
The second term here does not actually contribute to the volume: at least one of the differential operators $\mathcal{D}_{m,n}^{(ib)}$ is not affected by the derivative in $\hat{P}_1$ and can be integrated by parts as was described in \ref{subsec:IntegrationByParts}. Nevertheless, we leave it, since it improves the transformation rules of the field (see below). This field may also be represented as
\begin{equation}
\check{\Phi}'_{m,-n} \equiv \left.\frac{\partial}{\partial \hat{P}_1}\Phi_{\hat{P}_1}^{m,n}\right|_{\hat{P}_1 = \hat{P}_{m,-n}}, 
\end{equation}
where
\begin{equation}
\Phi_{\hat{P}_1}^{m,-n} \equiv (\hat{P_1} - \hat{P}_{m,-n})\Phi_{\hat{P}_1} -\frac{1}{B_{m,n}^{(ib)}}D^{(ib)}_{m,n}\bar{D}^{(ib)}_{m,n}\Phi_{\hat{P}_2(\hat{P}_1)}, \quad \text{with} \quad \hat{P}_2^2(\hat{P}_1) = \hat{P}_1^2 - mn.
\end{equation}
Note that due to HEM in timelike LFT we have $\Phi^{m,-n}_{\hat{P}_{m,-n}} = 0$.

In order to see how the generators of Virasoro algebra act on the field $\check{\Phi}'_{m,-n}$, we first act on $\Phi_{\hat{P}_1}^{m,n}$:
\begin{equation}\label{eq:MactionPhimnP1}
\begin{aligned}
& L_0 \Phi_{\hat{P}_1}^{m,-n}
=
\hat{\Delta}(\hat{P}_1) \Phi_{\hat{P}_1}^{m,-n}, & &\\
& L_r \Phi_{\hat{P}_1}^{m,-n}
=
-\frac{1}{\hat{B}_{m,n}} \bar{D}^{(ib)}_{m,n} L_r D^{(ib)}_{m,n}\Phi_{\hat{P}_2(\hat{P}_1)}, &\quad &0<r\leq mn,\\
& L_r \Phi_{\hat{P}_1}^{m,-n}
=
0, & &r>mn.
\end{aligned}
\end{equation}
Now we take the derivative at $\hat{P}_1 = \hat{P}_{m,-n}$. When doing so in the first equation in \eqref{eq:MactionPhimnP1} we take into account that $\Phi^{m,-n}_{\hat{P}_{m,-n}} = 0$:
\begin{equation}
L_0 \check{\Phi}'_{m,-n} = \hat{\Delta}_{m,-n} \check{\Phi}'_{m,-n}.
\end{equation}
This means that the field $\check{\Phi}'_{m,-n}$ transforms like a primary field under dilatation: the ``logarithmic behavior'' of the fields $\Phi'_{m,-n}$ and $\Phi'_{m,n}$ cancels in the definition of $\check{\Phi}'_{m,-n}$.

In the second equation in \eqref{eq:MactionPhimnP1} by definition of $D_{m,n}^{(ib)}$ the action of $L_r$ vanishes if $\hat{P}_2 = \hat{P}_{m,n}$. This means that as a result of the action we get a polynomial in $L_{k}$, the coefficients of which vanish at $\hat{P}_2 = \hat{P}_{m,n}$. Thus, when we take the derivative at this point we differentiate only the coefficients of the polynomial and not the field $\Phi_{\hat{P}_2}$. The result is a vector in the Verma module of $\Phi_{m,n}$. For example, 
\begin{equation}
L_1 \check{\Phi}_{1,-1} = - \frac{2\hat{P}_{1,1}}{B_{1,1}^{(ib)}}\cdot \bar{D}^{(ib)}_{1,1}  2\Phi_{1,1}, 
\qquad 
L_1 \check{\Phi}_{3,-1} = - \frac{2\hat{P}_{3,1}}{B_{3,1}^{(ib)}}\cdot \bar{D}^{(ib)}_{3,1}  (6L_{-1}^2 - 8b^2L_{-2})\Phi_{3,1}.
\end{equation}
Note that the $\bar{D}_{m,n}^{(ib)}$ remains intact. 

The action of both holomorphic and antiholomorphic generators on $\Phi^{m,n}_{\hat{P}_1}$ produces double zero. This means that
\begin{equation}
L_{r}\bar{L}_{s}\check{\Phi}'_{m,-n} = 0, \qquad r,s>0.
\end{equation}

\subsubsection{Constraints on \texorpdfstring{$V_{0,4}$}{V04}}
In order to derive some constraints on $v_{0,4}^{m,n}$ (and hence $V_{0,4}$), we take one of the fields, say $V_{P_2}$, to be degenerate. To utilize this limit, it is helpful to make it so that we integrate over the position of $V_{P_2}$, so that we can integrate by parts. This is achieved by the following global conformal transformation:
\begin{equation}
w(z) = \frac{z-z_1}{1-z_1}.
\end{equation}
Under this conformal transformation all the fields in the correlation function
\[
\left\langle V_{m,n}\check{\Phi}'_{m,-n}(z_1)
V_{P_2}\Phi_{iP_2}C\bar{C}\left(0\right) V_{P_3}\Phi_{iP_3}C\bar{C}(1)V_{P_4}\Phi_{iP_4}C\bar{C}(\infty)
\right\rangle
\]
transform like primary fields:
\begin{multline}
\left\langle V_{m,n}\check{\Phi}'_{m,-n}(z_1)
V_{P_2}\Phi_{iP_2}C\bar{C}\left(0\right) V_{P_3}\Phi_{iP_3}C\bar{C}(1)V_{P_4}\Phi_{iP_4}C\bar{C}(\infty)
\right\rangle
=
\\
=
|1-z_1|^{-2}\left\langle V_{m,n}\check{\Phi}'_{m,-n}(0)
V_{P_2}\Phi_{iP_2}C\bar{C}\left(-\frac{z_1}{1-z_1}\right) V_{P_3}\Phi_{iP_3}C\bar{C}(1)V_{P_4}\Phi_{iP_4}C\bar{C}(\infty)
\right\rangle.
\end{multline}
Now we make the change of the integration variable
\begin{equation}
z(z_1) = -\frac{z_1}{1-z_1} = 1 - \frac{1}{1-z_1}
\end{equation}
to get
\begin{multline}
\int d^2 z_1\left\langle V_{m,n}\check{\Phi}'_{m,-n}(z_1)
V_{P_2}\Phi_{iP_2}C\bar{C}\left(0\right) V_{P_3}\Phi_{iP_3}C\bar{C}(1)V_{P_4}\Phi_{iP_4}C\bar{C}(\infty)
\right\rangle
=
\\
=
\int d^2 z\, |1-z|^{-2}\left\langle V_{m,n}\check{\Phi}'_{m,-n}(0)
V_{P_2}\Phi_{iP_2}C\bar{C}\left(z\right) V_{P_3}\Phi_{iP_3}C\bar{C}(1)V_{P_4}\Phi_{iP_4}C\bar{C}(\infty)
\right\rangle
\end{multline}
Finally, we substitute the ghost correlation function:
\begin{equation}\label{eq:v04AfterPositionSwitch}
v_{0,4}^{m,n}(P_2, P_3, P_4) = \int d^2 z \left\langle V_{m,n}\Phi_{m,n}'(0)
V_{P_2}\Phi_{iP_2}\left(z\right) V_{P_3}\Phi_{iP_3}V_{P_4}\Phi_{iP_4}(\infty)
\right\rangle.
\end{equation}

In \ref{subsubsec:Spacelike4pointAnalyticStructure} we established that the spacelike LFT correlation function with two degenerate fields is zero for general case $P_3$, $P_4$. Because of that if we take the limit $P_2\to P_{r,s}$ in \eqref{eq:v04AfterPositionSwitch} we will not have the term with $\check{\Phi}'_{m,-n}$:
\begin{multline}\label{eq:Constraint1}
v_{0,4}^{m,n}(P_{r,s}, P_3, P_4) 
= 
\int d^2 z \left\langle V_{m,n}\check{\Phi}'_{m,-n}(0)
V'_{r,s}\Phi_{r,-s}\left(z\right) V_{P_3}\Phi_{iP_3}V_{P_4}\Phi_{iP_4}(\infty)
\right\rangle
 \stackrel{\eqref{eq:TimelikeHEM}, \eqref{eq:IBP2}, \eqref{eq:SpacelikeHEM}}{=}
\\
=
\int d^2 z \left\langle V_{m,n}\check{\Phi}'_{m,-n}(0)
V_{r,-s}\Phi_{r,s}\left(z\right) V_{P_3}\Phi_{iP_3}V_{P_4}\Phi_{iP_4}(\infty)
\right\rangle = v_{0,4}^{m,n}(P_{r,-s},P_3,P_4).
\end{multline}
In terms of the volume this constraint is
\begin{equation}
V_{0,4}(P_{m,n}, P_{r,s}, P_3, P_4) - V_{0,4}(P_{m,-n}, P_{r,s}, P_3, P_4) - V_{0,4}(P_{m,n}, P_{r,-s},P_3, P_4) + V_{0,4}(P_{m,-n}, P_{r,-s}, P_3, P_4) = 0.
\end{equation}

\subsubsection{Computation of \texorpdfstring{$V_{0,4}$}{V04} for polynomial case}

If we assume that $V_{0,4}$ depends on $\{P_k\}$ polynomially, the constraint \eqref{eq:Constraint1} together with \eqref{eq:SimpleCases} becomes strong enough to determine the volume uniquely. This is due to the fact that the set $\{P_{r,-s}\}$ is dense in the imaginary line. Because of that for each $P_2 \in i\mathbb{R}$ there exists a sequence $\{r_k, s_k\}$ such that 
\begin{equation}
v_{0,4}^{m,n}(P_2, P_3, P_4) = \lim\limits_{k\to\infty} v_{0,4}^{m,n}(P_{r_k,-s_k}, P_3, P_4) 
\stackrel{\eqref{eq:Constraint1}}{=} \lim\limits_{k\to\infty} v_{0,4}^{m,n}(P_{r_k,s_k}, P_3, P_4)
\end{equation}
When $k\to \infty$ we have $P_{r_k, s_k}\to i\infty$. Since $v_{0,4}^{m,n}(P, P_3, P_4)$ is a polynomial in $P$, it has a unique limit at $P\to i\infty$, which does not depend on the sequence $P_{r_k, s_k}$. From above we conclude that $v_{0,4}(P_2, P_3, P_4)$ does not depend on $P_2$. It follows that
\begin{equation}
\frac{\partial}{\partial P_2} V_{0,4}(P_{m,n}, P_2, P_3, P_4)
=
\frac{\partial}{\partial P_2} V_{0,4}(P_{m,-n}, P_2, P_3, P_4)
\end{equation}
Applying the above logic one more time we get the following:
\begin{equation}
\frac{\partial^2}{\partial P_1\partial P_2}V_{0,4}(P_1, P_2, P_3, P_4) = 0.
\end{equation}
From this and from the fact that $V_{0,4}(P_1, P_2, P_3, P_4)$ is a symmetric function it follows that
\begin{equation}
V_{0,4}(P_1, P_2, P_3, P_4) = f(P_1) + f(P_2) + f(P_3) + f(P_4),
\end{equation}
for some polynomial $f$. On the other hand, from the triality symmetry of the volume we get
\begin{multline}
f(P_1) + f(P_2) + f(P_3) + f(P_4) 
=
f\left(\frac{P_1 + P_2 + P_3 + P_4}{2}\right) + f\left(\frac{P_1 + P_2 - P_3 - P_4}{2}\right) +\\+ f\left(\frac{P_1 - P_2 + P_3 - P_4}{2}\right) + f\left(\frac{P_1 - P_2 - P_3 + P_4}{2}\right).
\end{multline}
The only entire function that satisfies this equation is a quadratic polynomial:
\begin{equation}
V_{0,4}(P_1, P_2, P_3, P_4) = c_1(b) + c_2(b) (P_1^2 + P_2^2 + P_3^2 + P_4^2).
\end{equation}
Since we know the value of $V_{0,4}(P_1, P_2, P_3, P_4)$ in two cases \eqref{eq:SimpleCases}, we can recover the two coefficients. The result is \eqref{eq:V04hypothesis}.

\subsection{Connection between \texorpdfstring{$V_{1,1}$}{V11} and \texorpdfstring{$V_{0,4}$}{V04}}
Following the same principles described in section \ref{sec:Liouville} one could define the LFT on a torus. For details we refer to \cite{Fateev:2009me,Hadasz:2009sw}. The first nontrivial volume on a torus is $V_{1,1}$:
\begin{equation}\label{eq:V11def}
V_{1,1}(P_1) 
=
8\int_{F} d^2\tau 
\langle B(0)\bar{B}(0)C(0)\bar{C}(0)\rangle_{\tau}
\langle V_{P_1}(0)\rangle_\tau 
\langle\Phi_{iP_1}(0)\rangle_{\tau}.
\end{equation}
Here $F$ is the same domain $\{|\Re\tau|<\frac12, |\tau|>1\}$ that we had on figure \ref{fig:domains}. For the justification of the factor $8$ we refer to \cite{Collier:2023cyw}. The ghost partition function~is 
\begin{equation}\label{TorusGhost}
\langle B(z)\bar{B}(z)C(w)\bar{C}(w)\rangle_{\tau} = |\eta(\tau)|^4.
\end{equation}
The LFT correlation functions may be written as
\begin{equation}\label{TorusSpacelikeLFT} 
\langle V_{P_1}(0)\rangle_{\tau} = \frac12\int_CdP\, C(-P, P_1, P) \left|q^{2P^2}\eta(\tau)^{-1}\mathcal{H}^{(b)}(P_1,P|q^2)\right|^2.
\end{equation}
\begin{equation}\label{TorusTimelikeLFT}
\langle V_{iP_1}(0)\rangle_{\tau} = \frac12\int_{\mathbb{R}+i\varepsilon}d\hat{P}\,(i\hat{P})^2 \frac1{C(-i\hat{P}, P_1, i\hat{P})} \left|q^{2\hat{P}^2}\eta(\tau)^{-1}\mathcal{H}^{(ib)}(iP_1,\hat{P}|q^2)\right|^2.
\end{equation}
Here by $\mathcal{H}$ we denote the toric elliptic conformal blocks and $q = e^{i\pi\tau}$. They are related to elliptic conformal blocks on a sphere as 
\begin{equation}
\mathcal{H}^{(b)}(P_1, P|q^2) = \mathfrak{H}^{(b/\sqrt{2})}\left(\left.\frac{i\sqrt{2}}{4b}, \frac{i\sqrt{2}}{4b},\frac{i\sqrt{2}}{4b}, \frac{P_1}{\sqrt{2}},\sqrt{2}P\right|q\right),
\end{equation}
or
\begin{equation}
\mathcal{H}^{(b)}(P_1, P|q^2) = \mathfrak{H}^{(\sqrt{2}b)}\left(\left.\frac{i\sqrt{2}b}{4}, \frac{i\sqrt{2}b}{4},\frac{i\sqrt{2}b}{4}, \frac{P_1}{\sqrt{2}},\sqrt{2}P\right|q\right).
\end{equation} 

The structure constants are also related to the corresponding structure constants on a sphere. One can prove using \eqref{eq:Upsilon*sqrt2} and \eqref{eq:Upsilon/sqrt2} that
\begin{equation} 
\frac{C_{b}(-P, P_1, P)}{C_{b}(-i\hat{P},P_1,i\hat{P})} = 16^{4(P^2+\hat{P}^2)} 
\frac{C_{b/\sqrt{2}}\left(\frac{i\sqrt{2}}{4b},\frac{i\sqrt{2}}{4b},\sqrt{2}P\right)C_{b/\sqrt{2}}\left(-\sqrt{2}P,\frac{P_1}{\sqrt{2}}, \frac{i\sqrt{2}}{4b}\right)}
{C_{b/\sqrt{2}}\left(\frac{i\sqrt{2}}{4b},\frac{i\sqrt{2}}{4b},i\sqrt{2}\hat{P}\right)C_{b/\sqrt{2}}\left(-i\sqrt{2}\hat{P},\frac{P_1}{\sqrt{2}}, \frac{i\sqrt{2}}{4b}\right)},
\end{equation}
or
\begin{equation}
\frac{C_{b}(-P, P_1, P)}{C_{b}(-i\hat{P},P_1,i\hat{P})} 
= 
16^{4(P^2+\hat{P}^2)} 
\frac{C_{\sqrt{2}b}\left(\frac{i\sqrt{2}b}{4},-\frac{i\sqrt{2}b}{4},\sqrt{2}P\right)C_{\sqrt{2}b}\left(-\sqrt{2}P,\frac{P_1}{\sqrt{2}}, \frac{-i\sqrt{2}b}{4}\right)}
{C_{\sqrt{2}b}\left(\frac{i\sqrt{2}b}{4},\frac{i\sqrt{2}b}{4},i\sqrt{2}\hat{P}\right)C_{\sqrt{2}b}\left(-i\sqrt{2}\hat{P},\frac{P_1}{\sqrt{2}}, \frac{i\sqrt{2}b}{4}\right)}.
\end{equation}
Using the relation between conformal blocks and structure constants and making the change of variables $\sqrt{2}P \to P$ and $\sqrt{2}\hat{P} \to \hat{P}$ one can see that
\begin{equation} 
V_{1,1}^{b}(P_1) 
=
\frac{1}{12} V_{0,4}^{b\sqrt{2}}\left(\frac{i\sqrt{2}b}{4}, \frac{i\sqrt{2}b}{4},\frac{i\sqrt{2}b}{4}, \frac{P_1}{\sqrt{2}}\right)
= 
\frac{1}{12} V_{0,4}^{b/\sqrt{2}}\left(\frac{i\sqrt{2}}{4b}, \frac{i\sqrt{2}}{4b},\frac{i\sqrt{2}}{4b}, \frac{P_1}{\sqrt{2}}\right),
\end{equation}
which is consistent with \eqref{eq:V04hypothesis}.
\section{Numerical results}\label{sec:Numerics}
From the hypothetical answer \eqref{eq:V04hypothesis}, which we proved for polynomial case in the previous section, we have the following conjecture for $v_{0,4}^{m,n}$ (see~\eqref{V04IBP.1}):
\begin{equation}\label{v04hypothesis}
	v_{0,4}^{m,n}(P_2, P_3, P_4) = -mn.
\end{equation}
Our goal here is to check this hypothesis for $(m,n) = (1,1)$ and $(m,n) = (2,1)$.

For simplicity we limit ourselves to the case $P_2 = P_3 = P_4$. Also, to reduce the number of terms we will omit the second term in the definition of $\check{\Phi}'_{m,-n}(z)$. We are free to do so, because it can be reduced to boundary terms. Hence, we are going to compute 
\begin{equation}\label{Numerics:1}
	\int_{\mathbb{C}} d^2z \langle V_{m,n}(z)V_{P_2}(0)V_{P_2}(1)V_{P_2}(\infty)\rangle \langle \Phi'_{m,-n}(z)\Phi_{iP_2}(0)\Phi_{iP_2}(1)\Phi_{iP_2}(\infty)\rangle.
\end{equation}
The first step is to reduce the integral over the complex plane to the integral over the domain $\Circled{1}$ (see fig.~\ref{fig:domainsz}). To do this we first replace $\Phi'_{m,-n}$ by $(P_1-P_{m,n})\Phi_{iP_1}$ to get:
\begin{multline}
\int_{\mathbb{C}} d^2z \langle V_{m,n}(z)V_{P_2}(0)V_{P_2}(1)V_{P_2}(\infty)\rangle \langle (P_1 - P_{m,n})
\Phi_{iP_1}(z)\Phi_{iP_2}(0)\Phi_{iP_2}(1)\Phi_{iP_2}(\infty)\rangle
=
\\
=
2\int_{\Circled{1}} d^2z
\langle V_{m,n}(z)V_{P_2}(0)V_{P_2}(1)V_{P_2}(\infty)\rangle \langle (P_1 - P_{m,n})\Phi_{iP_1}(z)\Phi_{iP_2}(0) \Phi_{iP_2}(1)\Phi_{iP_2}(\infty)\rangle
\times
\\
\times
\left(1+|z|^{4(\hat{\Delta}_1 + \Delta_{m,n} - 1)}+|1-z|^{4(\hat{\Delta}_1 + \Delta_{m,n} - 1)}\right).
\end{multline}
Now we take the derivative at $P_1 = P_{m,n}$:
\begin{multline}
\int_{\mathbb{C}} d^2z \langle V_{m,n}(z)V_{P_2}(0)V_{P_2}(1)V_{P_2}(\infty)\rangle \langle\Phi'_{m,-n}(z)\Phi_{iP_2}(0)\Phi_{iP_2}(1)\Phi_{iP_2}(\infty)\rangle
=
\\
=
2\int_{\Circled{1}} d^2z
\langle V_{m,n}(z)V_{P_2}(0)V_{P_2}(1)V_{P_2}(\infty)\rangle \left[
3\langle
\Phi'_{m,-n}(z)\Phi_{iP_2}(0)\Phi_{iP_2}(1)\Phi_{iP_2}(\infty)\rangle
- \right.
\\
- \left.
4P_{m,n} \langle \Phi_{m,-n}(z)\Phi_{iP_2}(0)\Phi_{iP_2}(1)\Phi_{iP_2}(\infty)\rangle\log|z(1-z)|^2\right].
\end{multline}
The next step is to substitute the $s$-channel decompositions of the correlation functions. When we do so we will have a term in which we differentiate $(P_1-P_{m,n})/\Upsilon_b(-2iP_1)$. We do not need to include it, since this term is proportional to
\begin{equation}
\int_{\mathbb{C}} d^2z \langle V_{m,n}(z)V_{P_2}(0)V_{P_2}(1)V_{P_2}(\infty)\rangle \langle\Phi_{m,-n}(z)\Phi_{iP_2}(0)\Phi_{iP_2}(1)\Phi_{iP_2}(\infty)\rangle = 0.
\end{equation}
The rest of the terms are
\begin{multline}
\int_{\Circled{1}}d^2z \sum_{k,l}
C_{m,n}^{k,l}(P_2)C(P_2+P_{k,l}, P_2, P_2) \left|\mathfrak{F}^{(b)}(P_{m,n}, P_2, P_2, P_2,P_2+P_{k,l}|z)\right|^2 
\times\hfill
\\
\times
\int_{\mathbb{R}+i\varepsilon} d\hat{P} (i\hat{P})^2 \frac{1}{C(-i\hat{P},P_2, P_2)}\frac{1}{-2i\Upsilon'(Q+2iP_{m,n})} \lim\limits_{P_1 \to P_{m,n}}\frac{\Upsilon_b(-2iP_1)}{C(P_1,P_2, i\hat{P})} 
\times \hfill
\\
\times
\left[\left|\mathfrak{F}^{(ib)}(iP_{m,n}, iP_2, iP_2, iP_2,\hat{P}|z)\right|^2 
\left(
3\left.\frac{\partial}{\partial P_1} \log \frac{\Upsilon_b(-2iP_1)}{C(P_1,P_2, i\hat{P})}\right|_{P_1 = P_{m,n}}  - 4P_{m,n}\log|z(1-z)|^2
\right) \right.
+
\\
+ \left.
3\mathfrak{F}^{(ib)}(iP_{m,n}, iP_2, iP_2, iP_2,\hat{P}|\bar{z})\left.\frac{\partial}{\partial P_1}\mathfrak{F}^{(ib)}(iP_1, iP_2, iP_2, iP_2,\hat{P}|z)\right|_{P_1 = P_{m,n}} + \cc\footnotemark\right]
\end{multline}\footnotetext{$\cc$ corresponds to the term similar to the one before it, but where the holomorphic and antiholomorphic conformal blocks are swapped.}
Now we make the coordinate transformation $z\to \tau$  (see~\eqref{eq:ZToTau}, \eqref{FblocksInTermsOfHBlocks} and \eqref{ProductOfTheta}) to get
\begin{multline}
\frac{6i}{\Upsilon'(Q+2iP_{m,n})}\sum_{k,l} C_{m,n}^{k,l}(P_2)
\int_{\mathbb{R}+i\varepsilon} d\hat{P} (i\hat{P})^2 
16^{2(P_2+P_{k,l})^2+2\hat{P}^2}
\frac{C(P_2+P_{k,l}, P_2, P_2)}{C(-i\hat{P},P_2, P_2)} 
 \lim\limits_{P_1 \to P_{m,n}}\frac{\Upsilon_b(-2iP_1)}{C(P_1,P_2, i\hat{P})} 
\times\hfill
\\
\times
\int_{F}d^2\tau\, |q|^{2(\hat{P}^2+(P_2+P_{k,l})^2)}
\left|\mathfrak{H}^{(b)}(P_{m,n}, P_2, P_2, P_2,P_2+P_{k,l}|q)\right|^2 
\times\hfill
\\
\times
\left[\left|\mathfrak{H}^{(ib)}(iP_{m,n}, iP_2, iP_2, iP_2,\hat{P}|q)\right|^2 
\left(
\left.\frac{\partial}{\partial P_1} \log \frac{\Upsilon_b(-2iP_1)}{C(P_1,P_2, i\hat{P})}\right|_{P_1 = P_{m,n}}  + 8P_{m,n}\log|2^\frac{1}3\eta(\tau)|^2
\right) \right.
+
\\
+ \left.
\mathfrak{H}^{(ib)}(iP_{m,n}, iP_2, iP_2, iP_2,\hat{P}|\bar{q})\left.\frac{\partial}{\partial P_1}\mathfrak{H}^{(ib)}(iP_1, iP_2, iP_2, iP_2,\hat{P}|q)\right|_{P_1 = P_{m,n}} + \cc\right]
\end{multline}
Here we also used $d^2z = d\Re z\cdot d\Im z\cdot4/\pi^2$ and $d^2\tau = d\Re\tau\cdot d\Im\tau$.

Next, we expand the integrand in powers of $q$ and take the integral in $\Im \tau$:
\begin{multline}\label{IntegralOverImTau1}
\int_{\cos \varphi}^{\infty} d\Im\tau |q|^{2(\hat{P}^2+(P_2+P_{k,l})^2)} q^{r}\bar{q}^s = \int_{\cos\varphi}^{\infty} d\Im\tau\, e^{-\pi(r+s+2\hat{P}^2+2(P_2+P_{k,l})^2)\Im\tau - i\pi (r-s)\sin\varphi} 
= 
\\
=
\frac{e^{-i\pi(r-s)\sin\varphi-\pi(r+s+2\hat{P}^2+2(P_2+P_{k,l})^2)\cos\varphi}}{\pi(r+s+2\hat{P}^2+2(P_2+P_{k,l})^2)
}.
\end{multline}
Here we parametrize $\Re\tau$ with
\begin{equation}
\Re\tau = -\sin\varphi, \qquad \varphi\in\left[-\frac{\pi}{6}, \frac{\pi}{6}\right].
\end{equation}
For the term with $\log|\eta(\tau)|^2$ we will also need
\begin{multline}\label{IntegralOverImTau2}
\int_{\cos \varphi}^{\infty} d\Im\tau |q|^{2(\hat{P}^2+(P_2+P_{k,l})^2)} q^{r}\bar{q}^s \Im\tau 
=
\\
=
\frac{e^{-i\pi(r-s)\sin\varphi-\pi(r+s+2\hat{P}^2+2(P_2+P_{k,l})^2)\cos\varphi}}{\pi^2(r+s+2\hat{P}^2+2(P_2+P_{k,l})^2)^2}\left(1+\pi(r+s+2\hat{P}^2+2(P_2+P_{k,l})^2)\cos\varphi\right). 
\end{multline}
In order for the above integrals to converge for all $r,s\in\mathbb{Z}_{\geq0}$ we need to have $|P_2|>|P_{m-1,n-1}|$. Otherwise, the right-hand sides of \eqref{IntegralOverImTau1} and \eqref{IntegralOverImTau2} define the analytic continuations of the integrals.  

The rest is done numerically. To compute coefficients in the $q$-series expansion of timelike LFT conformal blocks we use the Zamolodchikov recursion formula \eqref{NumericsHRecursion}. The spacelike LFT conformal blocks that we need here are known in closed form:
\begin{equation}
\left|\mathfrak{H}^{(b)}(P_{1,1},P_2,P_2,P_2, P_2|q)\right|^2 = |q|^{-2P_2^{2}}|\eta(\tau)|^{24P_2^2} = \prod_{j=1}^{\infty}\left|1-q^{2j}\right|^{24P_2^2}.
\end{equation}\vspace{-.75cm}
\begin{multline}
\left|\mathfrak{H}^{(b)}(P_{2,1},P_2,P_2,P_2, P_2 \pm P_{1,0}|q)\right|^2
=
\\
=
\left|\frac{\eta(\tau)^{24}}{q^2}\right|^{\frac{Q^{2}}{4}+P_{2,1}^2+P_{2}^2}
|\theta_3(q)^4|^{-Q^2-4P_{2,1}^2}
\left|\frac{z}{16q}\right|^{2b\left(\frac{Q}{2} \pm iP_{2}\right)}|1-z|^{2b\left(\frac{Q}{2} + iP_{2}\right)}
\left|{}_2F_1(A_\pm, B_{\pm}, C_{\pm}|z)\right|^2.
\end{multline}
Here
\begin{equation}
	A_\pm = \frac12 + 2ibP_2 \pm ibP_2, \qquad C_\pm = 2B_\pm = 1\pm 2ibP_2.
\end{equation}

To numerically compute the $\Upsilon_b$ function we use the method suggested in \cite{Ribault:2015sxa}. Namely, we use \mbox{eq-n.}~\eqref{UpsilonNumerics}.

The results are presented below. As we see, they are in good agreement with the hypothesis \eqref{v04hypothesis}.

 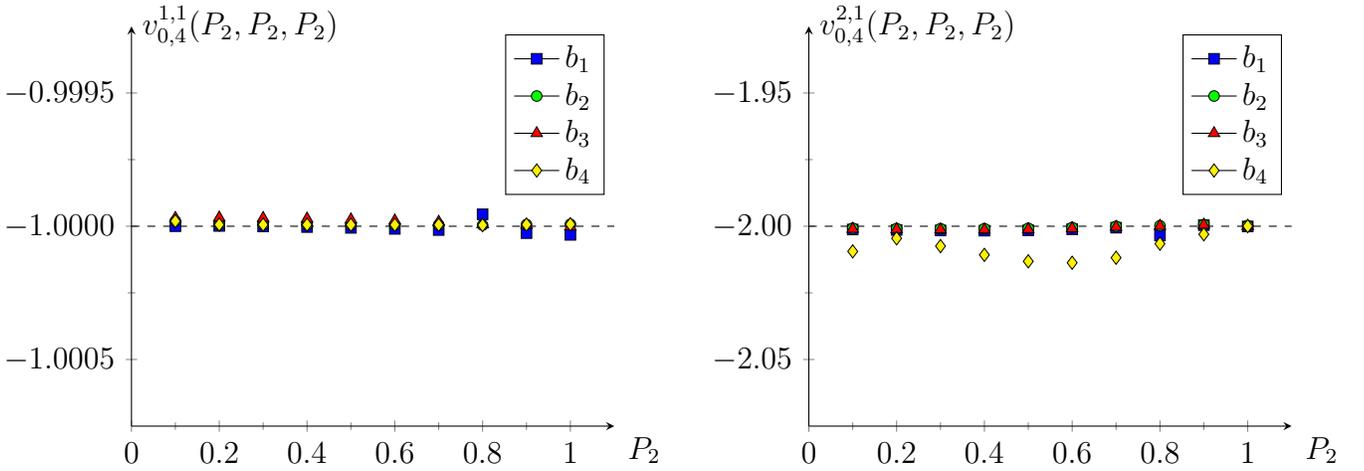
\begin{figure}[H]
 \begin{tikzpicture}[>=latex]
 	\begin{axis}[
 		width=8cm,
 		axis lines=center,
 		axis x line=bottom,
 		axis y line=left,
 		xlabel={$P_2$},
 		ylabel={$v_{0,4}^{1,1}(P_2, P_2, P_2)$},
 		x label style={anchor=north, xshift=.4cm},
 		y label style={anchor=west},
 		xmin=0, xmax=1.1,
 		ymin=-1.00075,ymax=-.99925,
 		xtick = {0,.2,...,1},
 		minor tick num=1,
 		ytick = {-.9995,-1,-1.0005},
 		yticklabels={$-0.9995$, $-1.0000$, $-1.0005$},
 		]

 		\addplot[
 		draw=none,mark=square*,mark options={color=blue, draw=black, ultra thin, mark size=2pt},
 		] table [x index=0, y index=1]{data/11data1.txt};
 		\addlegendentryexpanded{$b_1$}
 		\addplot[
 		draw=none,mark=*,mark options={color=green, draw=black, ultra  thin, mark size=2pt},
 		] table [x index=0, y index=1]{data/11data2.txt};
 		\addlegendentryexpanded{$b_2$}
 		\addplot[
 		draw=none,mark=triangle*,mark options={color=red, draw=black, ultra  thin, mark size=2.5pt},
 		] table [x index=0, y index=1]{data/11data3.txt};
 		\addlegendentryexpanded{$b_3$}
 		\addplot[
 		draw=none,mark=diamond*,mark options={color=yellow, draw=black, ultra  thin, mark size=2.5pt},
 		] table [x index=0, y index=1]{data/11data4.txt};
 		\addlegendentryexpanded{$b_4$}

 		\draw[dashed](0,-1) -- (1.1,-1);

 	\end{axis}
 \end{tikzpicture}
 ~
 \begin{tikzpicture}[>=latex]
 		\begin{axis}[
 			width=8cm,
 			axis lines=center,
 			axis x line=bottom,
 			axis y line=left,
 			xlabel={$P_2$},
 			ylabel={$v_{0,4}^{2,1}(P_2, P_2, P_2)$},
 			x label style={anchor=north, xshift=.4cm},
 			y label style={anchor=west},
 			xmin=0, xmax=1.1,
 			ymin=-2.075,ymax=-1.925,
 			xtick = {0,.2,...,1},
 			minor tick num=1,
 			ytick = {-1.95,-2,-2.05},
 			yticklabels={$-1.95$, $-2.00$, $-2.05$},
 			]

 			\addplot[
 			draw=none,mark=square*,mark options={color=blue, draw=black, ultra thin, mark size=2pt},
 			] table [x index=0, y index=1]{data/21data1.txt};
 			\addlegendentryexpanded{$b_1$}
 			\addplot[
 			draw=none,mark=*,mark options={color=green, draw=black, ultra  thin, mark size=2pt},
 			] table [x index=0, y index=1]{data/21data2.txt};
 			\addlegendentryexpanded{$b_2$}
 			\addplot[
 			draw=none,mark=triangle*,mark options={color=red, draw=black, ultra  thin, mark size=2.5pt},
 			] table [x index=0, y index=1]{data/21data3.txt};
 			\addlegendentryexpanded{$b_3$}
 			\addplot[
 			draw=none,mark=diamond*,mark options={color=yellow, draw=black, ultra  thin, mark size=2.5pt},
 			] table [x index=0, y index=1]{data/21data4.txt};
 			\addlegendentryexpanded{$b_4$}

 			\draw[dashed](0,-2) -- (1.1,-2);

 		\end{axis}
 	\end{tikzpicture}
 \caption{Numerical results for the $v_{0,4}^{1,1}(P_2, P_2, P_2)$ and $v_{0,4}^{2,1}(P_2, P_2, P_2)$. The hypothesis is $v_{0,4}^{m,n}(P_2, P_2, P_2) = -mn$, shown as a dashed line. The values of $b$ are as follows: $b_1 = \frac{1}{e}$, $b_2 = \frac{1}{\pi}$, $b_3 = \frac12 \left(\frac{1}{e} + \frac1{\pi}\right)$, $b_4 = \frac{e}{\pi}$.}
\end{figure}

\section{Conclusion}\label{sec:Conclusion}
In this paper we have studied correlation numbers in Virasoro minimal string \cite{Collier:2023cyw}. We were able to analytically confirm the formula  for the volumes $V_{0,4}$ and $V_{1,1}$ proposed  in \cite{Collier:2023cyw}. Unfortunately, our proof is not complete since it relies on the assumption of polynomial boundedness of the volumes -- an assumption that we cannot deduce from first principles. 

When trying to generalize our method to a more general case of $V_{g,n}$, certain problems arise. First, it is necessary to prove that the volumes defined by the formula \eqref{eq:Vgndef} have no singularities. The absence of  singularities at $P_k\rightarrow P_{m,n}$ can be proved  using HEM, similarly to what was done in subsection \ref{subsec:V04AnalyticStructure}.  On the other hand, the poles corresponding to the screening conditions in spacelike Liouville CFT \eqref{eq:npointpoles} have a completely different nature and in the absence of a triality equation this type of poles cannot be reduced to the previous one. It would be interesting to come up with an argument why the residues in such poles correspond to exact forms. We will consider this issue in a future publication.
\section*{Acknowledgments}
We thank Alexander Artemev, Lorenz Eberhardt and Maxim Kazarian for insightful discussions. This work was supported by Basis foundation. The work of D.K.  performed in Landau Institute has been supported by the Russian Science Foundation under the grant 23-12-00333.
\Appendix
\section{Special functions}
$\Upsilon_b(z)$ is a function that satisfies the following shift relations:
\begin{equation}
\Upsilon_b(z+b) = \gamma(bz) b^{1-2bz}\Upsilon_b(z), \qquad \Upsilon_b(z+b^{-1}) = \gamma(zb^{-1}) b^{2zb^{-1}-1} \Upsilon_b(z)
\end{equation}
together with normalization and reflection properties
\begin{equation}
\Upsilon_b\left(\frac{Q}{2}\right) = 1, \qquad \Upsilon_b(Q - z) = \Upsilon_b(z).
\end{equation}
Here 
\begin{equation}
\gamma(z) = \frac{\Gamma(z)}{\Gamma(1-z)}.
\end{equation}
Such a function exists and may be defined by the integral representation:
\begin{equation}
\log \Upsilon_b(z) = \int_0^{\infty} \frac{dt}{t}\left[\left(\frac{Q}{2} - z\right)^2e^{-t} - \frac{\sinh^2\left(\frac{Q}{2}-z\right)\frac{t}{2}}{\sinh\frac{bt}{2}\sinh \frac{t}{2b}}\right],
\end{equation}
which  converges for $0<\Re z<Q$.

$\Upsilon_b(z)$ is an entire function with the following zeroes:
\begin{equation}
\Upsilon_b(z) = 0, \quad \Longleftrightarrow\quad z \in \left(-b \mathbb{Z}_{\geq0}-b^{-1}\mathbb{Z}_{\geq0}\right)\cup\left(Q+b \mathbb{Z}_{\geq0}+b^{-1}\mathbb{Z}_{\geq0} \right).
\end{equation}
A more general shift relation for $r,s\in\mathbb{Z}_{\geq0}$ has the form
\begin{multline}
\Upsilon_b(z+rb+sb^{-1}) = \prod_{k=0}^{r-1} \gamma(zb+kb^2+s) \prod_{l=0}^{s-1}\gamma(zb^{-1}+lb^{-2}) 
\times
\\
\times
b^{r-s-2 rs-2z(rb-sb^{-1})-r(r-1)b^2 + s(s-1)b^{-2}}\Upsilon_b(z).
\end{multline}
Using the shift relation one can prove the following useful formula for $k,l\in\mathbb{Z}_{>0}$:
\begin{equation}\label{UsefulUpsilon}
\frac{\Upsilon(Q/2+z+iP_{k,-l})\Upsilon(Q/2-z+iP_{k,-l})}{\Upsilon(Q/2+z+iP_{k,l})\Upsilon(Q/2-z+iP_{k,l})} = \prod\limits_{p,q}\frac{1}{(iz + P_{p,q})^2},
\end{equation}
where
\begin{equation}\label{UpsilonPQSet}
p\in\{-k+1, -k+3,\ldots,k-3,k-1\}, \quad q\in\{-l+1, -l+3,\ldots,l-3,l-1\}
\end{equation}
$\Upsilon_b(z)$ is related to Barnes double gamma function $\Gamma_b(z)$:
\begin{equation}
\Upsilon_{b}(z)=\frac{1}{\Gamma_{b}(z)\Gamma_{b}(Q-z)}.
\end{equation}
Double argument formula:
\begin{equation}\label{UpsilonDoubleArgument}
\Upsilon_b(2z) = \frac{2^{4z\left(z-\frac{Q}2\right)+1}}{\Upsilon_b\left(\frac{b}{2}\right)\Upsilon_b\left(\frac{b^{-1}}{2}\right)}
\Upsilon_b\left(z\right)
\Upsilon_b\left(z + \frac{b}{2}\right)
\Upsilon_b\left(z + \frac{b^{-1}}{2}\right)
\Upsilon_b\left(z + \frac{Q}{2}\right).
\end{equation}
The following identities also hold:
\begin{equation}\label{eq:Upsilon/sqrt2}
\Upsilon_{\frac{b}{\sqrt{2}}}\left(x\sqrt{2}\right)
= 
\frac{\Upsilon_{\frac{b}{\sqrt{2}}}\left(\frac{b}{\sqrt{2}}\right)}{\Upsilon_b(b)\Upsilon_b\left(\frac{b}{2}\right)}2^{x\left(x-\frac{b}{2}-\frac{1}{b}\right)+\frac12}\Upsilon_b(x)\Upsilon_b\left(x+\frac{b}{2}\right) .
\end{equation}
\begin{equation}\label{eq:Upsilon*sqrt2}
\Upsilon_{b\sqrt{2}}\left(x\sqrt{2}\right)
= 
\frac{\Upsilon_{b\sqrt{2}}\left(\frac{b^{-1}}{\sqrt{2}}\right)}{\Upsilon_b(b^{-1})\Upsilon_b\left(\frac{b^{-1}}{2}\right)}2^{x\left(x-b-\frac{1}{2b}\right)+\frac12}\Upsilon_b(x)\Upsilon_b\left(x+\frac{b^{-1}}{2}\right) .
\end{equation}
Using shift relations one can show that
\begin{equation}
\Upsilon'_b(0) = -\Upsilon'_b(Q) = \Upsilon_b(b),
\end{equation}
and more generally for $r,s\in\mathbb{Z}_{\geq0}$
\begin{multline}
\Upsilon'_b(-rb-sb^{-1}) =  - \Upsilon'_b(Q+rb+sb^{-1}) =
\\
=
\prod_{k=1}^{r} \gamma(1+kb^2+s) \prod_{l=1}^{s}\gamma(1+lb^{-2}) b^{s-r-2rs-r(r+1)b^2 +s(s+1)b^{-2}}\Upsilon_b(b).
\end{multline}
Using the shift relations one can prove the following:
\begin{equation}\label{UsefulDeriv}
\frac{\Upsilon'_b(Q/2+iP_{r,s} + iP_{k,l})\Upsilon'_b(Q/2-iP_{r,s} + iP_{k,l})}
{\Upsilon'_b(Q/2 + iP_{r,s} + iP_{k,-l})\Upsilon'_b(Q/2-iP_{r, s} + iP_{k,-l})}
=
\prod\limits_{p,q}(P_{r,s} + P_{p,q})^2.
\end{equation}
with $r, s\in\mathbb{Z}>0$, $k,l$ in \eqref{FusionKLSet},  $p$, $q$ in \eqref{UpsilonPQSet}.
Also, using the shift relations one can prove that
\begin{multline}\label{UpsilonLogDeriv}
\left.\frac{\partial}{\partial P_1} 
\left[\log \Upsilon_b(P + iP_1)  + \log \Upsilon_b(P - iP_1)\right]
\right|_{P_1 = P_{m,n}}
=\hfill
\\
=
2i(mb-nb^{-1})\log b
-
ib\sum_{k=0}^{m-1} \left[\psi\left(bP+\left(k-\frac{m}{2}\right)b^2+\frac{n}{2}\right) + \psi\left(1-bP-\left(k-\frac{m}{2}\right)b^2-\frac{n}{2}\right)\right]
-
\\
-
ib^{-1}\sum_{l=0}^{n-1}\left[
\psi
\left(Pb^{-1}+\left(l-\frac{n}{2}\right)b^{-2} - \frac{m}{2}\right) 
+ 
\psi
\left(1-Pb^{-1}-\left(l-\frac{n}{2}\right)b^{-2} + \frac{m}{2}\right) 
\right]
\end{multline}

$\Upsilon_b(z)$ has another representation, as an infinite product:
\begin{equation}\label{UpsilonNumerics}
\Upsilon_b(z) = \lambda_b^{\left(\frac{Q}{2}-z\right)^2} \prod_{m,n=0}^{\infty}f\left(\frac{\frac{Q}{2}-z}{\frac{Q}{2}+mb+nb^{-1}}\right)\qquad \text{where}\qquad f(z) = (1-z^2)e^{z^2},
\end{equation}
where $\lambda_b$ depends only on $b$ and, if needed, may be recovered numerically from the integral representation. This representation allows to derive the following approximation \cite{Ribault:2015sxa}:
\begin{multline}
\Upsilon_b\left(\frac{Q}{2}+iP\right) \sim \lambda_b^{-P^2} \prod_{m=0}^M\prod_{n=0}^N \left(1 +\frac{P^2}{y_{m,n}^2} \right)\times
\\
\times
e^{-\sum_{m=0}^M\sum_{n=0}^N\frac{P^2}{y_{m,n^2}}
-Pb\sum_{m=0}^M\psi'\left(\frac{y_{m,N+1/2}}{P}\right)
-Pb^{-1}\sum_{m=0}^M\psi'\left(\frac{y_{M+1/2,n}}{P}\right)
+P^2\psi\left(\frac{y_{M+1/2,N+1/2}}{P}\right)},
\end{multline}
where 
\begin{equation}
\begin{aligned}
\psi''(y) &= \log\left(1 + \frac{1}{y^2}\right) - \frac{1}{y^2},\\
\psi'(y) &= y\log\left(1+\frac{1}{y^2}\right) + \frac{1}{y} + i\log\frac{y+i}{y-i}, \\
\psi(y) &= \frac32 +  \frac12(y^2-1)\log\left(1+\frac{1}{y^2}\right) + iy\log\frac{y+i}{y-i}.
\end{aligned}
\end{equation}

Elliptic theta functions are defined as follows:
\begin{equation}
\theta_2(q) = \sum_{n\in\mathbb{Z}}q^{\left(n+\frac12\right)^2}, \qquad \theta_3(q) = \sum_{n\in\mathbb{Z}}q^{n^2}, \qquad 
\theta_4(q) = \sum_{n\in\mathbb{Z}}(-1)^nq^{n^2}.
\end{equation}
Dedekind eta function is
\begin{equation}
\eta(\tau) = e^{\frac{i\pi\tau}{12}}\prod_{n=1}^{\infty}(1-e^{2i\pi n\tau}).
\end{equation}
They have the following properties:
\begin{equation}
\theta_2(q)^4 + \theta_4(q)^4 = \theta_3(q)^4, \qquad \theta_2(-q)^4 = -\theta_2(q)^4, \quad \theta_3(-q)^4 = \theta_4(q)^4.
\end{equation}
\begin{equation}\label{ProductOfTheta}
\theta_2(q)\theta_3(q)\theta_4(q) = 2\eta(\tau)^3, \quad q = e^{i\pi\tau}.
\end{equation}

\section{Conformal blocks}\label{appendix:Blocks}

Conformal block $\mathfrak{F}$ is a special function which is completely determined by the conformal symmetry. However, it is not known in a closed form. It may be written as a series expansion
\begin{equation}
    \mathfrak{F}^{(b)}(\{P_k\},P|z) = z^{\Delta(P) - \Delta(P_1) - \Delta(P_2)}\left(1 + \frac{(\Delta(P) + \Delta(P_1)- \Delta(P_2))(\Delta(P) + \Delta(P_3)- \Delta(P_4))}{2\Delta(P)}\,z + \ldots\right).
\end{equation}
In \cite{Zamolodchikov:1984eqp} Zamolodchikov discovered an efficient way to compute $\mathfrak{F}$:
\begin{equation}\label{FblocksInTermsOfHBlocks}
	\mathfrak{F}^{(b)}(\{P_k\},P|z)
	=
	(16q)^{P^{2}}z^{-\frac{Q^{2}}{4}-P_{1}^2-P_{2}^2}(1-z)^{-\frac{Q^{2}}{4}-P_{1}^2-P_{3}^2}\theta_{3}(q)^{-Q^{2}-4\sum_{i}P_{i}^2}
	\mathfrak{H}^{(b)}(\{P_k\},P|q),
\end{equation}
where
\begin{equation}
	q = e^{i\pi \tau}, \qquad \tau(z)=i\frac{K(1-z)}{K(z)},\qquad K(z)=\frac{\pi}{2}\,{}_{2}F_{1}\left(\frac{1}{2},\frac{1}{2};1\Big| z\right).
\end{equation}
Functions $\mathfrak{H}$ are called elliptic conformal blocks. For this function Zamolodchikov derived a recursion formula:
\begin{equation}\label{eq:Recursion.H}
\mathfrak{H}^{(b)}(\{P_k\},P|q) = 1 + \sum_{m,n = 1}^{\infty} (16q)^{mn} \frac{R_{m,n}(\{P_k\}) \mathfrak{H}^{(b)}(\{P_k\}, P_{m,-n}|q)}{P^2 - P_{m,n}^2}.
\end{equation}
Here 
\begin{equation}\label{eq:Recursion.R}
R_{m,n}(\{P_k\}) = -\frac12 \prod_{p,q} 
\left(P_1 + P_2 - P_{p,q}\right)
\left(P_1 - P_2 - P_{p,q}\right)
\left(P_3 + P_4 - P_{p,q}\right)
\left(P_3 - P_4 - P_{p,q}\right)
\prod_{k,l}(2P_{k,l})^{-1},
\end{equation}
and
\begin{equation}
\begin{aligned}
&p\in\{-m+1,-m+3, \ldots,m-3,m-1\}, \quad q\in\{-n+1,-n+3, \ldots,n-3,n-1\},\\
&k\in\{-m+1,-m+2, \ldots,m-1,m\}, \quad l\in\{-n+1,-n+2, \ldots,n-1,n\}, \quad (k,l)\notin\{(0,0),\, (m,n)\}.
\end{aligned}
\end{equation}
The recursion kernel $R_{m,n}(\{P_k\})$ has the following property:
\begin{equation}\label{Rproperty}
R_{m,n}(P_1, P_2, P_3, P_4) = (-1)^{mn}R_{m,n}(P_1, P_2, P_4, P_3) = (-1)^{mn}R_{m,n}(P_2, P_1, P_3, P_4). 
\end{equation}
From the recursion representation one may derive an explicit series expansion for $\mathfrak{H}$:
\begin{equation}\label{NumericsHRecursion}
\begin{aligned}
\mathfrak{H}^{(b)}(\{P_k\}, P|q) =1+ \sum_{N = 1}^{\infty} h_N^{(b)}(\{P_k\}, P) (16q)^{N},
\qquad 
h_N^{(b)}(\{P_k\}, P) = \sum_{m,n\colon mn \leqslant N} \frac{H_N^{m,n}(\{P_k\})}{P^2 - P_{m,n}^2},\\
H_N^{m,n}(\{P_k\})= R_{m,n}(\{P_k\})  \Big(\delta_{mn,N}+  \sum_{m',n'\colon m'n' \leqslant N-mn}   \frac{H_{N-mn}^{m',n'}(\{P_k\})}{P_{m,-n}^2 - P_{m',n'}^2} \Big) 
\end{aligned}
\end{equation}
We also note that the $s$-channel conformal blocks transform in a simple way under the transformation
\begin{equation}
z \to \frac{z}{z-1},
\end{equation}
which exchanges $1$ and $\infty$. In terms of the elliptic variables this transformation is $\tau \to \tau+1$ or $q \to -q$. The transformation rule for conformal blocks may be derived using eq-n \eqref{Rproperty}:
\begin{equation}\label{eq:1inftyexchangetransformH}
\mathfrak{H}^{(b)}\left(P_1, P_2, P_3, P_4, P |-q\right)
=
\mathfrak{H}^{(b)}(P_1, P_2, P_4, P_3, P |q)
\end{equation}
\begin{equation}\label{eq:1inftyexchangetransform}
\mathfrak{F}^{(b)}\left(P_1, P_2, P_3, P_4, P \bigg|\frac{z}{z-1}\right)
=
e^{i\pi (\Delta(P) - \Delta(P_1)-\Delta(P_2))}(1-z)^{2\Delta(P_1)}\mathfrak{F}^{(b)}(P_1, P_2, P_4, P_3, P |z)
\end{equation}


\bibliographystyle{MyStyle}
\bibliography{MyBib}

\providecommand{\href}[2]{#2}\begingroup\raggedright\begin{thebibliography}{10}

\bibitem{Collier:2023cyw}
S.~Collier, L.~Eberhardt, B.~M\"uhlmann and V.~A. Rodriguez, \emph{{The
  Virasoro minimal string}},
  \href{https://doi.org/10.21468/SciPostPhys.16.2.057}{\emph{SciPost Phys.}
  {\bfseries 16} (2024) 057}
  [\href{https://arxiv.org/abs/2309.10846}{{\ttfamily 2309.10846}}].

\bibitem{Polyakov:1981rd}
A.~M. Polyakov, \emph{{Quantum geometry of bosonic strings}},
  \href{https://doi.org/10.1016/0370-2693(81)90743-7}{\emph{Phys. Lett.}
  {\bfseries B103} (1981) 207}.

\bibitem{Knizhnik:1988ak}
V.~Knizhnik, A.~M. Polyakov and A.~Zamolodchikov, \emph{{Fractal Structure of
  2D Quantum Gravity}},
  \href{https://doi.org/10.1142/S0217732388000982}{\emph{Mod.Phys.Lett.}
  {\bfseries A3} (1988) 819}.

\bibitem{David:1988hj}
F.~David, \emph{{Conformal Field Theories Coupled to 2D Gravity in the
  Conformal Gauge}},
  \href{https://doi.org/10.1142/S0217732388001975}{\emph{Mod.Phys.Lett.}
  {\bfseries A3} (1988) 1651}.

\bibitem{Distler:1988jt}
J.~Distler and H.~Kawai, \emph{{Conformal Field Theory and 2D Quantum
  Gravity}}, \href{https://doi.org/10.1016/0550-3213(89)90354-4}{\emph{Nucl.
  Phys.} {\bfseries B321} (1989) 509}.

\bibitem{Belavin:1984vu}
A.~A. Belavin, A.~M. Polyakov and A.~B. Zamolodchikov, \emph{{Infinite
  Conformal Symmetry in Two-Dimensional Quantum Field Theory}},
  \href{https://doi.org/10.1016/0550-3213(84)90052-X}{\emph{Nucl. Phys. B}
  {\bfseries 241} (1984) 333}.

\bibitem{DiFrancesco:1993cyw}
P.~Di~Francesco, P.~H. Ginsparg and J.~Zinn-Justin, \emph{{2-D Gravity and
  random matrices}},
  \href{https://doi.org/10.1016/0370-1573(94)00084-G}{\emph{Phys. Rept.}
  {\bfseries 254} (1995) 1}
  [\href{https://arxiv.org/abs/hep-th/9306153}{{\ttfamily hep-th/9306153}}].

\bibitem{Zamolodchikov:2005fy}
{\relax Al}.~B. Zamolodchikov, \emph{{Three-point function in the minimal
  Liouville gravity}},
  \href{https://doi.org/10.1007/s11232-005-0003-3}{\emph{Theor. Math. Phys.}
  {\bfseries 142} (2005) 183}
  [\href{https://arxiv.org/abs/hep-th/0505063}{{\ttfamily hep-th/0505063}}].

\bibitem{Zamolodchikov:2003yb}
{\relax Al}.~B. Zamolodchikov, \emph{{Higher equations of motion in Liouville
  field theory}}, \href{https://doi.org/10.1142/S0217751X04020592}{\emph{Int.
  J. Mod. Phys.} {\bfseries A19S2} (2004) 510}
  [\href{https://arxiv.org/abs/hep-th/0312279}{{\ttfamily hep-th/0312279}}].

\bibitem{Belavin:2005yj}
A.~A. Belavin and A.~B. Zamolodchikov, \emph{{Moduli integrals and ground ring
  in minimal Liouville gravity}},
  \href{https://doi.org/10.1134/1.2045329}{\emph{JETP Lett.} {\bfseries 82}
  (2005) 7}.

\bibitem{Artemev:2022rng}
A.~Artemev and A.~Belavin, \emph{{Five-point correlation numbers in minimal
  Liouville gravity and matrix models}},
  \href{https://doi.org/10.1016/j.nuclphysb.2022.116019}{\emph{Nucl. Phys. B}
  {\bfseries 985} (2022) 116019}
  [\href{https://arxiv.org/abs/2207.01665}{{\ttfamily 2207.01665}}].

\bibitem{Artemev:2022sfi}
A.~Artemev and V.~Belavin, \emph{{Torus one-point correlation numbers in
  minimal Liouville gravity}},
  \href{https://doi.org/10.1007/JHEP02(2023)116}{\emph{JHEP} {\bfseries 02}
  (2023) 116} [\href{https://arxiv.org/abs/2210.14568}{{\ttfamily
  2210.14568}}].

\bibitem{Ribault:2015sxa}
S.~Ribault and R.~Santachiara, \emph{{Liouville theory with a central charge
  less than one}}, \href{https://doi.org/10.1007/JHEP08(2015)109}{\emph{JHEP}
  {\bfseries 08} (2015) 109}
  [\href{https://arxiv.org/abs/1503.02067}{{\ttfamily 1503.02067}}].

\bibitem{Mirzakhani:2006fta}
M.~Mirzakhani, \emph{{Simple geodesics and Weil-Petersson volumes of moduli
  spaces of bordered Riemann surfaces}},
  \href{https://doi.org/10.1007/s00222-006-0013-2}{\emph{Invent. Math.}
  {\bfseries 167} (2006) 179}.

\bibitem{Fateev:2009me}
V.~A. Fateev, A.~V. Litvinov, A.~Neveu and E.~Onofri, \emph{{Differential
  equation for four-point correlation function in Liouville field theory and
  elliptic four-point conformal blocks}},
  \href{https://doi.org/10.1088/1751-8113/42/30/304011}{\emph{J. Phys. A}
  {\bfseries 42} (2009) 304011}
  [\href{https://arxiv.org/abs/0902.1331}{{\ttfamily 0902.1331}}].

\bibitem{Hadasz:2009sw}
L.~Hadasz, Z.~Jaskolski and P.~Suchanek, \emph{{Modular bootstrap in Liouville
  field theory}},
  \href{https://doi.org/10.1016/j.physletb.2010.01.036}{\emph{Phys. Lett. B}
  {\bfseries 685} (2010) 79} [\href{https://arxiv.org/abs/0911.4296}{{\ttfamily
  0911.4296}}].

\bibitem{Ribault:2014hia}
S.~Ribault, \emph{{Conformal field theory on the plane}},
  \href{https://arxiv.org/abs/1406.4290}{{\ttfamily 1406.4290}}.

\bibitem{Collier:2024kmo}
S.~Collier, L.~Eberhardt, B.~M{\"u}hlmann and V.~A. Rodriguez, \emph{{Complex
  Liouville String}}, \href{https://doi.org/10.1103/k74n-s63l}{\emph{Phys. Rev.
  Lett.} {\bfseries 134} (2025) 251602}
  [\href{https://arxiv.org/abs/2409.17246}{{\ttfamily 2409.17246}}].

\bibitem{Collier:2024kwt}
S.~Collier, L.~Eberhardt, B.~M{\"u}hlmann and V.~A. Rodriguez, \emph{{The
  complex Liouville string: the worldsheet}},
  \href{https://doi.org/10.21468/SciPostPhys.19.2.033}{\emph{SciPost Phys.}
  {\bfseries 19} (2025) 033}
  [\href{https://arxiv.org/abs/2409.18759}{{\ttfamily 2409.18759}}].

\bibitem{Zamolodchikov:1995aa}
A.~B. Zamolodchikov and A.~B. Zamolodchikov, \emph{{Structure constants and
  conformal bootstrap in Liouville field theory}},
  \href{https://doi.org/10.1016/0550-3213(96)00351-3}{\emph{Nucl. Phys. B}
  {\bfseries 477} (1996) 577}
  [\href{https://arxiv.org/abs/hep-th/9506136}{{\ttfamily hep-th/9506136}}].

\bibitem{Guillarmou:2020wbo}
C.~Guillarmou, A.~Kupiainen, R.~Rhodes and V.~Vargas, \emph{{Conformal
  bootstrap in Liouville theory}},
  \href{https://doi.org/10.4310/ACTA.2024.v233.n1.a2}{\emph{Acta Math.}
  {\bfseries 233} (2024) 33}
  [\href{https://arxiv.org/abs/2005.11530}{{\ttfamily 2005.11530}}].

\bibitem{Kac:1978ge}
V.~G. Kac, \emph{{Contravariant Form for Infinite Dimensional Lie Algebras and
  Superalgebras}}, \href{https://doi.org/10.1007/3-540-09238-2_102}{\emph{Lect.
  Notes Phys.} {\bfseries 94} (1978) 441}.

\bibitem{Feigin:1981st}
B.~L. Feigin and D.~B. Fuks, \emph{{Invariant skew symmetric differential
  operators on the line and verma modules over the Virasoro algebra}},
  \href{https://doi.org/10.1007/BF01081626}{\emph{Funct. Anal. Appl.}
  {\bfseries 16} (1982) 114}.

\bibitem{Teschner:1995yf}
J.~Teschner, \emph{{On the Liouville three point function}},
  \href{https://doi.org/10.1016/0370-2693(95)01200-A}{\emph{Phys. Lett.}
  {\bfseries B363} (1995) 65}
  [\href{https://arxiv.org/abs/hep-th/9507109}{{\ttfamily hep-th/9507109}}].

\bibitem{Dorn:1992at}
H.~Dorn and H.~J. Otto, \emph{{On correlation functions for noncritical strings
  with $c\leq 1$ $d\geq1$}},
  \href{https://doi.org/10.1016/0370-2693(92)90116-L}{\emph{Phys. Lett.}
  {\bfseries B291} (1992) 39}
  [\href{https://arxiv.org/abs/hep-th/9206053}{{\ttfamily hep-th/9206053}}].

\bibitem{Dorn:1994xn}
H.~Dorn and H.~J. Otto, \emph{{Two and three point functions in Liouville
  theory}}, \href{https://doi.org/10.1016/0550-3213(94)00352-1}{\emph{Nucl.
  Phys.} {\bfseries B429} (1994) 375}
  [\href{https://arxiv.org/abs/hep-th/9403141}{{\ttfamily hep-th/9403141}}].

\bibitem{Zamolodchikov:1985ie}
{\relax Al}.~B. Zamolodchikov, \emph{{Conformal symmetry in two-dimensions: an
  explicit reccurence formula for the conformal partial wave amplitude}},
  \href{https://doi.org/10.1007/BF01214585}{\emph{Comm. Math. Phys.} {\bfseries
  96} (1984) 419}.

\bibitem{Artemev:2025pvk}
A.~Artemev, \emph{{x-y swap for a (2, 2p+1) minimal string}},
  \href{https://doi.org/10.1103/pd2y-j2x5}{\emph{Phys. Rev. D} {\bfseries 112}
  (2025) 046019} [\href{https://arxiv.org/abs/2506.09222}{{\ttfamily
  2506.09222}}].

\bibitem{Belavin:2006ex}
A.~A. Belavin and {\relax Al}.~B. Zamolodchikov, \emph{{Integrals over moduli
  spaces, ground ring, and four-point function in minimal Liouville gravity}},
  {\emph{Theor. Math. Phys.} {\bfseries 147} (2006) 729}
  [\href{https://arxiv.org/abs/hep-th/0510214}{{\ttfamily hep-th/0510214}}].

\bibitem{Feigin:1984}
B.~L. Feigin and D.~B. Fuchs, \emph{{Verma modules over the Virasoro algebra}},
  \href{https://doi.org/10.1007/BFb0099939}{\emph{Lect. Notes in Math}
  {\bfseries 1060} (1984) 230}.

\bibitem{Zamolodchikov:1984eqp}
A.~B. Zamolodchikov, \emph{{Conformal symmetry in two-dimensions: an explicit
  reccurence formula for the conformal partial wave amplitude}},
  \href{https://doi.org/10.1007/BF01214585}{\emph{Commun. Math. Phys.}
  {\bfseries 96} (1984) 419}.

\end{thebibliography}\endgroup
\end{document}